\documentclass[prb,preprintnumbers,amsmath,amssymb]{revtex4}
\usepackage{graphicx,color}
\usepackage{dcolumn}% Align table columns on decimal point
\usepackage{bm}% bold math
\newtheorem{df}{~~~{\bf Definition}}[section]%[sn]
\newtheorem{thm}[df]{~~~{\bf Theorem}}

\newtheorem{lm}[df]{~~~{\bf Lemma}}

\newcommand{\qed}{\raisebox{.6ex}{\fbox{\rule[0.0mm]{0mm}{0.8mm}}}}

\newcommand{\ket}[1]{%
           | #1 \rangle}

\newcommand{\bra}[1]{%
           \langle #1 |}

\newcommand{\braket}[2]{%
           \langle #1 | #2 \rangle}

\newcommand{\ketbra}[2]{%
           | #1 \rangle \langle #2 |}

\begin{document}

\preprint{}

\title{Zero energy resonance and the logarithmically slow decay \\
of unstable multilevel systems}

% \altaffiliation[Also at ]{Physics Department, XYZ University.}%Lines break automatically or
%can be forced with \\
\author{Manabu Miyamoto}%
 \email{miyamo@hep.phys.waseda.ac.jp }
\affiliation{%
Department of Physics, Waseda University, 3-4-1 Okubo, Shinjuku-ku,
Tokyo 169-8555,  Japan
}%

\date{\today}% It is always \today, today,

\begin{abstract}
The long time behavior of the reduced time evolution operator
for unstable multilevel systems is studied based on the N-level Friedrichs model in the presence of a zero energy resonance.
The latter means the divergence of the resolvent at zero energy.
Resorting to the technique developed by Jensen and Kato [Duke Math. J. {\bf 46}, 583 (1979)],
the zero energy resonance of this model is characterized 
by the zero energy eigenstate that does not belong to the Hilbert space.
It is then shown that for some kinds of the rational form factors
the logarithmically slow decay proportional to $(\log t)^{-1}$
of the reduced time evolution operator
can be realized.
\end{abstract}
% limit to 600 characters including the spaces

%\pacs{03.65.Db, 32.80.-t, 42.50.Md, 42.50.Vk}
%Initial states,\ Power decay law \sep
%Multilevel systems,\ Friedrichs model

\maketitle

%%%%%%%%%%%%%%%%%%%%%%%%%%%%%%%%%%%%%%%%%%%%%%%%%%%%%%%%%%%%%%%%%%%%

\section{Introduction} \label{sec:1}

The exponential decay of unstable systems 
has been a well-known %phenomenological
law since the early days of quantum theory.
The quantum description of those systems, however, allows 
deviation from exponential decay both at shorter and longer times \cite{Khalfin(1957)}
than those times over which the exponential decay law dominates. \cite{Fonda(1978),Nakazato(1996)}
The short time deviation was actually found in a quantum tunneling experiment,
\cite{Wilkinson(1997)}
while the long time deviation seems still not to have been detected in any quantum system. 
\cite{Greenland(1988)}
The main cause that hinders the detection 
is considered as the smallness of the deviation at such long times. 
%The main cause hindering the detection of such a long time deviation 
%is regarded as the smallness of 
%the remaining component of the unstable initial state at long times. 
\cite{Delgado(2006)}

In a recent study, a method enhancing the long time deviation was proposed.
\cite{Jittoh(2005)}
%To understanding the method, we should 
The decay of the unstable systems is theoretically modeled in 
the time evolution of the survival probability of unstable initial state. 
The survival probability is just the probability of finding the initial state 
in the state at a later time $t$. 
Since it is rewritten in a Fourier integral of the spectral function, 
its behavior at long times 
is determined by that of the spectral function near the threshold of the energy continuum. 
%and often follows the power decay law. 
\cite{Fonda(1978),Nakazato(1996)} 
The essential aspect of the method is then 
distorting the spectral function from the Breit-Wigner form and dislocating its peak
toward the threshold energy. 
Mathematically, this causes a divergence of the spectral function, i.e., 
the resolvent at the threshold. 
Then, it is expected that the exponential decay period disappears and 
the survival probability at long times is increased. 
A similar idea was also considered in a related context. \cite{Rzazewski(1982),Greenland(1988)}
In addition, in the analysis of the Friedrichs model \cite{Friedrichs(1948),Exner(1985)} 
that is often used for the study on the decays of the unstable systems, 
the survival probability at long times sometimes exhibits a power decay law 
slower than that in cases of no divergence. \cite{Kofman(1994),Lewenstein(2000),Nakazato(2003)}

These facts remind the author 
of the zero energy resonance proposed by Jensen and Kato. \cite{Jensen(1979)}
%However the relation to zero energy resonances is not yet clarified 
%in the previous studies. 
%in Refs. \makebox(10,1){\Large \cite{Kofman(1994),Lewenstein(2000),Nakazato(2003),Jittoh(2005)}}
%\hspace{5mm}.
According to them, %Ref. \makebox(8,1){\Large \cite{Jensen(1979)}}\hspace{1mm},
such zero energy singularities are classified by 
the zero energy eigenstates of the total Hamiltonian
that either belong to or do not belong to the Hilbert space.
%The case where the latter type of the eigenstate exists only
%is called the exceptional case of the first kind.
%That where the former one exists only is the exceptional case of the second kind,
%and that where both of them exist the exceptional case of the third kind.
%Furthermore, the case where no such an eigenstate exists is called the regular case.
%whereas
The cases where such eigenstates exist are called the exceptional cases; 
otherwise they are referred to as the regular case.
%and classified in the three types.
The result in Ref. \makebox(9,1){\Large \cite{Jensen(1979)}}\hspace{1mm}
is concerned with the three-dimensional system of the one particle in short-range potentials, 
and they proved that the time evolution operator asymptotically decreases as 
$O(t^{-1/2})$ for the exceptional cases, 
that is slower than $O(t^{-3/2})$ for the regular case. 
However, to the author's knowledge, 
the zero energy resonance for the Friedrichs model seems not to have been examined 
in the previous studies 
including Refs. \makebox(10,1){\Large 
\cite{Kofman(1994),Lewenstein(2000),Nakazato(2003),Jittoh(2005)}
}\hspace{10mm}, 
in spite of the wide applicability of the model to the various physical systems. 
\cite{Facchi(1998),Antoniou(2001),Kofman(1994),Rzazewski(1982),Nakazato(2003)}

In the present paper, 
we examine the zero energy singularities of the resolvent at the threshold energy 
for the Friedrichs model from the viewpoint of the zero energy resonance, \cite{Jensen(1979)} 
and clarify how the asymptotic behavior of the survival probability at long times is affected. 
The Friedrichs model\cite{Friedrichs(1948),Exner(1985)} 
describes the system of the finite discrete levels coupled with the continuous spectrum, 
in which the former can be interpreted as the unstable excited levels of atoms and the latter as 
the environmental electromagnetic fields. \cite{Kofman(1994),Facchi(1998),Antoniou(2001)}
We emphasize that the model is not restricted to the single level case
\cite{Friedrichs(1948),Exner(1985),Kofman(1994),Lewenstein(2000),Nakazato(2003),Rzazewski(1982),
Jittoh(2005),Facchi(1998),Antoniou(2001)}
but, rather, the $N$-level case, 
\cite{Exner(1985),Davies(1974),Antoniou(2003),Antoniou(2004),Miyamoto(2004),Miyamoto(2005)}
In addition, we assume that the square modulus of the form factors 
%restrict ourselves to a kind of the form factor that 
vanishes at zero energy with an integer power, \cite{Facchi(1998),Antoniou(2001),Seke(1994)}
however it is treated without restriction to a specific form to some extent. 
%
%The result in Ref. \makebox(9,1){\Large \cite{Jensen(1979)}}\hspace{1mm}
%is concerned with the three-dimensional system of the one particle in short-range potentials,
%and we need to modify the technique therein 
%appropriately to our model. %which is actually done in the later parts. 
%
Furthermore, since we only consider the initial state spanned by the discrete states, 
it is sufficient for us to see the reduced resolvent $\tilde{R}(z)$ 
that is just the restriction of the resolvent to the subspace spanned by the discrete states. 
Then, 
the Fourier integral of $\tilde{R}(z)$ that we call the reduced time evolution operator 
$\tilde{U}(t)$ 
enables us to calculate the survival probability. 
In fact 
it is expressed by the square modulus of the expectation value of $\tilde{U}(t)$ 
in a given initial state. 
We first study the zero energy eigenstates of the model 
which either belong to or do not belong to the Hilbert space.
%On the basis of this analysis, 
It is then possible to estimate correctly 
the asymptotic behavior of $\tilde{R}(z)$ at small energies
both in the regular case and the exceptional cases. 
The latter cases are examined in detail only for the first kind,
where only the zero energy eigenstate not belonging to the Hilbert space exists. 
On the basis of this analysis, 
we can derive the long-time asymptotic formula for $\tilde{U}(t)$ in those cases. 
In particular,
the logarithmic decay proportional to $(\log t)^{-1}$ of $\tilde{U}(t)$
is shown to occur in the exceptional case of the first kind for our form factors,
which is extremely slower than the power decays in the regular case
and in the exceptional case for another type of form factor. 
\cite{Kofman(1994),Lewenstein(2000),Nakazato(2003)}
These results are shown in Theorems \ref{thm:rffimLong} and \ref{thm:rffLong1st}.

The organization of the paper is as follows.
We first explain in Sec. \ref{sec:3} the $N$-level Friedrichs model 
with an appropriate Hilbert space, and then 
in Sec. \ref{sec:4} we introduce the reduced resolvent $\tilde{R}(z)$. 
%that plays a crucial role in the analysis of the long time behavior of $\tilde{U}(t)$.
%Here, the several properties of it are also shown.
Section \ref{sec:5} is devoted to the identification of
zero energy eigenstates in this model. %, in the context of Jensen and Kato.
%The case where no such an eigenstate exists is called the regular case,
%whereas
%the cases where such an eigenstate exists are the exceptional cases,
%that are classified in the three types.
%The case where the latter eigenstate only exists
%is called the exceptional case of the first kind,
%that where the former one only exists is the exceptional case of the second kind,
%and that where both of them exists the exceptional case of the third kind.
It is then possible to obtain
the asymptotic expansion of $\tilde{R}(z)$ at small energies
in Sec. \ref{sec:5.5}, where we examine the regular and the exceptional case of the first kind.
By making sure of the relation between $\tilde{R}(z)$ and $\tilde{U}(t)$ in Sec. \ref{sec:6},
the asymptotic formula for $\tilde{U}(t)$ in the regular and the exceptional case
of the first kind
are derived in Sec. \ref{sec:7} . Concluding remarks are given in Sec. \ref{sec:8}.

%%%%%%%%%%%%%%%%%%%%%%%%%%%%%%%%%%%%%%%%%%%%%%%%%%%%%%%%%%%%%%%%%%%%%%%%%%%%%%%
\section{Hilbert space and the $N$-level Friedrichs model}
\label{sec:3}

We shall use bracket notation; however it can be understood in a standard treatment
based on functional analysis as in Refs.
\makebox(6,1){\Large \cite{Exner(1985),Davies(1974)}}\hspace{7.5mm}.
The Hilbert space describing the unstable multilevel systems is here defined by
\begin{equation}
{\cal H}:=\mathbb{C}^N \oplus L^2 ((0,\infty )).
\label{eqn:5.10}
\end{equation}
A vector $\ket{c} \in \mathbb{C}^N$ is expressed by $\ket{c}=\sum_{n=1}^{N} c_n \ket{n}$,
where $\ket{n}$'s are the orthonormal basis of $\mathbb{C}^N$,
so that $\braket{n}{n'}=\delta_{nn'}$, where $\delta_{nn'}$ is Kronecker's delta.
$L^2 ((0,\infty ))$ is the Hilbert space of
the square-integrable complex function $\ket{f}$ of the variable $\omega$ defined on $(0,\infty )$, i.e.,
\begin{equation}
\ket{f} \in L^2 ((0,\infty )) \Leftrightarrow \int_0^{\infty} |f(\omega)|^2 d\omega <\infty.
\label{eqn:5.10b}
\end{equation}
In a standard notation using
the (generalized) eigenstate $\ket{\omega}$ of the multiplication operator by $\omega$,
$\ket{f}$ is nothing more than
\begin{equation}
\ket{f}=\int_{0}^{\infty} f(\omega ) \ket{\omega} d\omega,
\label{eqn:5.15}
\end{equation}
where $\braket{\omega}{\omega'}=\delta (\omega -\omega')$ and
$\delta (\omega -\omega')$ is Dirac's delta.
%However, the present study can be developed without introducing such notations.
Then, an arbitrary vector $\ket{\Psi} \in {\cal H}$
composed of $\ket{c} \in \mathbb{C}^N$ and $\ket{f} \in
L^2 ((0,\infty ))$
is denoted by
\begin{equation}
\ket{\Psi}:=
%(\{ c_n \}_{n=1}^{N}, f), ~~{\rm or}~~
| c \rangle +\ket{f} ,
%+ \int_0^{\infty} f(\omega) | \omega \rangle d\omega ,
\label{eqn:5.20}
\end{equation}
and the inner product between any two vectors $\ket{\Psi}$ and
$\ket{\Phi} \in {\cal H}$ is defined by \cite{innerproduct}
\begin{equation}
\braket{\Phi}{\Psi}
:=\braket{d}{c}+ \braket{g}{f}
=\sum_{n=1}^{N} d_n^* c_n + \int_0^{\infty} g^* (\omega ) f(\omega) d\omega ,
\label{eqn:5.30}
\end{equation}
where ($^*$) denotes the complex conjugate and
$\ket{\Phi}=\ket{d}+\ket{g}$
with $\ket{d} \in \mathbb{C}^N$ and $\ket{g} \in L^2 ((0,\infty ))$.
In particular, the associated norm of $\ket{\Psi}$ is
$\| \Psi \| := \sqrt{\braket{\Psi}{\Psi}}$, 
which is ensured to be finite for all $\ket{\Psi} \in \cal{H}$.

Let us now introduce the $N$-level Friedrichs model
for a description of the decay of the unstable multilevel systems.
The Hamiltonian $H$ of this model is defined by
\begin{equation}
H:=H_0 + \lambda V,
\label{eqn:5.60}
\end{equation}
where
$H_0$ is the free part and $V$
the interaction part of $H$, respectively,
and $\lambda \in \mathbb{R}$ is the coupling constant.
$H_0$ is defined by
\begin{equation}
H_0 :=\sum_{n=1}^N \omega_n \ketbra{n}{n}+\int_{0}^{\infty} \omega \ketbra{\omega}{\omega}d\omega,
\label{eqn:5.40}
\end{equation}
where $\omega_n \in \mathbb{R}$ with $\omega_1 \leq \omega_2 \leq \cdots \leq \omega_N$,
and its action is prescribed by
$H_0 \ket{\Psi}=\sum_{n=1}^N \omega_n c_n \ket{n}+ \omega \ket{f}$
for any $\ket{\Psi}=\ket{c}+\ket{f} \in D(H_0)$.
$D(H_0)$ is the domain of $H_0$ defined by
$
D(H_0 ):=\left\{ \ket{\Psi} \in {\cal H} ~\left|~ %\omega f(\omega ) \in L^2 ((0, \infty )),
\int_{0}^{\infty} |\omega f(\omega )|^2 d\omega < \infty \right. \right\}
$,
and then the self-adjointness of $H_0$ is guaranteed.
The interaction part $V$ is defined by
\begin{equation}
V:=\sum_{n=1}^{N} \int_{0}^{\infty}
\bigl[ v_n^*(\omega) \ketbra{n}{\omega} +v_n(\omega) \ketbra{\omega}{n} \bigr] d\omega,
\label{eqn:5.70}
\end{equation}
where we assumed that $\ket{v_n} \in L^2 ((0, \infty ))$. 
\cite{threshold}
We call the $L^2$-functions $v_n(\omega)$ the form factors of the system under consideration. 
The action of $V$ is then given by
$V \ket{\Psi}=\sum_{n=1}^N \braket{v_n}{f} \ket{n} + \sum_{n=1}^{N} c_n \ket{v_n}$
for any $\ket{\Psi}\in {\cal H}$.
Note that since $D(V)=\cal{H}$ and $V$ is a bounded self-adjoint operator,
$H$ is self-adjoint with the domain $D(H)=D(H_0)\cap D(V)=D(H_0)$.

In the whole of the paper,
we will restrict ourselves to the special kind of the form factor:
Suppose that the product $v_m^* (\omega) v_n (\omega)$
between an arbitrary pair of $v_m^* (\omega)$ and $v_n (\omega)$
is written in a rational function, i.e.,
it is expressed by
\begin{equation}
v_m^* (\omega) v_n (\omega) =\frac{\pi_{mn} (\omega) }{ \rho_{mn} (\omega)},
\label{eqn:formfactor1}
\end{equation}
where $\pi_{mn} (\omega)$ and $\rho_{mn} (\omega)$ are the polynomials 
of the degree $M_{mn}$ and $N_{nm}$, respectively, 
and we assume that $\rho_{mn} (\omega)$ has no zeros in $[0, \infty )$. 
It is also assumed that $M_{mn}+2 \leq N_{mn}$ and $\pi_{mn} (0)=0$. 
The former condition ensures that $v_m^* (\omega) v_n (\omega)$ is integrable in $[0, \infty )$ 
and $\lim_{\omega\to\infty} v_m^* (\omega) v_n (\omega)=0$, 
while the latter condition implies that 
the rational function $v_m^* (\omega) v_n (\omega)=O(\omega)$ as $\omega\to +0$. 
The form factors with such properties are often found in actual systems
involving the process of the spontaneous emission of photons from the hydrogen atom,
\cite{Facchi(1998),Seke(1994)}
and quantum dots. \cite{Antoniou(2001)}
We do not treat the algebraic form factor 
that behaves as $O(\omega^{1/2})$ as $\omega\to +0$ instead, 
associated with 
the photodetachment of electrons from the negative ion
\cite{Rzazewski(1982),Haan(1984),Lewenstein(2000),Nakazato(2003)}
and the spontaneous emission from the atoms in the photonic crystals;\cite{Kofman(1994)} 
however, the discussion developed in the following could be easily extended to such a case.

%%%%%%%%%%%%%%%%%%%%%%%%%%%%%%%%%%%%%%%%%%%%%%%%%%%%%%%%%%%%%%%%%%%%%%%%%%%%%%%
\section{Reduced resolvent for the $N$-level Friedrichs model}
\label{sec:4}

In the following, we introduce the reduced resolvent that is simply the restriction
of the resolvent of $H$ to the $N$ dimensional subspace $\mathbb{C}^N \oplus \{ 0 \}$.
Since we only consider the initial state belonging to this subspace,
this restriction is sufficient for our study.
In a technical sense, this treatment corresponds to the appropriate choice
of a weighted Sobolev space. \cite{Jensen(1979),Murata(1982)}
In the later sections, we do not distinguish the vector in $\mathbb{C}^N$ from
that in $\mathbb{C}^N \oplus \{ 0 \}$.
After introducing the reduced resolvent,
we see the existence of the boundary values of
the reduced resolvent on the positive real line. %in Subsec. \ref{subsec:4.2}.
The large-energy behavior of the reduced resolvent is also examined, %in Subsec. \ref{subsec:4.3},
which is necessary for a rigorous estimation of the long time behavior of
the reduced time evolution operator.

\subsection{Reduced resolvent}
%\label{subsec:}

The resolvent of $H_0$ and that of $H$ are defined by 
$R_0(z ) =(H_0 -z )^{-1}$ and $R(z ) =(H -z )^{-1}$, respectively, 
where we assume that $z \in \mathbb{C}\backslash (\sigma (H_0) \cup \sigma (H))$.
$\sigma (H_0)$ (or $\sigma (H)$) is the spectrum of $H_0$ (or $H$),
i.e., the set of the singular points of $R_0(z)$ (or R(z)).
Then, we have
\begin{eqnarray}
R(z ) - R_0(z )
&=& -R_0(z )  V R(z )
\label{eqn:2.70} \\
&=& -R_0(z ) V R_0(z ) +R_0(z ) V R_0(z )  V R(z ) .
\label{eqn:2.40}
\end{eqnarray}
From Eq. (\ref{eqn:2.70}), one obtains the equation $R(z)= (1+R_0(z ) V )^{-1} R_0(z )$,
which is the starting point of the asymptotic expansion of $R(z)$
for the short-range potential systems. \cite{Jensen(1979)}
On the other hand, we instead start from Eq. (\ref{eqn:2.40}) to obtain
\begin{equation}
[H_0 -z - V R_0(z )  V ] R(z ) = 1 -V R_0(z ) .
\label{eqn:2.110}
\end{equation}
This equation can be solved for our model if 
we confine ourselves to the state subspace $\mathbb{C}^N \oplus \{ 0 \}$.
\cite{Exner(1985)}
In fact, from the fact that $\bra{n} V R_0(z ) \ket{n'} =0$
for any $\ket{n}$ and $\ket{n'} \in \mathbb{C}^N \oplus \{ 0 \}$,
Eq. (\ref{eqn:2.110}) reads
\begin{equation}
\sum_{m=1}^N
[(\omega_{n} -z ) \delta_{n m} -\lambda^2 S_{n m}(z) ]
\tilde{R}_{m n'}(z )
=\delta_{nn'} ,
\label{eqn:.20}
\end{equation}
where $S (z ) $ and $\tilde{R} (z ) $ are
the $N \times N$ matrix defined
with the matrix components
\begin{equation}
S_{mn}  (z )
:=
\bra{m} V R_0(z ) V \ket{n}
=
\int_{0}^{\infty}
\frac{v_m^* (\omega) v_n (\omega)}{\omega -z}d \omega,
\mbox{ and }
\tilde{R}_{mn}  (z ):=\bra{m} R(z ) \ket{n}.
\label{eqn:.40}
\end{equation}
We call $S(z) $ and $\tilde{R}(z)$ the {\it self energy} and the {\it reduced resolvent}, 
respectively.
Note that $S(z)$ can be analytically defined for all $z \in \mathbb{C}\backslash [0, \infty)$.
For a later convenience, we also introduce the matrix ${K_0}$ and $K(z)$ by
\begin{equation}
{K_0}_{mn}:=\bra{m} H_0 \ket{n}=\omega_n  \delta_{m n},
\quad \mbox{and} \quad
K_{mn}(z):=[K_0 -\lambda^2 S(z)]_{mn},
\label{eqn:.55}
\end{equation}
respectively. Then, Eq. (\ref{eqn:.20}) is equivalent to
\begin{equation}
[K(z)-z]
\tilde{R} (z ) =1, ~~~
\forall z \in \mathbb{C}\backslash (\sigma (H_0) \cup \sigma (H)),
\label{eqn:.60}
\end{equation}
which implies that
${\rm det}[K(z)-z]
{\rm det} [\tilde{R} (z ) ] =1 $,
so that
${\rm det}[K(z)-z] \neq 0$ and
${\rm det} [\tilde{R} (z ) ] \neq 0 $
for all $z \in \mathbb{C}\backslash (\sigma (H_0) \cup \sigma (H))$.
%
%because both ${\rm det}[K(z)-z]$
%and ${\rm det} [\tilde{R} (z ) ] $ are definite
%for all $z \in \mathbb{C}\backslash [\sigma (H_0) \cup \sigma (H)]$.
%
Thus, the inverse of $K(z)-z$
exists, and we have
\begin{equation}
\tilde{R} (z )
=
[K(z)-z]^{-1}, ~~~
\forall z \in \mathbb{C}\backslash (\sigma (H_0) \cup \sigma (H)).
\label{eqn:.70}
\end{equation}

\subsection{The boundary values of $\tilde{R} (z )$ and its large energy behavior}
\label{subsec:4.2}

From the assumption on the form factors, every $v_m^* (\omega )v_n (\omega )$ is 
continued to the whole complex plane as a meromorphic function which we merely denote 
as $v_m^* (z)v_n (z)$. 
It may have a finite number of poles.
Then, it follows from Lemma \ref{lm:formfactor1} that
${S} (z)$ can be reduced to the form
\begin{equation}
{S} (z)
=
S(0) + A(z) -(\log (-z)) {\mit \Gamma}(z) ,
\label{eqn:.76}
\end{equation}
where we choose %$-z=|z|e^{i(\theta-\pi)}$ when $z=|z|e^{i\theta}$.
${\rm arg} (-z) ={\rm arg} (z)-\pi$ and $0<{\rm arg} (z)<2\pi$.
The matrix ${\mit \Gamma}(z)$ is defined with the components
\begin{equation}
{\mit \Gamma}_{mn} (z):=v_m^* (z)v_n (z),
\label{eqn:.100}
\end{equation}
and satisfies ${\mit \Gamma}(z)\to 0$ as $z\to 0$ in $\mathbb{C}$.
$S(0)$ is the limit of $S(z)$ as $z\to 0$ in $\mathbb{C}\backslash [0, \infty)$,
which turns out to be unique. Indeed, as we see 
from the Appendix in Ref. \makebox(0,1){\Large \cite{Miyamoto(2005)}}\hspace{4mm},
$S_{mn}(0)=\int_{0}^{\infty}v_m^* (\omega) v_n (\omega)/\omega  \ d \omega$.
$A(z)$ is then defined through Eq. (\ref{eqn:.76}) and becomes
a Hermitian matrix for real $\omega$,
whose components are the rational functions of $z$
without any singularity on $[0, \infty )$.
By definition, $A(z)$ satisfies $A(z)\to 0$ as $z\to 0$.
One sees that the boundary values of $S (z )$
at the half line $(0,\infty )$ exist and satisfy \cite{Exner(1985)}
\begin{equation}
\lim_{\epsilon \to +0}
S (\omega \pm i\epsilon)
=
{D} (\omega ) \pm \pi i {\mit \Gamma}(\omega ),
\label{eqn:.80a}
\end{equation}
where
\begin{equation}
{D} (\omega )
:=
S(0) + A(\omega) -(\log \omega) {\mit \Gamma}(\omega) .
\label{eqn:.80b}
\end{equation}
The matrix ${D} (\omega )$ is just
of the components
\begin{equation}
{D}_{mn} (\omega ):=P \int_{0}^{\infty}
\frac{v_m^* (\omega' )v_n (\omega' )}{\omega'-\omega} d\omega' ,
\label{eqn:.90}
\end{equation}
where $P$ denotes the principal value of the integral.
Note that both ${D} (\omega )$ and ${\mit \Gamma}(\omega )$ are Hermitian
matrices and ${\mit \Gamma}(\omega ) \geq 0$.

In all the discussion developed in the following, we assume that
\begin{equation}
{\rm det}[K^\pm (\omega)-\omega]\neq 0, ~~~ \forall \omega >0,
\label{eqn:.110}
\end{equation}
where we introduced
\begin{equation}
K^\pm (\omega):= \lim_{\epsilon \to +0} K(\omega \pm i\epsilon )
={K_0} -\lambda^2 {D} (\omega )\mp \lambda^2 \pi i {\mit \Gamma}(\omega ),
 ~~~ \forall \omega >0.
\label{eqn:.112}
\end{equation}
It is worth noting that condition (\ref{eqn:.110}) is
equivalent to the requirement of no positive eigenvalues of $H_0$,
whose eigenstates are normalizable.
Indeed, if
${\rm det}[K^\pm (\omega)-\omega]=0
$ for some $\omega >0$, there is a non-zero vector
$\ket{\eta}=\sum_{n=1}^N \eta_n \ket{n} \in \mathbb{C}^N$ such that
$[K^\pm (\omega)-\omega] \ket{\eta}=0 $.
Since both ${D} (\omega )$ and ${\mit \Gamma}(\omega )$ are Hermitian matrices,
the latter equation implies that
\begin{equation}
\bra{\eta} [{K_0}-\omega -\lambda^2 {D} (\omega )]
\ket{\eta} =0 ~~
\mbox{and} ~~
\bra{\eta} {\mit \Gamma}(\omega ) \ket{\eta}
=\left|\sum_{n=1}^{N} v_n(\omega)\eta_n \right|^2=0 .
\label{eqn:.167}
\end{equation}
Note that the latter relation means that
${\mit \Gamma}(\omega ) \ket{\eta}=0$ because ${\mit \Gamma}(\omega ) \geq 0$.
Thus, Eq. (\ref{eqn:.167}) implies that
${\mit \Gamma}(\omega ) \ket{\eta} =0$ and
$[{K_0} -\lambda^2 {D} (\omega )]\ket{\eta}=\omega \ket{\eta}$, i.e.,
\begin{equation}
\sum_{n=1}^{N} v_n(\omega)\eta_n =0,
~~ \mbox{and} ~~
\sum_{n=1}^{N}
[\omega_{m}\delta_{mn} -\lambda^2 {D}_{mn} (\omega )]
\eta_n =\omega \eta_m ,
\label{eqn:.169}
\end{equation}
for all $m=1, \ldots, N$.
This is merely the condition for the existence of a positive
eigenvalue $\omega$ of $H$. \cite{Miyamoto(2005)}

\begin{lm}%[Boundary values]
\label{lm:100}
Under the assumption (\ref{eqn:.110}), it holds that
$\tilde{R}^{\pm}(\omega)
:=\lim_{\epsilon \to +0} \tilde{R}(\omega \pm i \epsilon )
$
exists for all $\omega >0$ and
$\tilde{R}^{\pm}(\omega) =[K^\pm (\omega)-\omega]^{-1}$.
\end{lm}

{\sl Proof} :
Under the assumption (\ref{eqn:.110}),
$[K^\pm (\omega)-\omega]^{-1} $ exists. Then %for any $\ket{y} \in \mathbb{C}^N$
\begin{eqnarray}
&&
\|
[K^\pm (\omega)-\omega]^{-1} -\tilde{R}(\omega \pm i \epsilon )
\|
\nonumber \\
&\leq&
\|
[K^\pm (\omega)-\omega]^{-1}
\|
\|
\pm i\epsilon +\lambda^2 S (\omega \pm i \epsilon)
-\lambda^2 {D} (\omega )
\mp \lambda^2 \pi i {\mit \Gamma}(\omega )
\|
\|
\tilde{R}(\omega \pm i \epsilon )
\| .
\label{eqn:.120}
\end{eqnarray}
Note that for any nonzero $\ket{y} \in \mathbb{C}^N$ ($\neq 0$)
there is a nonzero $\ket{x} \in \mathbb{C}^N$ such that
$\ket{y}=[K(\omega \pm i \epsilon )-\omega \mp i \epsilon]\ket{x}$.
We then obtain
\begin{eqnarray}
\frac{
\|
\tilde{R}(\omega \pm i \epsilon )\ket{y}
\|
}
{\| y\|}
%&=&
%\frac{\| x \|}{\|[K(\omega \pm i \epsilon )-\omega \mp i \epsilon]\ket{x} \| }
%\nonumber \\
&\leq&
\frac{\| x \|}
{
\bigl|
\|
[K^\pm (\omega)-\omega] \ket{x}
\|
-
\|
[\pm i\epsilon +\lambda^2 S (\omega \pm i \epsilon)
-\lambda^2 {D} (\omega )
\mp \lambda^2 \pi i {\mit \Gamma}(\omega )]\ket{x}
\|
\bigr|
}
\nonumber \\
\!&\leq&\!
\biggl[
\inf_{\ket{x} \neq 0, \ket{x}\in \mathbb{C}^N}\! \!
\frac{
\|[K^\pm (\omega)-\omega]\|
}{\| x \|}
\nonumber\\
&&
-
\|
\pm i\epsilon +\lambda^2 S (\omega \pm i \epsilon)
-\lambda^2 {D} (\omega )
\mp \lambda^2 \pi i {\mit \Gamma}(\omega )
\|
\biggr]^{-1} \hspace{-2mm},
%\nonumber \\
%&&
\label{eqn:.130}
\end{eqnarray}
which implies that
\begin{equation}
%\| \tilde{R}(\omega \pm i \epsilon ) \|
%\leq
\mathop{\overline{\lim}}_{\epsilon \to +0}
\|
\tilde{R}(\omega \pm i \epsilon )
\|
%=
%\mathop{\overline{\lim}}_{\epsilon \to +0}
%\sup_{\ket{y} (\neq 0)}
%\frac{
%\|
%\tilde{R}(\omega \pm i \epsilon )\ket{y}
%\|
%}
%{\| y\|}
\leq
\left[
\inf_{\ket{x} \neq 0, \ket{x}\in \mathbb{C}^N}
\frac{
\|K^\pm (\omega)-\omega\|
}{\| x \|}
\right]^{-1} < \infty ,
\label{eqn:.140}
\end{equation}
where
the norm of an $N\times N$ matrix $A$ is defined by
$\| A\|=
\sup_{\ket{x} \neq 0, \ket{x}\in \mathbb{C}^N} \| A\ket{x}\| /\| x\|$.
In Eq. (\ref{eqn:.130}), we used the fact that
there is some $\epsilon_0 >0$ such that
for any positive $\epsilon < \epsilon_0$
and for any non zero $\ket{x} \in \mathbb{C}^N$
\begin{eqnarray}
\frac{
\|
[K^\pm (\omega)-\omega] \ket{x}\|
}{\| x \|}
&\geq&
\inf_{\ket{x} \neq 0, \ket{x}\in \mathbb{C}^N}
\frac{
\|[K^\pm (\omega)-\omega] \ket{x}\|
}{\| x \|}
\nonumber \\
&>&
\|
\pm i\epsilon +\lambda^2 S (\omega \pm i \epsilon)
-\lambda^2 {D} (\omega )
\mp \lambda^2 \pi i {\mit \Gamma}(\omega )
\|
\nonumber \\
&\geq&
\frac{
\|
[\pm i\epsilon +\lambda^2 S (\omega \pm i \epsilon)
-\lambda^2 {D} (\omega )
\mp \lambda^2 \pi i {\mit \Gamma}(\omega )
]\ket{x}
\|
}{\| x \|},
\label{eqn:.150}
\end{eqnarray}
where the assumption (\ref{eqn:.110}) is taken into account.
Thus, by using Eq. (\ref{eqn:.140}), Eq. (\ref{eqn:.120}) leads us to
\begin{equation}
\lim_{\epsilon \to +0}
\|
[K^\pm (\omega)-\omega]^{-1}
-\tilde{R}(\omega \pm i \epsilon )
\|
=0,
\label{eqn:.160}
\end{equation}
which completes the proof of the lemma.
\qed

%\subsection{Large energy behavior of $\tilde{R}^\pm (\omega) $}
%\label{subsec:4.3}

\begin{lm}%[Jensen and Kato, Lemma 10.1]
\label{lm:Large-omega}
:
Under the assumption (\ref{eqn:.110}),
$\tilde{R}^\pm (\omega)$ is $r$-times differentiable in
$\omega \in (0, \infty)$, and it behaves as
\begin{equation}
\frac{d^r \tilde{R}^\pm (\omega)}{d\omega^r}
=O(\omega^{-r-1}) \mbox{ as } \omega \to \infty .
\label{eqn:Large10}
\end{equation}
\end{lm}

{\sl Proof} :
We first show the statement for $r=0$.
From the assumption on the form factors and Lemma \ref{lm:formfactor1}, one sees that
\begin{equation}
\lim_{\omega \to \infty} {D} (\omega) =0 \mbox{ and }
\lim_{\omega \to \infty} {\mit \Gamma} (\omega) =0.
\label{eqn:Large40}
\end{equation}
Since from the assumption (\ref{eqn:.110})
$K^\pm(\omega)-\omega$ is invertible
for all $\omega >0$,
it holds that there is some positive $\bar{\omega}  > \omega_N$
such that for any $\omega > \bar{\omega}$,
%$\omega -\omega_N -\lambda^2 \| {D} (\omega) \pm \pi i {\mit \Gamma}(\omega) \| >0$, so that
\begin{eqnarray}
\frac{
\|
\tilde{R}^\pm (\omega)\ket{y}
\|
}
{\| y\|}
&\leq&
\frac{\| x \|}
{\displaystyle
%\bigl|
\|
({K_0}-\omega)\ket{x}
\|
-\lambda^2
\|
[{D} (\omega) \pm \pi i {\mit \Gamma}(\omega)]\ket{x}
\|
%\bigr|
}
\label{eqn:Large30a}\\
&\leq&
\frac{1}
{
\omega -\omega_N -\lambda^2
\|
{D} (\omega) \pm \pi i {\mit \Gamma}(\omega)
\|
}
=
O(\omega^{-1}),
\label{eqn:Large30}
\end{eqnarray}
where %we used that $\min_n \{ \omega-\omega_n \} =\omega-\omega_N $.
the last inequality is obtained as follows:
we can choose some positive $\bar{\omega}  > \omega_N$
such that for any $\omega > \bar{\omega}
$\begin{equation}
\frac{
\|
[{K_0}-\omega]\ket{x}
\|
}{\| x \|}
\geq
\min_n \{ \omega-\omega_n \} =\omega-\omega_N
>
\lambda^2 \|{D} (\omega) \pm \pi i {\mit \Gamma}(\omega) \|
\geq
\lambda^2
\frac{
\|
[{D} (\omega)
\pm \lambda^2 \pi i {\mit \Gamma}(\omega)
]\ket{x}
\|
}{\| x \|} ,
\label{eqn:Large50}
\end{equation}
where Eq. (\ref{eqn:Large40}) was used.
Thus Eq. (\ref{eqn:Large30}) reads just as Eq. (\ref{eqn:Large10}) does for $r=0$.
In the case of $r\geq 1$, we first note that from our assumptions on the form factors and
Lemma \ref{lm:formfactor1} again, $A(\omega ) $ and ${\mit \Gamma}(\omega ) $,
which are connected through ${D} (\omega )=S_0 + A(\omega ) -\log \omega {\mit \Gamma}(\omega )$,
also satisfy
\begin{equation}
\frac{d^r A(\omega )}{d\omega^r }= O(\omega^{-1-r}), ~~
\frac{d^r \log \omega {\mit \Gamma}(\omega )}{d\omega^r }= O(\omega^{-1-r} \log \omega),
\label{eqn:Large80}
\end{equation}
as $\omega \to \infty$, where we used the estimation that
$\frac{d^r {\mit \Gamma}(\omega )}{d\omega^r }= O(\omega^{-1-r})$.
Thus, for $r=1$, we have
\begin{equation}
\frac{d \tilde{R}^{\pm}(\omega )}{d\omega}
=
\tilde{R}^{\pm}(\omega )
\frac{d}{d\omega}
\left[
\omega + \lambda^2 {D}(\omega )
\pm \lambda^2 \pi i {\mit \Gamma}(\omega )
\right]
\tilde{R}^{\pm}(\omega )
%
%\label{eqn:Large90a} \\
%&=&
%\tilde{R}^{\pm}(\omega )
%\left[
%1 + \lambda^2 \frac{d {D}(\omega )}{d\omega}
%\pm \lambda^2 \pi i \frac{d {\mit \Gamma}(\omega )}{d\omega}
%\right]
%\tilde{R}^{\pm}(\omega )
%
= O(\omega^{-2}),
\label{eqn:Large90b}
\end{equation}
as $\omega \to \infty$, where Eq. (\ref{eqn:Large10}) for $r=0$ was used.
For $r\geq1$, we obtain
\begin{equation}
\frac{d^r \tilde{R}^{\pm}(\omega )}{d\omega^r}
=
\sum_{j=1}^{r}
{\sum_{\{s_i \}_{i=1}^{j} }}' %\ (\sum_{i=1}^{j} s_i =s)}
a^{(r)} (\{s_i \}_{i=1}^{j} )
\left\{
\prod_{i=1}^{j}
\tilde{R}^{\pm}(\omega )
\frac{d^{s_i}}{d\omega^{s_i}}
\left[
\omega + \lambda^2 {D}(\omega )
\pm \lambda^2 \pi i {\mit \Gamma}(\omega )
\right]
\right\}
\tilde{R}^{\pm}(\omega ),
\label{eqn:Large100}
\end{equation}
%as $\omega \to \infty$.
where $a^{(r)} (\{s_i \}_{i=1}^{j} ) $ is an appropriate positive integer.
Note that the symbol ($\ '$) means that
the summation over $\{s_i \}_{i=1}^{j}$ is taken under the condition that
$s_i \geq 1$ for all $i$ %is distributed over $1, 2, \ldots$,
%a positive integer
and $\sum_{i=1}^{j} s_i =r$.
If $r=1$, Eq. (\ref{eqn:Large100}) reproduces Eq. (\ref{eqn:Large90b})
with $a^{(1)} (\{s_i \}_{i=1}^{1} ) =1$.
In the general case, if Eq. (\ref{eqn:Large100}) holds
for $r=k$, then its derivative is made up of a linear combination of
\begin{equation}
\left\{
\prod_{i=1}^{j+1}
\tilde{R}^{\pm}(\omega )
\frac{d^{s_i}}{d\omega^{s_i}}
\left[
\omega + \lambda^2 {D}(\omega )
\pm \lambda^2 \pi i {\mit \Gamma}(\omega )
\right]
\right\} \tilde{R}^{\pm}(\omega ),
%\mbox{ with }  \sum_{i=1}^{j+1} s_i =k+1 ,
\label{eqn:Large110}
\end{equation}
where $\sum_{i=1}^{j+1} s_i =k+1$ for $1 \leq j \leq k$, and
\begin{equation}
\left\{
\prod_{i=1}^{j}
\tilde{R}^{\pm}(\omega )
\frac{d^{s_i}}{d\omega^{s_i}}
\left[
\omega + \lambda^2 {D}(\omega )
\pm \lambda^2 \pi i {\mit \Gamma}(\omega )
\right]
\right\} \tilde{R}^{\pm}(\omega ),
%\mbox{ with }  \sum_{i=1}^{j} s_i =k+1 ,
%= O(\omega^{-2}).
\label{eqn:Large120}
\end{equation}
where $\sum_{i=1}^{j} s_i =k+1$ for $1 \leq j \leq k$.
On the other hand, they are actually included in the right-hand side (rhs) of
Eq. (\ref{eqn:Large100}) for $r=k+1$. Thus Eq. (\ref{eqn:Large100}) is
valid for all integer $r \geq 1$.
Let us now evaluate the asymptotic behavior of
$d^r \tilde{R}^{\pm}(\omega )/d\omega^r$ for large $\omega$.
One can see that
the summand for $j=r$ in Eq. (\ref{eqn:Large100}),
where all $s_i =1$, contributes
$O(\omega^{-r-1})$ to $d^r \tilde{R}^{\pm}(\omega )/d\omega^r$,
while the other summands for $j<r$ specified by $\{s_i \}_{i=1}^{j}$
contribute $O(\omega^{-r-1-2s_0 }(\log \omega )^{s_0}))$ at most,
where $s_0$ is a number of $s_i$ satisfying $s_i \geq 2$ and never vanishes for $j<r$.
Therefore, the summand dominating for large $\omega$ is
that for $j=r$. Since we recursively show
$a^{(r)} (\{s_1 \}_{i=1}^{r} ) =r!$, which never vanishes, the statement is proved. 
\qed

%%%%%%%%%%%%%%%%%%%%%%%%%%%%%%%%%%%%%%%%%%%%%%%%%%%%%%%%%%%%%%%%%%%%%%%%%%%%%%%%%%%%%%%%%%

\section{Classification of the zero-energy singularity of $\tilde{R}^\pm (\omega ) $}
\label{sec:5}

%\subsection{Classification of the singularity of $\tilde{R} (z ) $ at $z =0$}
%\label{subsec:}

In order to prescribe the zero energy resonance in the $N$-level Friedrichs model,
we should identify the zero energy eigenstates in this model
which either belong to or do not belong to ${\cal H}$.
In the case of the short-range potential systems, \cite{Jensen(1979)}
this task needs some elaborate examination with an appropriately extended Hilbert space.
On the other hand, in our case,
it is rather easily performed, as is seen in the following.

Let us first
see whether the eigenvector $\ket{\psi} \in \mathbb{C}^N$ of ${K_0} -\lambda^2 S(0)$,
belonging to the zero eigenvalue,
can be actually extended to the eigenvector of $H$ belonging to the zero eigenvalue of $H$.
If $\ket{\Psi}=\ket{\psi}+\ket{f} \in D(H) \subset {\cal H}$ is a zero eigenvector of $H$,
it should satisfy $H\ket{\Psi}=0$, or equivalently \cite{Miyamoto(2005)}
\begin{equation}
\omega_n \psi_n +\lambda \braket{v_n}{f} =0 ~\mbox{for}~ n=1, \ldots, N,~~~\mbox{and}~~~
\omega f(\omega)+ \lambda \sum_{n=1}^{N}\psi_n v_n (\omega)=0.
\label{eqn:.333}
\end{equation}
The latter equation of Eq. (\ref{eqn:.333}) is immediately solved as
\begin{equation}
f(\omega)=-\lambda \frac{\sum_{n=1}^{N}\psi_n v_n (\omega)}{\omega},
\label{eqn:.335}
\end{equation}
which should be square integrable
because we intend to find $\ket{\Psi}$ in ${\cal H}$.
If this is the case, $\omega f(\omega) \in L^2 ((0, \infty))$, i.e.,
$\ket{\Psi} \in D(H)$ is ensured,
and the substitution of Eq. (\ref{eqn:.335}) into $\braket{v_n}{f}$ is safely done.
Then,
we find that the former equation of Eq. (\ref{eqn:.333}) is nothing more than
\begin{equation}
({K_0} -\lambda^2 S(0) )\ket{\psi}=K(0)\ket{\psi}=0,
\label{eqn:.336}
\end{equation}
where $K(0):=K^\pm (0)={K_0} -\lambda^2 S(0)$.
However, it is noted that such an $f(\omega)$ associated with $\ket{\psi}$ is not necessarily
square integrable. %and it needs to be examined in detail.
Hence, we shall decompose the zero eigenspace of $K(0)$, denoted by
$M=\{ \ket{\psi} \in \mathbb{C}^N |~ K(0)\ket{\psi}=0 \}$,
into two kinds of subspaces:
$M_1=(M_0 \oplus M_2)^{\perp}$ 
and $M_2 =\{ \ket{\psi} \in M | f(\omega) \in L^2 ((0, \infty)) \}$.
Here $M_0=M^{\perp}$, and
$D^{\perp}$ denotes the orthogonal complement of the subspace $D$.
In short we have $\mathbb{C}^N = M_0 \oplus M_1 \oplus M_2$.
Then, as is expected from the definition, we have
\begin{equation}
M_1 \subset \{ \ket{\psi} \in M | f(\omega) \notin L^2 ((0, \infty)) \} .
\label{eqn:.325}
\end{equation}
Note that in general the subset on the rhs of the above is not a subspace.
We call $0$ the {\it zero energy resonance} (or merely {\it zero resonance}) 
of $H$ if $M_1$ is not empty.
We also introduce the projection operators $Q_0$, $Q_1$, and $Q_2$, associated with
$M_0$, $M_1$, and $M_2$, respectively.
What we next do is to introduce the terminology 
following the study of Jensen and Kato. \cite{Jensen(1979)}

\begin{df}%[Regular case]
\label{df:regular}
{\rm
We call the system a {\it regular} case if it holds that
$0 \notin \sigma (K(0))$, i.e.,
\begin{equation}
{\rm det}[K(0)]
\neq 0.
\label{eqn:.300}
\end{equation}
In this case, $0$ is said to be a {\it regular} point for $H$.
}
\end{df}

\begin{df}%[Exceptional cases]
\label{df:exceptional}
{\rm 
We call the system the {\sl exceptional} case if, instead of Eq. (\ref{eqn:.300}), it holds that
$0 \in \sigma (K(0))$, i.e.,
\begin{equation}
{\rm det}[K(0)]
= 0.
\label{eqn:.330}
\end{equation}
In particular, if $0$ is a resonance but not an eigenvalue
($Q_1 \neq 0$, $Q_2 =0$), $0$ is said to be an {\it exceptional} point for $H$
of the {\sl first kind}.
If $0$ is not a resonance, but an eigenvalue ($Q_1 = 0$, $Q_2 \neq 0$),
$0$ is said to be an exceptional point
of the {\sl second kind}.
If $0$ is both a resonance and an eigenvalue ($Q_1 \neq 0$, $Q_2 \neq 0$),
$0$ is said to be an exceptional point
of the {\it third kind}.
}
\end{df}

We here remark that in general a non-trivial solution of Eq. (\ref{eqn:.336}) does not exist, 
however we can find a special case where such a solution surely exists. 
Suppose that $N_+$ eigenvalues $\omega_n$ of $H_0$ are positive, and 
all form factors $v_n (\omega)$ satisfying the assumption (\ref{eqn:formfactor1}) 
are linearly independent. 
Then increasing $\lambda$ gradually form $0$ to $\infty$, 
we can find some critical values of $\lambda$ 
for which $K(0)$ has the zero eigenvalue. 
Let us denote the $n$-th eigenvalue of $K(0)$ by $\kappa_n (0)$ where 
$\kappa_1 (0)\leq \kappa_2 (0) \leq \cdots \leq \kappa_N (0)$. 
Then, $\kappa_n (0)$ turns out to satisfy the inequality
\begin{equation}
\omega_n -\lambda^2 \sigma_N (0) \leq \kappa_n (0) \leq \omega_n -\lambda^2 \sigma_1 (0),
\label{eqn:zeroeigenvalue}
\end{equation}
where both of $\sigma_1 (0)$ and $\sigma_N (0)$ 
are positive constants and ensured not to vanish.\cite{Miyamoto(2005)} 
Thus, for a sufficiently small $|\lambda|$ 
$\kappa_n (0)$ for each $n\geq N-N_+ +1$ should be positive, 
while for a sufficiently large $|\lambda|$ they should be negative. 
Furthermore, one easily sees that all $\kappa_n (0)$ are continuous functions of $\lambda^2$. 
Therefore, we conclude from the intermediate value theorem 
that there is at least one critical value of $\lambda$ to make 
$\kappa_n (0)=0$ for each $n\geq N-N_+ +1$. 
We can actually find such special values of $\lambda$ in Fig. 1 
depicted in Ref. \makebox(8,1){\Large \cite{Miyamoto(2005)}}\hspace{1mm}. 
The example mentioned here could be treated in a more general way 
with resort to the analytic Fredholm theorem \cite{the analytic Fredholm theorem,Rauch(1978)}
which tells us at most a finite number of the critical values exists.

It is also worth remarking that 
the existence of the zero energy eigenstates that either belong to or not to 
the Hilbert space necessarily prescribes the small energy behavior of 
the form factors in the following way. 
Remember that under the assumption on the form factors, 
${\mit \Gamma}(\omega )$ defined by Eq. (\ref{eqn:.100}) has an asymptotic form like
\begin{equation}
{\mit \Gamma}(\omega ) = \sum_{n=1}^{N} \omega^n {\mit \Gamma}_n +O(\omega^{N+1}),
\label{eqn:formfactor2}
\end{equation}
as $\omega\to +0$. 
Then, if $\ket{\psi} \in M_1$ exists, it should satisfy 
\begin{equation}
\bra{\psi} {\mit \Gamma}_1 \ket{\psi} \neq 0. 
\label{eqn:.340}
\end{equation}
In fact, if $\bra{\psi} {\mit \Gamma}_1 \ket{\psi} = 0$, we see that 
$f(\omega)$ in Eq. (\ref{eqn:.335}) has to satisfy 
$|f(\omega)|^2=\lambda^2\bra{\psi}{\mit \Gamma}(\omega)\ket{\psi}/\omega^2=O(1)$ 
as $\omega \to +0$, however 
which concludes that $f(\omega)$ is square integrable. 
This contradicts the assumption that $\ket{\psi} \in M_1$. 
In order to make the condition (\ref{eqn:.340}) be satisfied, 
at least ${\mit \Gamma}_1$ should not vanish identically. 
We can find such form factors in the physical systems for 
the spontaneous emission process of photons from the Hydrogen atom 
\cite{Facchi(1998),Seke(1994)} and the quantum dot. \cite{Antoniou(2001)} 
On the other hand, the discussion mentioned here immediately implies the fact that 
if $\ket{\psi} \in M_2$ exists, this time it should satisfy
\begin{equation}
\bra{\psi} {\mit \Gamma}_1 \ket{\psi} = 0, 
\label{eqn:.350}
\end{equation}
which just ensures the requirement that $f(\omega) \in L^2 ((0, \infty))$.
However, note that Eq. (\ref{eqn:.350}) does not imply ${\mit \Gamma}_1=0$ identically 
and only requires ${\mit \Gamma}_1=0$ on the subspace $M_2$.

\subsection{The small-energy behavior of $\tilde{R}^\pm (\omega) $ in the regular case}
%\label{subsec:}

In this case, the same as in Eq. (\ref{eqn:.160}), we can show that
$\tilde{R}^{\pm}(0)
=(K(0))^{-1}$.
Furthermore, we can choose some positive $\omega_0 >0$
such that
\begin{equation}
\| (K(0))^{-1}\|
\| [\omega +\lambda^2 {D} (\omega )
\pm \lambda^2 \pi i {\mit \Gamma}(\omega )
-\lambda^2 S(0)
] \|
< 1,
\label{eqn:.170}
\end{equation}
for all positive $\omega < \omega_0$. Then,
$\tilde{R}^{\pm}(\omega )$ is expanded as a Neumann series,
\begin{equation}
\tilde{R}^{\pm}(\omega )
=
\{
K(0)
[
1- (K(0))^{-1}
[\omega +\lambda^2 {D} (\omega )
\pm \lambda^2 \pi i {\mit \Gamma}(\omega )
-\lambda^2 S(0)
]]
\}^{-1}
=
\lim_{N\to \infty}S_N(\omega),
\label{eqn:.180}
\end{equation}
where
\begin{equation}
S_N(\omega)
=
\sum_{j=0}^N
\{
(K(0))^{-1}
[\omega +\lambda^2 {D} (\omega )
\pm \lambda^2 \pi i {\mit \Gamma}(\omega )
-\lambda^2 S(0)
]\}^j
(K(0))^{-1}
,
\label{eqn:.185}
\end{equation}
for all positive $\omega < \omega_0$ with
$\{
(K(0))^{-1}
[\omega +\lambda^2 {D} (\omega )
\pm \lambda^2 \pi i {\mit \Gamma}(\omega )
-\lambda^2 S(0)
]\}^0 =1
$.
Under our assumptions on the form factors,
$A(\omega ) $ defined in Eq. (\ref{eqn:.80b}) is asymptotically expanded as
\begin{equation}
A(\omega) = \sum_{n=1}^{N} \omega^n A_n +O(\omega^{N+1}),
\label{eqn:.220}
\end{equation}
as $\omega \to 0$. By using Eqs. (\ref{eqn:formfactor2}) and (\ref{eqn:.220}), 
it also follows that 
\begin{equation}
{D} (\omega )
=
S(0)
-\omega \log \omega {\mit \Gamma}_1
+\omega A_1 + O(\omega^2 \log \omega ) , %\mbox{ as } \omega \to +0,
\label{eqn:.230}
\end{equation}
as $\omega \to +0$. 
Then, Eq. (\ref{eqn:.180}) tells us the dominant asymptotic behavior of $\tilde{R}^{\pm}(\omega )$
becomes
\begin{equation}
\tilde{R}^{\pm}(\omega ) =(K(0))^{-1} +O(\omega \log \omega),
\label{eqn:.240}
\end{equation}
as $\omega \to +0$, where $(K(0))^{-1}$ never vanishes in the regular case.

\subsection{The small-energy behavior of $\tilde{R}(z) $
in the exceptional case of the first kind}
\label{subsec:5.C}

In the exceptional case of the first kind,
from the definition, $Q_1 \neq 0$ while $Q_2 = 0$, so that $Q_0 +Q_1 =1$.
Then, $\tilde{R} (z)$ is divided into the following four terms,
\begin{equation}
\tilde{R}(z)
=
Q_0 \tilde{R}(z) Q_1
+Q_0 \tilde{R}(z) Q_0
+Q_1 \tilde{R}(z) Q_0
+Q_1 \tilde{R}(z) Q_1 .
\label{eqn:5.C.1.30}
\end{equation}
We now introduce the four matrices,
\begin{equation}
E_{kl}(z)
=Q_k [K(z) -z]Q_l
=
Q_k [K_0 -z -\lambda^2 [S(0)+A(z)-(\log z) {\mit \Gamma}(z)+i\pi {\mit \Gamma}(z)]]Q_l,
\label{eqn:5.C.1.40}
\end{equation}
where $k, l =0, 1$, and $\log(-z) -\log z =-i \pi$ is used.
From the relation that
$[K(z) -z]\tilde{R}(z) =1$,
they satisfy
\begin{equation}
E_{k0}Q_0 \tilde{R}Q_l + E_{k1}Q_1 \tilde{R}Q_l =Q_k \delta_{kl},
\label{eqn:5.C.1.50}
\end{equation}
for $k,l=0,1$.
To solve the above equations we need to check whether
$E_{11}$ and $E_{00}$ are invertible in the subspaces $M_1$ and $M_0$, respectively.
By using Eq. (\ref{eqn:formfactor2}), $E_{11}(z)$ is rewritten as
\begin{equation}
E_{11}(z)
=
\lambda^2z  (\log z) Q_1 {\mit \Gamma}_1 Q_1
-z Q_1
-\lambda^2 Q_1 \{ A(z) -(\log z) [{\mit \Gamma}(z) -z {\mit \Gamma}_1 ]
+i\pi {\mit \Gamma}(z)\} Q_1 ,
\label{eqn:5.C.1.60b}
\end{equation}
where $Q_1K(0)Q_1 =0$ is used.
Note that $A(z)=O(z)$ and ${\mit \Gamma}(z) -z {\mit \Gamma}_1 =O(z^2 )$
for our form factors,
so that all terms excepting the first one of the rhs of Eq. (\ref{eqn:5.C.1.60b}) are
of the order of $O(z)$.
Furthermore, the exceptional case of the first kind imposes the fact that
$Q_1 {\mit \Gamma}_1Q_1 \neq 0$ [see Eq. (\ref{eqn:.340})], and
$Q_1 {\mit \Gamma}_1Q_1$ is positive definite in $M_1$ and thus invertible in $M_1$.
Hence, $E_{11}(\omega) $ is invertible for sufficiently small
$|z|>0$, and the inverse can be expanded by the Neumann series as,
\begin{eqnarray}
\hspace*{-5mm}
E_{11}^{-1}(z)
&=&
\sum_{j=0}^{\infty} (\tilde{E}_{11}(z))^j
\frac{1}{\lambda^2z  \log z} (Q_1 {\mit \Gamma}_1 Q_1 )^{-1}
\label{eqn:5.C.1.90a}\\
&=&
\frac{1}{\lambda^2z  \log z} (Q_1 {\mit \Gamma}_1 Q_1 )^{-1}
 +O(z^{-1} (\log z)^{-2})=O(z^{-1} (\log z)^{-1}) ,
\label{eqn:5.C.1.90b}
\end{eqnarray}
for small $|z|$, where we define
\begin{equation}
\tilde{E}_{11}(z)
:=
\frac{1}{\lambda^2z  \log z} (Q_1 {\mit \Gamma}_1 Q_1 )^{-1}
\bigl\{ z Q_1
+\lambda^2 Q_1 \{ A(z) -(\log z) [{\mit \Gamma}(z) -z {\mit \Gamma}_1 ]
+ \pi i {\mit \Gamma}(z) \} Q_1
\bigr\} ,
%=O((\log z )^{-1} ).
\label{eqn:5.C.1.90c}
\end{equation}
which behaves as $1/\log z$ as $z\to0$. For $E_{00}$ we have,
\begin{equation}
E_{00}(z)
=
Q_0 K(0)Q_0
-Q_0 [z +\lambda^2 [A(z)-(\log z) {\mit \Gamma}(z)
+ i\pi  {\mit \Gamma}(z)]]Q_0 ,
\label{eqn:5.C.1.100}
\end{equation}
where the first term of the above is invertible in $M_0$,
and the last term vanishes as $z \to 0$.
Hence, $E_{00}(z) $ is invertible in $M_0$
for sufficiently small $|z|>0$ and is expanded as
\begin{equation}
E_{00}^{-1}(z) \!
=\!
-\sum_{j=0}^{\infty}
\{[Q_0 K(0)Q_0 ]^{-1}
[z +\lambda^2 [A(z)-(\log z) {\mit \Gamma}(z)
+ i\pi {\mit \Gamma}(z)]]Q_0 \}^j
[Q_0 K(0)Q_0 ]^{-1}
=
O(1) ,
\label{eqn:5.C.1.110b}
\end{equation}
for small $|z|$.
Furthermore, we obtain
\begin{equation}
E_{kl}(z) =
-\lambda^2 Q_k [A(z)-(\log z) {\mit \Gamma}(z) +\pi i {\mit \Gamma}(z) ]Q_l =O(z \log z),
\label{eqn:5.C.1.130}
\end{equation}
for $k\neq l$ as $z \to 0$, because
$Q_0 K(0)Q_1=Q_1 K(0)Q_0=0$.
%Note that $\omega^{-1} (\log \omega)^{-1}$ is integrable around $\omega=0$
%because $d \log |\log \omega |/d\omega =\omega^{-1} (\log \omega)^{-1}$.
Solving Eq. (\ref{eqn:5.C.1.50}), we obtain \cite{MatrixAnalysis}
\begin{eqnarray}
Q_0\tilde{R}Q_0
&=&
(E_{00}-E_{01}E_{11}^{-1}E_{10})^{-1} =O(1),
\label{eqn:5.C.1.120a}\\
Q_0\tilde{R}Q_1
&=&
-E_{00}^{-1}E_{01} Q_1\tilde{R}Q_1
=
-E_{00}^{-1}E_{01} (E_{11}-E_{10}E_{00}^{-1}E_{01})^{-1}
=O(1),
\label{eqn:5.C.1.120b}\\
Q_1\tilde{R}Q_0
&=&
-Q_1\tilde{R}Q_1 E_{10} E_{00}^{-1}
=
-(E_{11}-E_{10}E_{00}^{-1}E_{01})^{-1} E_{10} E_{00}^{-1}
=O(1),
\label{eqn:5.C.1.120c}\\
Q_1\tilde{R}Q_1
&=&
(E_{11}-E_{10}E_{00}^{-1}E_{01})^{-1} =O(z^{-1} (\log z)^{-1}).
\label{eqn:5.C.1.120d}
\end{eqnarray}
%for small $|z|$, where we abbreviated the sign ($\pm$).
It is worth noting that 
since the relation $[K(z) -z]\tilde{R}(z) =1$
is analytically continued to the second Riemann sheet
through the cut $[0, \infty)$, 
the above-mentioned results are also valid for such a continued region and
the estimations obtained here can be applied without any corrections.

When we consider the small energy behavior of $\tilde{R}^- (\omega)$,
it is convenient to expand $E_{11}$, differently from Eq. (\ref{eqn:5.C.1.60b}),
as
\begin{eqnarray}
E_{11}(z)
&=&
\lambda^2z (\log z -2\pi i) Q_1 {\mit \Gamma}_1 Q_1
\nonumber\\
&&
-z Q_1
-\lambda^2 Q_1 \{ A(z) -(\log z -2\pi i)[{\mit \Gamma}(z) -z {\mit \Gamma}_1 ]
-i\pi {\mit \Gamma}(z)\} Q_1 .
\label{eqn:5.C.1.125b}
\end{eqnarray}
All the above-obtained results are only changed by replacing the term
$\log z$ with $\log z -2\pi i$.
Then, we can obtain from Eq. (\ref{eqn:5.C.1.60b})
\begin{equation}
E_{11}^+ (\omega)
=
\lim_{\epsilon \to +0}E_{11}(\omega+i\epsilon)
=
\lambda^2 (\log \omega) Q_1 {\mit \Gamma}(\omega) Q_1
-\omega Q_1 -\lambda^2 Q_1 [A(\omega)-i\pi {\mit \Gamma}(\omega)]Q_1 ,
\label{eqn:5.C.1.127a}
\end{equation}
while from Eq. (\ref{eqn:5.C.1.125b})
\begin{equation}
E_{11}^- (\omega)
=
\lim_{\epsilon \to +0}E_{11}(\omega-i\epsilon)
=
\lambda^2 (\log \omega) Q_1 {\mit \Gamma}(\omega) Q_1
-\omega Q_1 -\lambda^2 Q_1 [A(\omega)+i\pi {\mit \Gamma}(\omega)]Q_1 .
\label{eqn:5.C.1.127b}
\end{equation}

\subsection{The small-energy behavior of $\tilde{R}(z)$
in the exceptional case of the second kind}
\label{subsec:5.D}

In the exceptional case of the second kind,
it follows that $Q_1 = 0$, $Q_2 \neq 0$, and $Q_0 +Q_2 =1$.
Let us consider the asymptotic behavior of the reduced resolvent at small energies,
%from which $-\tilde{P}_0/z$ is extracted,
%\begin{equation}
%\tilde{R}(z) +\frac{\tilde{P}_0}{z}
%=
%\frac{1}{K(z) -z} +\frac{\tilde{P}_0}{z}
%\label{eqn:5.C.2.20}
%\end{equation}
%Then,
which %$\tilde{R} (\omega) +\tilde{P}_0/\omega$
is written in the following form,
$\tilde{R} (z)=\sum_{k,l=0,2}Q_k \tilde{R} (z) Q_l $.
We now introduce the four matrices again,
\begin{equation}
E_{kl}(z)
=Q_k [K(z) -z]Q_l ,
\label{eqn:5.C.2.40}
\end{equation}
where $k, l =0, 2$.
From the relation that
$[K(z) -z]\tilde{R}(z) =1$,
they satisfy that
\begin{equation}
E_{k0} Q_0 \tilde{R} Q_l + E_{k2} Q_2 \tilde{R} Q_l =Q_k \delta_{kl},
\label{eqn:5.C.2.50}
\end{equation}
for $k, l=0, 2$.
This time,
$E_{22}$ and $E_{00}$ are invertible in $M_2$ and $M_0$, respectively.
In fact, from Eqs. (\ref{eqn:formfactor2}) and (\ref{eqn:.220}) we have
\begin{equation}
E_{22}(z) =-z Q_2(1+\lambda^2 A_1) Q_2 -\lambda^2 Q_2[A(z)-z A_1 -(\log z) {\mit \Gamma}(z)
+ i\pi {\mit \Gamma}(z)]Q_2 ,
\label{eqn:5.C.2.60b}
\end{equation}
where $Q_2K(0)Q_2 =0$ was used.
Note that since $A(z)-z A_1=O(z^2)$ and $Q_2 {\mit \Gamma}_1 Q_2=0$
[see Eq. (\ref{eqn:.350})],
%for our form factors,
the second term of the rhs of Eq. (\ref{eqn:5.C.2.60b}) is of the order of
$O(z^2 \log z)$.
Furthermore, since $Q_2 A_1 Q_2 \geq 0$ from $Q_2 {\mit \Gamma}_1 Q_2=0$
and Lemma \ref{lm:1st_order},
$Q_2(1+\lambda^2 A_1)Q_2 >0$ and invertible in $M_2$.
These facts bring us the fact that $E_{22}(z) $ is invertible in $M_2$ for sufficiently small
$z>0$, that is %and the inverse can be expanded by the Neumann series as,
\begin{eqnarray}
E_{22}^{-1}(z)
&=&
-\sum_{j=0}^{\infty}
(-\tilde{E}_{22}(z) )^j \frac{1}{z}[Q_2 (1+\lambda^2 A_1 )Q_2 ]^{-1}
\label{eqn:5.C.2.90a}\\
&=&
\frac{1}{z}[Q_2 (1+\lambda^2 A_1 )Q_2 ]^{-1}+O(\log z)=O(z^{-1}),
\label{eqn:5.C.2.90b}
\end{eqnarray}
%for small $|z|$.
where
\begin{equation}
\tilde{E}_{22}(z)
:=
\frac{1}{z}[Q_2 (1+\lambda^2 A_1 )Q_2  ]^{-1}
\lambda^2 Q_2[A(z)-z A_1 -(\log z) {\mit \Gamma}(z)+ i\pi {\mit \Gamma}(z)]Q_2.
\label{eqn:5.C.2.90c}
\end{equation}
For $E_{00}$, we next have
\begin{equation}
E_{00}(z)
=
Q_0 K(0)Q_0
-Q_0 [z +\lambda^2 [A(z)-(\log z) {\mit \Gamma}(z)
+ i\pi {\mit \Gamma}(z)]]Q_0,
\label{eqn:5.C.2.100}
\end{equation}
where the first term of the above is invertible in $M_0$,
and the last term vanishes as $|z| \to 0$.
Hence, $E_{00}(z) $ is invertible in $M_0$
for sufficiently small $|z|>0$, and the inverse is obtained as a Neumann series.
On the other hand, $E_{20}$ and $E_{02}$ behave as
\begin{equation}
E_{kl}(z) =
Q_k [-z(1+\lambda^2 A_1 )
-\lambda^2 [A(z)-z A_1-(\log z) {\mit \Gamma}(z)
+ \pi i {\mit \Gamma}(z)] ]Q_l =O(z),
\label{eqn:5.C.2.130}
\end{equation}
for small $|z|$ where $k\neq l$. %for the form factors of the first kind.
Solving Eqs. (\ref{eqn:5.C.2.50}) as in Eqs. (\ref{eqn:5.C.1.120a}) to (\ref{eqn:5.C.1.120d}),
one sees that
$%\begin{equation}
Q_0\tilde{R}Q_0=O(1),
%\label{eqn:5.C.2.120a}
$ %\end{equation}
for $k,l=0,2$, except
\begin{equation}
Q_2\tilde{R}Q_2
=
(E_{22}-E_{20}E_{00}^{-1}E_{02})^{-1} =O(z^{-1}),
\label{eqn:5.C.2.120d}
\end{equation}
as $z\to 0$.
In particular, the last equation is expanded as,
\begin{equation}
Q_2 \tilde{R} (z) Q_2
=
\sum_{j=0}^{\infty}
\Bigl[ E_{22}^{-1}E_{20}E_{00}^{-1}E_{02} \Bigr]^j
E_{22}^{-1}
%=
%E_{22}^{-1} +O(z)
%\label{eqn:5.C.2.140b}\\
%&=&
=
-\frac{1}{z}[Q_2 (1+\lambda^2 A_1 )Q_2 ]^{-1} +O(\log z),
\label{eqn:5.C.2.140d}
\end{equation}
for small $|z|$, where we used Eq. (\ref{eqn:5.C.2.90b}).
%Therefore, $\tilde{R} (z)+[z Q_2 (1+\lambda^2 A_1 )Q_2 ]^{-1}$ has at most a singularity
%of $O(\log z)$.

We now remark that the zero energy eigenspace of $H$
denoted by ${\cal N}_0$ is completely characterized by $M_2$.
That is, there is a bijection from $M_2\oplus \{0\}$ to ${\cal N}_0$.
From the discussion concerning Eqs. (\ref{eqn:.333}), (\ref{eqn:.335}), and (\ref{eqn:.336}),
for any $\ket{\Psi} \in {\cal M}_0$, there is a vector $\ket{\psi}\in M_2\oplus \{0\}$ such that
\begin{equation}
\ket{\Psi}=\ket{\psi}-\lambda \int_0^\infty \frac{\sum_{n=1}^N v_n (\omega)\psi_n}{\omega}
\ket{\omega}d\omega
=
[1-\lambda R_0 (0)V]\ket{\psi},
\label{eqn:5.C.2.150}
\end{equation}
where %$\ket{\psi}$ is regarded as a vector in $M_2\oplus \{0\}$ and
$V$ is restricted to $\mathbb{C}^N \oplus \{0\}$ and
$R_0 (0)$ is the (unbounded) multiplication operator of $1/\omega$ in $L^2((0,\infty))$.
Then we see $V\ket{\psi} \in D(R_0 (0))$ because $\ket{\psi}\in M_2\oplus \{0\}$.
Thus $1-\lambda R_0 (0)V$ is well defined as an operator from $M_2\oplus \{0\}$ to ${\cal H}$.
Now, Eq. (\ref{eqn:5.C.2.150}) tells us that
$1-\lambda R_0 (0)V$ is a surjection from $M_2\oplus \{0\}$ to ${\cal N}_0$.
On the other hand, for any $\ket{\Psi} \in {\cal N}_0$, if $\ket{\Psi}=0$, i.e.,
$0=\braket{\Psi}{\Psi}$, Eq. (\ref{eqn:5.C.2.150}) implies that
$0=\braket{\Psi}{\Psi}\geq \braket{\psi}{\psi}$. Therefore,
$1-\lambda R_0 (0)V$ is also an injection from $M_2\oplus \{0\}$ to ${\cal N}_0$,
and the proof is completed.

\subsection{The small-energy behavior of $\tilde{R}^\pm (\omega) $
in the exceptional case of the third kind}
\label{subsec:5.E}

In the exceptional case of the third kind,
from the definition, $Q_1 \neq 0$, $Q_2 \neq 0$, and $Q_0 +Q_1 +Q_2 =1$.
The reduced resolvent is written in the form,
$\tilde{R}^\pm (\omega) %+\frac{\tilde{P}_0}{\omega}
=
\sum_{k,l=0}^{2}
Q_k \tilde{R}^\pm (\omega) Q_l
$.
This time, we need nine matrices,
\begin{equation}
E_{kl}^{\pm}(\omega) =Q_k [K^\pm (\omega) -\omega]Q_l ,
\label{eqn:5.C.3.40}
\end{equation}
for $k, l = 0, 1, 2$.
From the relation that
$[K^\pm (\omega) -\omega]\tilde{R}^\pm (\omega) =1$,
they satisfy that
\begin{equation}
E_{k0}^{\pm} Q_0 \tilde{R}^\pm Q_l + E_{k1}^{\pm} Q_1 \tilde{R}^\pm Q_l
+ E_{k2}^{\pm} Q_2 \tilde{R}^\pm Q_l = Q_k \delta_{kl},
\label{eqn:5.C.3.50}
\end{equation}
for $k, l = 0, 1, 2$.
The asymptotic behaviors of $E_{kl}^{\pm}(\omega) $
are essentially examined
in the preceding subsections, except for $E_{12}^{\pm}(\omega)$ and $E_{21}^{\pm}(\omega)$.
Then, $E_{12}^{\pm}(\omega)$ becomes
\begin{eqnarray}
E_{12}^{\pm}(\omega)
&=&-\omega \lambda^2 Q_1 A_1 Q_2
\nonumber \\
&&
-\lambda^2 Q_1 \bigl[A(\omega)-\omega A_1 -(\log \omega) [{\mit \Gamma}(\omega)-\omega {\mit \Gamma}_1 ]
\pm \pi i [{\mit \Gamma}(\omega)-\omega {\mit \Gamma}_1 ] \bigr]Q_2
\label{eqn:5.C.3.60b}\\
&=&-\omega \lambda^2 Q_1 A_1 Q_2+O(\omega^2 \log \omega ),
\label{eqn:5.C.3.60c}
\end{eqnarray}
where $Q_1K(0)Q_2 =0$, $Q_1 Q_2 =0$,
and ${\mit \Gamma}_1 Q_2 =0$ are used. The last relation follows from the fact that $Q_2 {\mit \Gamma}_1 Q_2 =0$ and
${\mit \Gamma}_1 \geq 0$.
In addition, since $Q_1 {\mit \Gamma} (\omega)Q_2=O(\omega^2)$,
we see that $Q_1 A_1 Q_2=\int_0^\infty Q_1 {\mit \Gamma} (\omega) Q_2 \omega^{-2} d\omega$.
By the same way, we also see that
\begin{equation}
E_{21}^{\pm}(\omega)
=-\omega \lambda^2 Q_1 A_1 Q_2+O(\omega^2 \log \omega ).
\label{eqn:5.C.3.60d}
\end{equation}
To solve Eqs. (\ref{eqn:5.C.3.50}), let us now put the $N\times N$ matrix $\bf E$ as
\begin{equation}
{\bf E}=
\left[
\begin{array}{ccc}
E_{00}&E_{01}&E_{02} \\
E_{10}&E_{11}&E_{12} \\
E_{20}&E_{21}&E_{22}
\end{array}
\right] ,
\label{eqn:5.C.3.65a}
\end{equation}
and partition it into
%\begin{equation}
${\bf A}=
\left[
\begin{array}{cc}
E_{00}&E_{01} \\
E_{10}&E_{11}
\end{array}
\right]
$, %~~~
${\bf B}=
\left[
\begin{array}{c}
E_{02} \\
E_{12}
\end{array}
\right]
$, %~~~
${\bf C}=
\left[
\begin{array}{cc}
E_{20}&E_{21}
\end{array}
\right]
$, %~~~
${\bf D}=
\left[
\begin{array}{c}
E_{22}
\end{array}
\right]$. %
%\label{eqn:5.C.3.65b}
%\end{equation}
Then, from the inverse matrix formula again,
${\bf E}^{-1}$ ($=\tilde{R}$) is expressed as \cite{MatrixAnalysis}
\begin{equation}
{\bf E}^{-1}=
\left[
\begin{array}{cc}
[{\bf A}-{\bf B}{\bf D}^{-1}{\bf C}]^{-1}
& -{\bf A}^{-1}{\bf B}[{\bf D}-{\bf C}{\bf A}^{-1}{\bf B}]^{-1}\\
-[{\bf D}-{\bf C}{\bf A}^{-1}{\bf B}]^{-1} {\bf C} {\bf A}^{-1}
&[{\bf D}-{\bf C}{\bf A}^{-1}{\bf B}]^{-1}
\end{array}
\right] .
\label{eqn:5.C.3.67}
\end{equation}
The validities of ${\bf A}^{-1}$ and ${\bf D}^{-1}$ are already ensured in
the exceptional cases of the first and second kinds, respectively.
Then, one sees that
since
${\bf A}^{-1}=O(\omega^{-1} (\log \omega)^{-1})$,
${\bf B}=O(\omega)$,
${\bf C}=O(\omega)$, and
${\bf D}^{-1}=O(\omega^{-1})$,
it holds that ${\bf A}^{-1}{\bf B}{\bf D}^{-1}{\bf C}=O((\log \omega)^{-1})$.
Thus, $[{\bf A}-{\bf B}{\bf D}^{-1}{\bf C}]^{-1}$ exists for small $\omega$
and $[{\bf A}-{\bf B}{\bf D}^{-1}{\bf C}]^{-1}=O(\omega^{-1}(\log \omega)^{-1})$.
We also show that $[{\bf D}-{\bf C}{\bf A}^{-1}{\bf B}]^{-1}$ exists for small $\omega$
and $[{\bf D}-{\bf C}{\bf A}^{-1}{\bf B}]^{-1}=O(\omega^{-1})$.
To obtain the asymptotic forms of the matrix components of ${\bf E}^{-1}$ explicitly,
some redundant calculation is required; however, it could be achieved by 
a manner as similar to that used in the preceding subsections.

%%%%%%%%%%%%%%%%%%%%%%%%%%%%%%%%%%%%%%%%%%%%%%%%%%%%%%%%%%%%%%%%%%%%%%%%%%%%%%%%%%%%%%%%%%
\section{Asymptotic expansion of the reduced resolvent at small $z$ }
\label{sec:5.5}

We examine the small-energy behavior of the reduced resolvent
only for the regular case and the exceptional case of the first kind.
This analysis is crucial for determining the asymptotic behavior of
the reduced time evolution operator at long times.

\subsection{The regular case}
%\label{subsec:}

Here, we introduce $\tilde{A}(\omega ) := \omega/\lambda^2 +A(\omega ) $
and suppose that $\tilde{A}(\omega )$ and ${\mit \Gamma}(\omega ) $ behave as
\begin{equation}
\tilde{A}(\omega ) := \frac{1}{\lambda^2} \omega +
A(\omega ) = \sum_{n=n_a}^{n_a+N} \omega^n \tilde{A}_n +O(\omega^{n_a+N+1}), ~~~
{\mit \Gamma}(\omega )
=\sum_{n=n_b}^{n_b+N} \omega^n {\mit \Gamma}_n +O(\omega^{n_b+N+1}),
\label{eqn:rff.220}
\end{equation}
as $\omega \to 0$, respectively,
that is, $\tilde{A}_{n} =0$ for all $n< n_a$ and ${\mit \Gamma}_{n_b}=0$ for all $n< n_b$, 
while $\tilde{A}_{n_a}\neq0$ and ${\mit \Gamma}_{n_b}\neq0$.
%and $n_a$ and $n_b$ are some positive integers.
%which implies that $A_n =0$ for all $n=1, \ldots, n_a$ and
%${\mit \Gamma}_n =0$ for all $n=1, \ldots, n_b$.
Then, we obtain
\begin{equation}
\frac{1}{\lambda^2} \omega +
{D} (\omega )
=
S(0) +\tilde{A}(\omega )-(\log \omega){\mit \Gamma}(\omega)
=
S(0)
-\omega^{n_b} \log \omega {\mit \Gamma}_{n_b}
+\omega^{n_a} \tilde{A}_{n_a} + O(\omega h(\omega) ) , %\mbox{ as } \omega \to +0,
\label{eqn:rff.230}
\end{equation}
as $\omega \to +0$, where
\begin{equation}
h(\omega)
=
\left\{\begin{array}{cc}
\omega^{n_b} \log \omega & (n_b \leq n_a) \\
\omega^{n_a}  & (n_b > n_a)
\end{array}\right. .
\label{eqn:rff.245}
\end{equation}

It is important to note that the values of two parameters $n_a$ and $n_b$ 
are not determined independently. We shall here consider $n_b$ as a controllable one. 
We first note that if $n_b \geq 2$ then $n_a =1$ should be concluded, 
because from Lemma \ref{lm:1st_order} we have $A_1 >0$, so that 
$\tilde{A}_1 =1/\lambda^2  +A_1 >0$ holds. 
Therefore, the conditions $n_b \leq n_a$ and $n_b > n_a$ 
can be realized only in the situations 
\begin{equation}
n_b =1 ~\mbox{and}~ n_a \geq 1,\quad\mbox{and}\quad n_b \geq 2~ \mbox{and}~ n_a = 1,
\label{eqn:rff.246}
\end{equation}
respectively.

\begin{lm} \label{lm:rffremainder}
: Assume that $0$ is a regular point for $H$.
Then the $r$-th derivative of $\tilde{R}^\pm (\omega)$ asymptotically
behaves as
\begin{equation}
%\left\|
\frac{d^r \tilde{R}^\pm (\omega)}{d\omega^r}
%\right\|
=
\left\{
\begin{array}{cc}
O(1) & (r=0)\\
O(\omega^{1-r}(\log \omega)^{\theta(1-r)}) & (r\geq1)
\end{array}
\right. ,
%=O(\omega^{-r} [h(\omega)]^{2-[2-r]^+}) ,
%
\mbox{ or }
\left\{
\begin{array}{cc}
O(1) & (r=0)\\
%O(\omega^{n_a -1}) & (r=1)\\
O(\omega^{[1-r]^+}) &(1\leq r < n_b) \\
%O(\omega^{[2n_a-r]^+}) &(2\leq r\leq n_a+n_b-1) \\
O(\omega^{n_b-r} (\log \omega)^{\theta(n_b-r)}) &(n_b\leq r \leq 2n_b)
%O(\omega^{n_a+n_b-r} (\log \omega)^{\theta(n_a+n_b-r)}) &(n_a+n_b\leq r \leq 2n_b)
\end{array}
\right. ,
\label{eqn:rffSmall3}
\end{equation}
for $n_b =1$, or $n_b \geq 2$, 
%for $n_b \leq n_a$ or $n_b > n_a$
respectively, as $\omega \to 0$,
where $[x]^+ =\max \{x, 0 \}$
and $\theta(x)=1$ for $x\geq 0$ or $0$ for $x<0$.
In addition, the $r$-th derivative of $\tilde{R}^\pm (\omega)$
is approximated by that of a finite series
\begin{equation}
(K(0))^{-1}
+
(K(0))^{-1}
\left[
-\omega^{n_b} (\log \omega) \lambda^2 {\mit \Gamma}_{n_b}
+\omega^{n_a} \lambda^2 \tilde{A}_{n_a} \pm \lambda^2  \pi i \omega^{n_b} {\mit \Gamma}_{n_b}
\right]
(K(0))^{-1} ,
\label{eqn:rffSmall5}
\end{equation}
that is, it is shown that
\begin{eqnarray}
&&
\Biggl\|
\frac{d^r}{d\omega^r}
\biggl\{
\tilde{R}^\pm (\omega)
\nonumber \\
&&
-
(K(0))^{-1}
-
(K(0))^{-1}
\left[
-\omega^{n_b} (\log \omega) \lambda^2 {\mit \Gamma}_{n_b}
+\omega^{n_a} \lambda^2 \tilde{A}_{n_a} \pm \lambda^2  \pi i \omega^{n_b} {\mit \Gamma}_{n_b}
\right]
(K(0))^{-1}
\biggr\}
\Biggr\|
\nonumber \\
&=&
\begin{array}{cc}
O(\omega^{2-r} (\log \omega)^{1+\theta(2-r)})
& (r \geq 0)
\end{array}
%
%\left\{
%\begin{array}{cc}
%O(\omega^{2-r} (\log \omega)^{1+\theta(2-r)})
%& (r \geq 0, n_b=1) \\
%O(\omega^{n_b+1-r} (\log \omega)^{\theta(n_b+1-r)})
%& (r \geq 0, n_b\geq2)
%\end{array}
%\right. ,
%
\nonumber \\
&&
\mbox{ or }
\left\{
\begin{array}{cc}
O(\omega^{[2-r]^+}) & (0\leq r\leq n_b) \\
O(\omega^{n_b+1-r} (\log \omega)^{\theta(n_b+1-r)}) & (n_b+1 \leq r \leq 2n_b)
\end{array}
\right. ,
%
%\left\{
%\begin{array}{cc}
%O(\omega^{[n_a+1-r]^+}) & (0\leq r\leq n_b) \\
%O(\omega^{n_b+1-r} (\log \omega)^{\theta(n_b+1-r)}) & (r\geq n_b+1)
%\end{array}
%\right. ,
%
\label{eqn:rffSmall7}
\end{eqnarray}
for $n_b =1$, or $n_b \geq 2$, 
%for $n_b \leq n_a$ or $n_b > n_a$
respectively, as $\omega \to 0$. Here, $n_a$ is restricted to the condition (\ref{eqn:rff.246}). 
\end{lm}

{\sl Proof} :
The left-hand side (lhs) of Eq. (\ref{eqn:rffSmall7}) is written as follows:
\begin{eqnarray}
&&\hspace*{-10mm}
\Biggl\|
\frac{d^r}{d\omega^r}
\biggl\{
\tilde{R}^\pm (\omega)
\nonumber \\
&&\hspace*{-10mm}
-
(K(0))^{-1}
-
(K(0))^{-1}
\left[
-\omega^{n_b} (\log \omega) \lambda^2 {\mit \Gamma}_{n_b}
+\omega^{n_a} \lambda^2 \tilde{A}_{n_a} \pm \lambda^2  \pi i \omega^{n_b} {\mit \Gamma}_{n_b}
\right]
(K(0))^{-1}
\biggr\}
%+ O(\omega^{1+p} )
\Biggr\|
\nonumber \\
&&\hspace*{-10mm}
\leq
\left\|
\frac{d^r}{d\omega^r}
\bigl\{
\tilde{R}^\pm (\omega)-S_1 (\omega)
\bigr\}
\right\|
\nonumber \\
&&\hspace*{-10mm}
~~~+
\Biggl\|
\frac{d^r}{d\omega^r}
\biggl\{
S_1 (\omega)
\nonumber \\
&&\hspace*{-10mm}
~~~-
(K(0))^{-1}
-
(K(0))^{-1}
\left[
-\omega^{n_b} (\log \omega) \lambda^2 {\mit \Gamma}_{n_b}
+\omega^{n_a} \lambda^2 \tilde{A}_{n_a} \pm \lambda^2  \pi i \omega^{n_b} {\mit \Gamma}_{n_b}
\right]
(K(0))^{-1}
\bigg\}
%+ O(\omega^{1+p} )
\Biggr\| ,
\label{eqn:rffSmall50}
\end{eqnarray}
where $S_N(\omega)$ is defined by Eq. (\ref{eqn:.185}).
When $r=0$, the first term on the rhs of the above is estimated from the special case of the below
for $N=1$,
\begin{equation}
\left\|
\tilde{R}^\pm (\omega)-S_N (\omega)
\right\|
\! \leq \!
\frac{
\|
\omega +\lambda^2 {D} (\omega )
\pm \lambda^2 \pi i {\mit \Gamma}(\omega )
-\lambda^2 S(0)
\|^{N+1}
\|
(K(0))^{-1}
\|^{N+2}
}
{
1-\|(K(0))^{-1}\|
\| [\omega +\lambda^2 {D} (\omega )
\pm \lambda^2 \pi i {\mit \Gamma}(\omega )
-\lambda^2 S(0)
]\|
}
=O(h(\omega)^{N+1} ),
\label{eqn:rffSmall60c}
\end{equation}
as $\omega \to 0$. When $r\geq 1$, instead we have
\begin{eqnarray}
&&
\left\|
\frac{d^r}{d\omega^r}
\bigl\{
\tilde{R}^\pm (\omega)
-
S_1 (\omega)
\bigr\}
\right\|
\nonumber \\
%&=&
%\left\|
%\frac{d^r \tilde{R}^\pm (\omega)}{d\omega^r}
%-
%(K(0))^{-1}
%\frac{d^r }{d\omega^r}
%[\omega +\lambda^2 {D} (\omega )
%\pm \lambda^2 \pi i {\mit \Gamma}(\omega )
%-\lambda^2 S(0)
%]
%(K(0))^{-1}
%\right\|
%\nonumber \\
%%\label{eqn:rffSmall70a} \\
&\leq&
\left\|
\sum_{j=2}^{r}
{\sum_{\{s_i \}_{i=1}^{j} }}' %\ (\sum_{i=1}^{j} s_i =s)}
a^{(r)} (\{s_i \}_{i=1}^{j} )
\left\{
\prod_{i=1}^{j}
\tilde{R}^{\pm}(\omega )
\frac{d^{s_i}}{d\omega^{s_i}}
\left[
\omega + \lambda^2 {D}(\omega )
\pm \lambda^2 \pi i {\mit \Gamma}(\omega )
\right]
\right\}
\tilde{R}^{\pm}(\omega )
\right\|
\nonumber \\
&&
+
\left\|
\tilde{R}^{\pm}(\omega )
\frac{d^r }{d\omega^r}
\left[
\omega + \lambda^2 {D}(\omega )
\pm \lambda^2 \pi i {\mit \Gamma}(\omega )
\right]
\tilde{R}^{\pm}(\omega )
-
\frac{d^r}{d\omega^r}
S_1 (\omega)
\right\| ,
\label{eqn:rffSmall70b}
\end{eqnarray}
where Eq. (\ref{eqn:Large100}) is used, and here $s_i \geq 1$ and $\sum_{i=1}^{j} s_i =r$
should be satisfied.
Note that the first term on the rhs of Eq. (\ref{eqn:rffSmall70b})
appears only for $r\geq 2$, which is estimated in the following.
In the following estimations, we temporarily forget the restriction (\ref{eqn:rff.246}) 
%on the pair of $n_a$ and $n_b$,
and consider the two general cases: $n_b \leq n_a$ and $n_b > n_a$.
In the case of $n_b \leq n_a$, we can obtain for $r\geq 2$
\begin{eqnarray}
&&
\left\|
\sum_{j=2}^{r}
{\sum_{\{s_i \}_{i=1}^{j} }}' %\ (\sum_{i=1}^{j} s_i =s)}
a^{(r)} (\{s_i \}_{i=1}^{j} )
\left\{
\prod_{i=1}^{j}
\tilde{R}^{\pm}(\omega )
\frac{d^{s_i}}{d\omega^{s_i}}
\left[
\omega + \lambda^2 {D}(\omega )
\pm \lambda^2 \pi i {\mit \Gamma}(\omega )
\right]
\right\}
\tilde{R}^{\pm}(\omega )
\right\|
\nonumber \\
&&\leq
%\sum_{j=2}^{r}
%{\sum_{\{s_i \}_{i=1}^{j} }}' %\ (\sum_{i=1}^{j} s_i =s)}
%a^{(r)} (\{s_i \}_{i=1}^{j} )
%\left\|
%\tilde{R}^{\pm}(\omega )
%\right\|^{j+1}
%\prod_{i=1}^{j}
%\left\|
%\frac{d^{s_i}}{d\omega^{s_i}}
%\left[
%\omega + \lambda^2 {D}(\omega )
%\pm \lambda^2 \pi i {\mit \Gamma}(\omega )
%\right]
%\right\|
%\nonumber \\
%&&=
\sum_{j=2}^{r}
{\sum_{\{s_i \}_{i=1}^{j} }}' %\ (\sum_{i=1}^{j} s_i =s)}
a^{(r)} (\{s_i \}_{i=1}^{j} )
\left\|
\tilde{R}^{\pm}(\omega )
\right\|^{j+1}
O(\omega^{jn_b-r})
\prod_{i=1}^{j}
O((\log \omega)^{\theta(n_b-s_i)})
\nonumber \\
&&=
O(\omega^{2n_b-r} (\log \omega)^{\theta(n_b+1-r)+\theta(2n_b+1-r)}),
\label{eqn:rffSmall80a}
\end{eqnarray}
as $\omega \to 0$.
For $n_a < n_b$,
\begin{eqnarray}
&&%\hspace*{-5mm}
\left\|
\sum_{j=2}^{r}
{\sum_{\{s_i \}_{i=1}^{j} }}' %\ (\sum_{i=1}^{j} s_i =s)}
a^{(r)} (\{s_i \}_{i=1}^{j} )
\tilde{R}^{\pm}(\omega )
\prod_{i=1}^{j}
\left\{
\frac{d^{s_i}}{d\omega^{s_i}}
\left[
\omega + \lambda^2 {D}(\omega )
\pm \lambda^2 \pi i {\mit \Gamma}(\omega )
\right]
\tilde{R}^{\pm}(\omega )
\right\}
\right\|
\nonumber \\
&&=
\left\{
\begin{array}{cc}
O(\omega^{[2n_a-r]^+}) &(2\leq r\leq n_a+n_b-1) \\
O(\omega^{n_a+n_b-r} (\log \omega)^{\theta(n_a+n_b-r)})
&(n_a+n_b\leq r \leq 2n_b)
\end{array}
\right. ,
\label{eqn:rffSmall80b}
\end{eqnarray}
as $\omega \to 0$.
We here used that
\begin{eqnarray}
&&
\frac{d^r}{d\omega^r }
[
\omega+\lambda^2 {D}(\omega )\pm \lambda^2 \pi i {\mit \Gamma}(\omega )
-\lambda^2 S(0)
]
\nonumber\\
&&
=
O(\omega^{n_b-r} (\log \omega)^{\theta(n_b-r)})
%=O(\omega^{-r} h(\omega) )
, \mbox{ or }
%\label{eqn:rffSmall98a}
%\end{equation}
%for $n_b\leq n_a$, or
%\begin{equation}
%\frac{d^r}{d\omega^r }
%[
%\omega+\lambda^2 {D}(\omega )\pm \lambda^2 \pi i {\mit \Gamma}(\omega )-\lambda^2 S(0)
%]
%=
%\left\{
%\begin{array}{cc}
%O(\omega^{n_a-r}) & (r< n_a) \\
%O(1) & (n_a \leq r< n_b) \\
%O(\log \omega) & (r= n_b) \\
%O(\omega^{n_b-r}) & (r> n_b )
%\end{array}
%\right. %\mbox{for $n_b> n_a$},
%=
\left\{
\begin{array}{cc}
O(\omega^{[n_a-r]^+}) & (0\leq r< n_b) \\
O(\omega^{n_b-r} (\log \omega)^{\theta(n_b-r)}) & (r\geq n_b)
\end{array}
\right. %\mbox{for $n_b> n_a$},
%=O(\omega^{-r} h(\omega) )
,
\label{eqn:rffSmall98b}
\end{eqnarray}
for $n_b\leq n_a$, or $n_a< n_b$, respectively, as $\omega \to 0$.
Eq. (\ref{eqn:rffSmall98b}) follows from
\begin{equation}
\frac{d^r \tilde{A}(\omega )}{d\omega^r }
%=
%\left\{
%\begin{array}{cc}
%O(\omega^{n_a-r}) & (r\leq n_a) \\
%O(1) & (r> n_a )
%\end{array}
%\right.
=O(\omega^{[n_a-r]^+}) ,
~~~
\frac{d^r {\mit \Gamma}(\omega )}{d\omega^r }
%=
%\left\{
%\begin{array}{cc}
%O(\omega^{n_b-r}) & (r\leq n_b) \\
%O(1) & (r> n_b )
%\end{array}
%\right.
=O(\omega^{[n_b-r]^+}) ,~~~
%\label{eqn:rffSmall97}
%\end{equation}
%and
%\begin{equation}
\frac{d^r (\log \omega) {\mit \Gamma}(\omega )}{d\omega^r }
%=
%\left\{
%\begin{array}{cc}
%O(\omega^{n_b-r} \log \omega) & (r\leq n_b) \\
%O(\omega^{n_b-r} ) & (r>n_b)
%\end{array}
%\right.
=O(\omega^{n_b-r} (\log \omega)^{\theta(n_b-r)}) ,
\label{eqn:rffSmall90b}
\end{equation}
as $\omega \to 0$.
Incorporating Eqs. (\ref{eqn:rffSmall80a}), (\ref{eqn:rffSmall80b}),
and Eq. (\ref{eqn:rffSmall98b}), with
\begin{eqnarray}
&&
\left\|
\frac{d^r \tilde{R}^\pm (\omega)}{d\omega^r}
\right\|
\nonumber\\
&\leq&
\left\|
\sum_{j=2}^{r}
{\sum_{\{s_i \}_{i=1}^{j} }}' %\ (\sum_{i=1}^{j} s_i =s)}
a^{(r)} (\{s_i \}_{i=1}^{j} )
\left\{
\tilde{R}^{\pm}(\omega )
\prod_{i=1}^{j}
\frac{d^{s_i}}{d\omega^{s_i}}
\left[
\omega + \lambda^2 {D}(\omega )
\pm \lambda^2 \pi i {\mit \Gamma}(\omega )
\right]
\right\}
\tilde{R}^{\pm}(\omega )
\right\|
\nonumber \\
&&
+
\left\|
\tilde{R}^{\pm}(\omega )
\frac{d^r }{d\omega^r}
\left[
\omega + \lambda^2 {D}(\omega )
\pm \lambda^2 \pi i {\mit \Gamma}(\omega )
\right]
\tilde{R}^{\pm}(\omega )
\right\| ,
\label{eqn:rffSmall99b}
\end{eqnarray}
we have
\begin{eqnarray}
&&\hspace*{-10mm}
\left\|
\frac{d^r \tilde{R}^\pm (\omega)}{d\omega^r}
\right\|
\nonumber\\
&&\hspace*{-10mm}
=
\left\{
\begin{array}{cc}
O(1) & (r=0) \\
O(\omega^{n_b-r}(\log \omega)^{\theta(n_b-r)})) & (r\geq 1 )
\end{array}
\right.
\mbox{ or }
\left\{
\begin{array}{cc}
O(1) & (r=0) \\
O(\omega^{[n_a-r]^+}) & (1\leq r< n_b) \\
O(\omega^{n_b-r} (\log \omega)^{\theta(n_b-r)}) & (n_b \leq r \leq 2n_b)
\end{array}
\right. ,
\label{eqn:rffSmall99c}
\end{eqnarray}
for $n_b\leq n_a$, or $n_a< n_b$, respectively, as $\omega \to 0$.
Then, the first part of the statement can be shown under the restriction (\ref{eqn:rff.246}).
Let us next examine the second term on the rhs of Eq. (\ref{eqn:rffSmall70b}),
which reads for $r\geq 1$,
\begin{eqnarray}
&&
\Biggl\|
\tilde{R}^{\pm}(\omega )
\frac{d^r }{d\omega^r}
\left[
\omega + \lambda^2 {D}(\omega )
\pm \lambda^2 \pi i {\mit \Gamma}(\omega )
\right]
\tilde{R}^{\pm}(\omega )
-
\frac{d^r }{d\omega^r}
S_1(\omega)
\Biggr\|
\nonumber \\
&&\leq
2 \left\|
\tilde{R}^{\pm}(\omega )
-(K(0))^{-1}
\right\|
\left\|
\frac{d^r }{d\omega^r}
\left[
\omega + \lambda^2 {D}(\omega )
\pm \lambda^2 \pi i {\mit \Gamma}(\omega )
\right]
\right\|
\left\|
\tilde{R}^{\pm}(\omega )
\right\|
\nonumber \\
&&=
O(\omega^{2n_b-r} (\log \omega)^{1+\theta(n_b-r)}),
\mbox{ or }
\left\{
\begin{array}{cc}
O(\omega^{n_a+[n_a-r]^+}) & (r< n_b) \\
O(\omega^{n_a+n_b-r} (\log \omega)^{\theta(n_b-r)}) & (r\geq n_b)
\end{array}
\right. ,
\label{eqn:rffSmall100}
\end{eqnarray}
for $n_b \leq n_a$ or $n_b > n_a$ respectively,
as $\omega \to 0$.
We here used Eq. (\ref{eqn:rffSmall60c}) with $N=0$.
Therefore, substituting Eqs. (\ref{eqn:rffSmall80a}),
(\ref{eqn:rffSmall80b}), and (\ref{eqn:rffSmall100}) into
Eq. (\ref{eqn:rffSmall70b}), one has  for $r\geq 1$,
\begin{eqnarray}
&&\hspace*{-12mm}
\left\|
\frac{d^r}{d\omega^r}
\bigr\{ \tilde{R}^\pm (\omega)- S_1 (\omega) \bigr\}
\right\|
\nonumber\\
&&\hspace*{-12mm}
=O(\omega^{2n_b-r} (\log \omega)^{1+\theta(n_b+1-r)}) ,
\mbox{or}
\left\{
\begin{array}{cc}
O(\omega^{[2n_a-r]^+}) &(r\leq n_a+n_b-1) \\
O(\omega^{n_a+n_b-r} (\log \omega)^{\theta(n_a+n_b-r)})
&(n_a+n_b\leq r \leq 2n_b)
\end{array}
\right. \label{eqn:rffSmall110}
\end{eqnarray}
for $n_b \leq n_a$ or $n_b > n_a$, respectively, as $\omega \to 0$.
Note that this estimation is also valid for $r=0$ because it reproduces
Eq. (\ref{eqn:rffSmall60c}) for $N=1$.

Let us now evaluate the last term in Eq. (\ref{eqn:rffSmall50}).
%
%For $r=0$, it follows that
%\begin{equation}
%\left\|
%S_1 (\omega)
%-
%(K(0))^{-1}
%-
%(K(0))^{-1}
%\left[
%-\omega^{n_b} (\log \omega) \lambda^2 {\mit \Gamma}_{n_b}
%+\omega^{n_a} \lambda^2 \tilde{A}_{n_a} \pm \lambda^2  \pi i \omega^{n_b} {\mit \Gamma}_{n_b}
%\right]
%(K(0))^{-1}
%\right\|
%=O(\omega h(\omega)) .
%\label{eqn:rffSmall120}
%\end{equation}
%See Eqs. (\ref{eqn:rff.230}).
%
For $r\geq 0$, we have
\begin{eqnarray}
&&
\Biggl\|
\frac{d^r}{d\omega^r}
\biggr\{
S_1 (\omega)
\nonumber\\
&&
-
(K(0))^{-1}
-
(K(0))^{-1}
\left[
-\omega^{n_b} (\log \omega) \lambda^2 {\mit \Gamma}_{n_b}
+\omega^{n_a} \lambda^2 \tilde{A}_{n_a} \pm \lambda^2  \pi i \omega^{n_b} {\mit \Gamma}_{n_b}
\right]
(K(0))^{-1}
\bigg\}
\Biggr\|
\nonumber \\
&&\leq
\left\|
(K(0))^{-1}
\right\|^2
\nonumber \\
&&
~~~\times
\left\|
\frac{d^r }{d\omega^r}
\left[
-(\log \omega) \lambda^2 ({\mit \Gamma}(\omega)- \omega^{n_b} {\mit \Gamma}_{n_b} )
+\lambda^2 (\tilde{A}(\omega)- \omega^{n_a} \tilde{A}_{n_a} ))
\pm \lambda^2  \pi i
\left(
{\mit \Gamma}(\omega)- \omega^{n_b} {\mit \Gamma}_{n_b}
\right)
\right]
\right\|
\nonumber \\
%\label{eqn:rffSmall130a} \\
&&=
O(\omega^{n_b+1-r} (\log \omega)^{\theta(n_b+1-r)}),
\mbox{ or }
\left\{
\begin{array}{cc}
O(\omega^{[n_a+1-r]^+}) & (0\leq r\leq n_b) \\
O(\omega^{n_b+1-r} (\log \omega)^{\theta(n_b+1-r)}) & (r\geq n_b+1)
\end{array}
\right. ,
\label{eqn:rffSmall130b}
\end{eqnarray}
for $n_b \leq n_a$ or $n_b > n_a$ respectively,
as $\omega \to 0$ for any $r\geq 0$.
%which is also valid for $r=0$ because of Eq. (\ref{eqn:rffSmall120}).
We here used that for $r\geq 0$,
\begin{equation}
\frac{d^r ({\mit \Gamma}(\omega)- \omega^{n_b} {\mit \Gamma}_{n_b})}{d\omega^r}
=O(\omega^{[n_b+1-r]^+}) ,~~
%=
%\left\{
%\begin{array}{cc}
%O(\omega^{n_b+1-r}) & (r\leq n_b+1) \\
%O(1) & (r> n_b+1 )
%\end{array}
%\right. ,
\frac{d^r (\log \omega) ({\mit \Gamma}(\omega)- \omega^{n_b} {\mit \Gamma}_{n_b})}{d\omega^r }
=
O(\omega^{n_b+1-r} (\log \omega)^{\theta(n_b+1-r)})
\label{eqn:rffSmall140}
\end{equation}
as $\omega \to 0$, and so forth.
%These follows from the facts that ${\mit \Gamma}(\omega)$ is expanded like
%${\mit \Gamma}(\omega)= {\mit \Gamma}(0) + \omega B^{(1)}(0)
%+(\omega^2 /2)\int_{0}^{1} t B^{(2)}((1-t)\omega) dt$ with ${\mit \Gamma}(0)=0$,
%$B^{(1)}(0) ={\mit \Gamma}_1$ and $\sup_{\omega >0} \|B^{(j)}(\omega)\| < \infty $,
%and those for $A(\omega)$ and $C^{(l)}(\omega)$.
%Here $B^{(j)}(\omega)$ denotes $d^j {\mit \Gamma}(\omega)/d\omega^j$ and so forth.
Thus,
setting Eqs. (\ref{eqn:rffSmall110}) and (\ref{eqn:rffSmall130b})
into Eq. (\ref{eqn:rffSmall50}), we conclude that
\begin{eqnarray}
&&%\hspace*{-10mm}
\Biggl\|
\frac{d^r}{d\omega^r}
\biggl\{
\tilde{R}^\pm (\omega)
\nonumber\\
&&
-
(K(0))^{-1}
-
(K(0))^{-1}
\left[
-\omega^{n_b} (\log \omega) \lambda^2 {\mit \Gamma}_{n_b}
+\omega^{n_a} \lambda^2 \tilde{A}_{n_a} \pm \lambda^2  \pi i \omega^{n_b} {\mit \Gamma}_{n_b}
\right]
(K(0))^{-1}
\biggr\}
%+ O(\omega^{1+p} )
\Biggr\|
\nonumber \\
&&%\hspace*{-10mm}
=
\left\{
\begin{array}{cc}
O(\omega^{2n_b-r} (\log \omega)^{1+\theta(n_b+1-r)})
& (r \geq 0, n_b=1) \\
O(\omega^{n_b+1-r} (\log \omega)^{\theta(n_b+1-r)})
& (r \geq 0, n_b\geq2)
\end{array}
\right. ,
\nonumber\\
&&
~~~
\mbox{ or }
\left\{
\begin{array}{cc}
O(\omega^{[n_a+1-r]^+}) & (0\leq r\leq n_b) \\
O(\omega^{n_b+1-r} (\log \omega)^{\theta(n_b+1-r)}) & (n_b+1 \leq r \leq 2n_b)
\end{array}
\right. ,
\label{eqn:rffSmall60}
\end{eqnarray}
for $n_b \leq n_a$ or $n_b > n_a$, respectively,
as $\omega \to 0$.
By taking into account the restriction (\ref{eqn:rff.246}), we can show the last part of the lemma.
\qed

To estimate the long time behavior of the reduced time evolution operator,
the above-mentioned lemma seems not precisely appropriate 
because the reduced time evolution operator is obtained from
the Fourier transform of the imaginary part of $\tilde{R}^+ (\omega)$,
not from $\tilde{R}^+ (\omega)$ itself, which is explained in the next section. %Sec. \ref{sec:6}.
Hence, the following lemma is more appropriate for our purpose.

\begin{lm} \label{lm:rffimremainder}
: Assume that $0$ is a regular point for $H$.
Then the $r$-th derivative of
${\rm Im}\tilde{R}^+ (\omega):=(\tilde{R}^+ (\omega) -\tilde{R}^- (\omega))/2i$
is approximated by that of
\begin{eqnarray}
(K(0))^{-1}
\lambda^2 \pi
\omega^{n_b} ({\mit \Gamma}_{n_b} +\omega {\mit \Gamma}_{n_b +1} +\omega^2 {\mit \Gamma}_{n_b +2})
(K(0))^{-1} ,
\label{eqn:rffimSmall5}
\end{eqnarray}
in the sense that for $0\leq r \leq n_b+1$ the remainder is estimated as,
\begin{eqnarray}
&&
\left\|
\frac{d^r}{d\omega^r}
\bigl\{
{\rm Im}\tilde{R}^+ (\omega)-
(K(0))^{-1}
\lambda^2 \pi
\omega^{n_b} ({\mit \Gamma}_{n_b} +\omega {\mit \Gamma}_{n_b +1} +\omega^2 {\mit \Gamma}_{n_b +2})
(K(0))^{-1}
\bigr\}
\right\|
\nonumber\\
&&
=
O(\omega^{2-r}\log \omega)
%
%\left\{
%\begin{array}{cc}
%O(\omega^{n_b +3-r}) & (n_b \geq 4)\\
%O(\omega^{2n_b-r}\log \omega) & (n_b \leq 3)
%\end{array}
%\right.
%
, ~\mbox{or} ~
O(\omega^{1+n_b-r})
%
%\left\{
%\begin{array}{cc}
%O(\omega^{n_b +3-r}) & (n_a \geq 3)\\
%O(\omega^{n_a+n_b-r}) & (n_a \leq 2)
%\end{array}
%\right.
%
,
\label{eqn:rffimSmall60}
\end{eqnarray}
for $n_b=1$, or $n_b \geq 2$, respectively, as $\omega \to 0$. 
%for $n_b \leq n_a$ or $n_a < n_b $, respectively, as $\omega \to 0$.
For $r=n_b +2$, the estimation is replaced by
$O(\omega^{-1})$ for $n_b=1$, $O(\log \omega)$ for $n_b = 2$, or $O(1)$ for $n_b \geq 3$, 
respectively, as $\omega \to 0$.
\end{lm}

{\sl Proof} :
Since ${\rm Im}\tilde{R}^+ (\omega)=\lambda^2 \pi \tilde{R}^+ (\omega) {\mit \Gamma}(\omega)
\tilde{R}^- (\omega)$, one has
%Let us first consider the following quantity again, like that
\begin{eqnarray}
&&
\left\|
\frac{d^r}{d\omega^r}
\bigl\{
{\rm Im}\tilde{R}^+ (\omega)
-
(K(0))^{-1}
\lambda^2 \pi
\omega^{n_b} ({\mit \Gamma}_{n_b} +\omega {\mit \Gamma}_{n_b +1} +\omega^2 {\mit \Gamma}_{n_b +2})
(K(0))^{-1}
\bigr\}
\right\|
\label{eqn:rffimSmall40a} \\
&\leq&
\lambda^2 \pi
\sum_{\scriptsize
\begin{array}{c}
s\geq0, t\geq0, u\geq0, \\
(s+t+u=r)
\end{array}}^{r}
[F_1(s,t,u)+F_2(s,t,u)+F_3(s,t,u)] ,
\label{eqn:rffimSmall40c}
\end{eqnarray}
with
\begin{eqnarray}
&\displaystyle F_1(s,t,u)=C_{stu}
\left\|
\frac{d^s \tilde{R}^+ (\omega)}{d\omega^s}
\right\|
\left\|
\frac{d^t {\mit \Gamma}(\omega)}{d\omega^t}
\right\|
\left\|
\frac{d^u}{d\omega^u}
\bigl\{ \tilde{R}^- (\omega)-(K(0))^{-1} \bigr\}
\right\| ,&
\label{eqn:rffimSmall41a}\\
%\end{equation}
%\begin{equation}
&\displaystyle F_2(s,t,u)=C_{stu}
\left\|
\frac{d^s \tilde{R}^+ (\omega)}{d\omega^s}
\right\|
\left\|
\frac{d^t}{d\omega^t}
\bigl\{
{\mit \Gamma}(\omega)
-
\omega^{n_b}({\mit \Gamma}_{n_b} +\omega {\mit \Gamma}_{n_b +1} +\omega^2 {\mit \Gamma}_{n_b +2})
\bigr\}
\right\|
&
\nonumber\\
&\hspace*{-57mm}
\displaystyle
\times
\left\|
\frac{d^u(K(0))^{-1}}{d\omega^u}
\right\|,&
\label{eqn:rffimSmall41b}\\
%\end{equation}
%\begin{equation}
&\displaystyle
F_3(s,t,u)=C_{stu}
\left\|
\frac{d^s}{d\omega^s}
\bigl\{
\tilde{R}^+ (\omega) -(K(0))^{-1}
\bigr\}
\right\|
\left\|
\frac{d^t}{d\omega^t}\omega^{n_b} ({\mit \Gamma}_{n_b} +\omega {\mit \Gamma}_{n_b +1} +\omega^2 {\mit \Gamma}_{n_b +2})
\right\|
&
\nonumber\\
&\hspace*{-67mm}
\displaystyle
\times
\left\|
\frac{d^u(K(0))^{-1}}{d\omega^u}
\right\| ,
&
\label{eqn:rffimSmall41c}
\end{eqnarray}
where $C_{stu}$'s are appropriate constants.
%Note that
%$\left\|
%$d^r (K(0))^{-1} /d\omega^r = \delta_{0r}(K(0))^{-1}$
%\right\|
%=O(1)$ for $r=0$ or $0$ $r\geq 1$,
%as $\omega \to 0$.
For $1\leq r\leq n_b +1$,
the summation of the first summand in Eq. (\ref{eqn:rffimSmall40c}) can be estimated as
\begin{eqnarray}
&&
\sum_{\scriptsize
\begin{array}{c}
s\geq0, t\geq0, u\geq0, \\
(s+t+u=r)
\end{array}}^{r}
F_1(s,t,u)
\nonumber\\
&&
=
F_1(0,r,0)
+
\sum_{t\geq 0,u\geq 1}^r
F_1(0,t,u)
+
\sum_{s\geq 1,t\geq 0}^r
F_1(s,t,0)
+
\sum_{\min\{ s, u\}\geq 1}^r F_1(s,t,u)
\nonumber \\
&&
=
O(F_1(0, r-1, 1))
=
O(\omega^{2n_b -r} \log \omega ) ,~~
\mbox{or}~~
O(\omega^{n_a+n_b -r}) ,
\label{eqn:rffimSmall47a}
\end{eqnarray}
for $n_b \leq n_a$ or $n_b > n_a$, respectively, as $\omega \to 0$.
Note that from Eq. (\ref{eqn:rffSmall60c}) for $N=0$ this estimation is valid for $r=0$ too.
For $r=n_b+2$, it is estimated as
\begin{equation}
\left\{
\begin{array}{cc}
O(\omega^{n_b-1}\log \omega) & (n_b\geq 2)\\
O((\log \omega)^2 ) & (n_b=1)
\end{array}
\right. ,~~
\mbox{or}~~
O(\omega^{n_a-1}) ,
\label{eqn:rffimSmall47b}
\end{equation}
for $n_b \leq n_a$ or $n_b > n_a$, respectively, as $\omega \to 0$.
The summation of the second summand in Eq. (\ref{eqn:rffimSmall40c}) for $0\leq r\leq n_b +3$
is also estimated as
\begin{equation}
\lambda^2 \pi
\sum_{\scriptsize
\begin{array}{c}
s\geq0, t\geq0, u\geq0, \\
(s+t+u=r)
\end{array}}^{r}
F_2 (s,t,u)
=
\sum_{(s+t=r)}^r
F_2(s, t, 0)
=
O(F_2(0, r, 0))
=
O(\omega^{n_b +3 -r}),
\label{eqn:rffimSmall47c}
\end{equation}
both for $n_b \leq n_a$ and for $n_b > n_a$, as $\omega \to 0$.
The summation of the last summand in Eq. (\ref{eqn:rffimSmall40c}) for $0\leq r\leq n_b +1$
is estimated as
\begin{eqnarray}
\lambda^2 \pi
\sum_{\scriptsize
\begin{array}{c}
s\geq0, t\geq0, u\geq0, \\
(s+t+u=r)
\end{array}}^{r}
F_3 (s,t,0)
&=&
\sum_{(s+t=r)}^r
F_3(s, t, 0)
=
O(F_3(1, r-1, 0))
\nonumber\\
&=&
O(\omega^{2n_b -r } \log \omega), ~
\mbox{or} ~
O(\omega^{n_a +n_b -r}),
\label{eqn:rffimSmall47d}
\end{eqnarray}
for $n_b \leq n_a$ or $n_a < n_b $,
respectively, as $\omega \to 0$.
For $r=n_b +2$, the estimation is replaced by
\begin{equation}
O(\omega^{n_b -2}(\log \omega)^{\theta(n_b-2)}) ,
\mbox{ or } ~
\left\{
\begin{array}{cc}
O(\omega^{[n_a -2]^+}) & (n_b \geq 3)\\
O(\log \omega) & (n_b = 2)
\end{array}
\right. ,
\label{eqn:rffimSmall50a}
\end{equation}
for $n_b \leq n_a$ or $n_a < n_b $, respectively,
as $\omega \to 0$. %where $[x]^- = \min \{0, x \}$.
Then, by summarizing the above-noted estimations from Eqs.
(\ref{eqn:rffimSmall47a}) to (\ref{eqn:rffimSmall50a}),
and by taking into account the restriction (\ref{eqn:rff.246}) again,
the proof of the lemma is completed.
\qed

%%%%%%%%%%%%%%%%%%%%%%%%%%%%%%%%%%%%%%%%%%%%%%%%%%%%%%%%%%%%%%%%%%%%%%%%%%%%%%%%%%%%%%%%%%
\subsection{The exceptional case of the first kind}
%\label{subsec:}

In this case, we first remember that from the discussion around Eq. (\ref{eqn:.340}) 
it necessarily holds that ${\mit \Gamma}_1\neq 0$, i.e., $n_b=1$ in Eqs. (\ref{eqn:rff.220}).

\begin{lm} \label{lm:1stremainder+}
: Assume that $0$ is an exceptional point of the first kind for $H$.
Then the $0$-th and the first derivative of $\tilde{R} (z)$
are approximated by those of a finite series
\begin{equation}
\frac{1}{\lambda^2 z \log z}( Q_1 {\mit \Gamma}_1 Q_1 )^{-1}
+
\frac{1}{\lambda^4 z (\log z)^2}( Q_1 {\mit \Gamma}_1 Q_1 )^{-1}
(Q_1+ \lambda^2 Q_1 A_1 Q_1 + \lambda^2  \pi i Q_1 {\mit \Gamma}_1 Q_1 )
( Q_1 {\mit \Gamma}_1 Q_1 )^{-1} ,
\label{eqn:1stSmall5}
\end{equation}
that is, it is shown that
\begin{eqnarray}
&&
\Biggl\|
\frac{d^r}{dz^r}
\Biggl[
\tilde{R} (z)
-
\frac{1}{\lambda^2 z \log z}( Q_1 {\mit \Gamma}_1 Q_1 )^{-1}
\nonumber \\
&&
-
\frac{1}{\lambda^4 z (\log z)^2}( Q_1 {\mit \Gamma}_1 Q_1 )^{-1}
(Q_1+ \lambda^2 Q_1 A_1 Q_1 + \lambda^2  \pi i Q_1 {\mit \Gamma}_1 Q_1 )
( Q_1 {\mit \Gamma}_1 Q_1 )^{-1}
\Biggr]
\Biggr\|
\nonumber \\
&&=
%\left\{
%\begin{array}{cc}
O(z^{-1} (\log z)^{-3}) ~~\mbox{for}~~r=0,~~~\mbox{or}~~~
%& (r=0) \\
O(z^{-2} (\log z)^{-3}) ~~\mbox{for}~~r=1,
%& (r=1)
%\end{array}
%\right. ,
\label{eqn:1stSmall7}
\end{eqnarray}
as $z \to 0$.
\end{lm}

{\sl Proof} :
Let us first consider the quantity that
\begin{eqnarray}
&&
\Biggl\|
\frac{d^r}{dz^r}
\Biggl[
\tilde{R} (z)
-
\frac{1}{\lambda^2 z \log z}( Q_1 {\mit \Gamma}_1 Q_1 )^{-1}
\nonumber \\
&&
-
\frac{1}{\lambda^4 z (\log z)^2}( Q_1 {\mit \Gamma}_1 Q_1 )^{-1}
(Q_1+ \lambda^2 Q_1 A_1 Q_1 + \lambda^2  \pi i Q_1 {\mit \Gamma}_1 Q_1 )
( Q_1 {\mit \Gamma}_1 Q_1 )^{-1}
\Biggr]
\Biggr\|
\nonumber \\
&&\leq
\left\|
\frac{d^r }{dz^r}
\left[
\tilde{R} (z) - Q_1 \tilde{R} (z) Q_1
\right]
\right\|
%\nonumber \\
%&&~~~
+
\left\|
\frac{d^r}{dz^r}
\Biggl[
Q_1 \tilde{R} (z) Q_1
-
\frac{1}{\lambda^2 z \log z}( Q_1 {\mit \Gamma}_1 Q_1 )^{-1}
\right.
\nonumber \\
&&~~~
\left.
-
\frac{1}{\lambda^4 z (\log z)^2}( Q_1 {\mit \Gamma}_1 Q_1 )^{-1}
(Q_1+ \lambda^2 Q_1 A_1 Q_1 + \lambda^2  \pi i Q_1 {\mit \Gamma}_1 Q_1 )
( Q_1 {\mit \Gamma}_1 Q_1 )^{-1}
\Biggr]
\right\| .
\label{eqn:1stSmall50}
\end{eqnarray}
For $r= 0$, the first term on the rhs of the above is estimated as follows:
%we have that for a small positive $z $ ($< z_0$)
\begin{equation}
\left\|
\tilde{R} (z)-Q_1 \tilde{R} (z) Q_1
\right\|
\leq
\left\|
Q_0 \tilde{R} (z) Q_0
\right\|
+
\left\|
Q_0 \tilde{R} (z) Q_1
\right\|
+
\left\|
Q_1 \tilde{R} (z) Q_0
\right\|
=O(1),
\label{eqn:1stSmall60}
\end{equation}
as $z \to 0$,
where Eqs. (\ref{eqn:5.C.1.120a})--(\ref{eqn:5.C.1.120c}) are used.
For $r=1$, one obtains
\begin{equation}
\left\|
\frac{d\tilde{R} (z)}{dz}
-
\frac{dQ_1 \tilde{R} (z) Q_1}{dz}
\right\|
\leq
\left\|
\frac{dQ_0 \tilde{R} (z) Q_0}{dz}
\right\|
+
\left\|
\frac{dQ_0 \tilde{R} (z) Q_1}{dz}
\right\|
+
\left\|
\frac{dQ_1 \tilde{R} (z) Q_0}{dz}
\right\| =O(z^{-1}).
\label{eqn:1stSmall70}
\end{equation}
In fact, by using expression (\ref{eqn:5.C.1.120a})
the first term on the rhs of the above is estimated as follows:
\begin{eqnarray}
\frac{d Q_0 \tilde{R} (z) Q_0}{dz}
&=&
-
Q_0 \tilde{R} (z) Q_0
\left( \frac{d E_{00}}{dz} -\frac{d E_{01}}{dz} E_{11}^{-1} E_{10}
-E_{01} \frac{dE_{11}^{-1} }{dz} E_{10}
-E_{01} E_{11}^{-1} \frac{dE_{10}}{dz}
\right)
\nonumber \\
&&
\times
Q_0 \tilde{R} (z) Q_0
\label{eqn:1stSmall71b}\\
&=&
O(\log z).
\label{eqn:1stSmall71c}
\end{eqnarray}
Four derivatives in Eq. (\ref{eqn:1stSmall71b})
have the same order, which can be shown from the use of
Eqs. (\ref{eqn:5.C.1.60b}), (\ref{eqn:5.C.1.90b}),
(\ref{eqn:5.C.1.100}), and (\ref{eqn:5.C.1.130}): We here note that
\begin{equation}
\frac{dE_{00}}{dz}
=O(\log z),~~~
\frac{dE_{01}}{dz}
=O(\log z),~~~
\frac{dE_{10}}{dz}
=O(\log z),~~~
\label{eqn:1stSmall72}
\end{equation}
and from Eqs. (\ref{eqn:5.C.1.60b}) and (\ref{eqn:5.C.1.90b})
\begin{eqnarray}
\frac{dE_{11}^{-1} }{dz}
&=&
E_{11}^{-1}
\left\{
\frac{d}{dz}
Q_1 \left\{
z +\lambda^2 [ A(z) -\log z {\mit \Gamma}(z) + \pi i {\mit \Gamma}(z) ]
\right\} Q_1
\right\}
E_{11}^{-1}
\label{eqn:1stSmall73a}\\
&=&
E_{11}^{-1}
Q_1 \left\{
1 +\lambda^2 \left[ \frac{dA(z)}{dz}-{\mit \Gamma}(z)/z
-\log z \frac{d{\mit \Gamma}(z)}{dz}
+ \pi i \frac{d{\mit \Gamma}(z)}{dz} \right]
\right\} Q_1
E_{11}^{-1}
\label{eqn:1stSmall73b}\\
&=&
O(z^{-2} (\log z)^{-1} ).
\label{eqn:1stSmall73c}
\end{eqnarray}
In the same way, the second term on the rhs of Eq. (\ref{eqn:1stSmall70})
is also estimated as follows:
\begin{eqnarray}
\frac{d Q_0 \tilde{R} (z) Q_1}{dz}
&=&
-\left(
\frac{dE_{00}^{-1}}{dz} E_{01}
+
E_{00}^{-1} \frac{dE_{01}}{dz}
\right)
Q_1\tilde{R}Q_1
+
E_{00}^{-1} E_{01}
Q_1\tilde{R}Q_1
\nonumber \\
&&
\times
\left( \frac{d E_{11}}{dz} -\frac{d E_{10}}{dz} E_{00}^{-1} E_{01}
-E_{10} \frac{dE_{00}^{-1} }{dz} E_{01}
-E_{10} E_{00}^{-1} \frac{dE_{01}}{dz}
\right)
Q_1\tilde{R}Q_1
\label{eqn:1stSmall74b}\\
&=&
%O(\log z )
%+
%O(1 z^{-1} (\log z )^{1-1})
%+
%O(z^{1-2} (\log z )^{1-2+1})
%=
O(z^{-1}),
\label{eqn:1stSmall74c}
\end{eqnarray}
where we used Eq. (\ref{eqn:5.C.1.120b}) and the fact that
\begin{equation}
\frac{dE_{00}^{-1} }{dz}=O(\log z),~~~
\frac{dE_{11}}{dz}=O(\log z).
\label{eqn:1stSmall75}
\end{equation}
In a similar manner, we can also show
\begin{equation}
\frac{d Q_1 \tilde{R} (z) Q_0}{dz}=O(z^{-1}).
\label{eqn:1stSmall76}
\end{equation}

Let us next consider the last term in Eq. (\ref{eqn:1stSmall50}). For $r=0$, it reads
\begin{eqnarray}
&&
\left\|
Q_1 \tilde{R} (z) Q_1
-
\frac{1}{\lambda^2 z \log z}( Q_1 {\mit \Gamma}_1 Q_1 )^{-1}
\right.
\nonumber \\
&&
\left.
-
\frac{1}{\lambda^4 z (\log z)^2}( Q_1 {\mit \Gamma}_1 Q_1 )^{-1}
(Q_1+ \lambda^2 Q_1 A_1 Q_1 + \lambda^2  \pi i Q_1 {\mit \Gamma}_1 Q_1 )
( Q_1 {\mit \Gamma}_1 Q_1 )^{-1}
\right\|
\label{eqn:1stSmall77a}\\
%&=&
%\left\|
%(E_{11}-E_{10} E_{00}^{-1} E_{01})^{-1}
%-
%\frac{1}{\lambda^2 z \log z}( Q_1 {\mit \Gamma}_1 Q_1 )^{-1}
%\right.
%\nonumber \\
%&&
%\left.
%-
%\frac{1}{\lambda^4 z (\log z)^2}( Q_1 {\mit \Gamma}_1 Q_1 )^{-1}
%(Q_1+ \lambda^2 Q_1 A_1 Q_1 + \lambda^2  \pi i Q_1 {\mit \Gamma}_1 Q_1 )
%( Q_1 {\mit \Gamma}_1 Q_1 )^{-1}
%\right\|
%\label{eqn:1stSmall77b}\\
&\leq&
\left\|
E_{11}^{-1}
-
\frac{1}{\lambda^2 z \log z}( Q_1 {\mit \Gamma}_1 Q_1 )^{-1}
\right.
\nonumber \\
&&
\left.
-
\frac{1}{\lambda^4 z (\log z)^2}( Q_1 {\mit \Gamma}_1 Q_1 )^{-1}
(Q_1+ \lambda^2 Q_1 A_1 Q_1 + \lambda^2  \pi i Q_1 {\mit \Gamma}_1 Q_1 )
( Q_1 {\mit \Gamma}_1 Q_1 )^{-1}
\right\|
\nonumber \\
&&
+\| (Q_{11} -E_{11}^{-1}E_{10} E_{00}^{-1} E_{01})^{-1}E_{11}^{-1} -E_{11}^{-1} \|
\label{eqn:1stSmall77c}\\
&\leq&
\biggl\|
\tilde{E}_{11}(z)\frac{1}{\lambda^2 z \log z}( Q_1 {\mit \Gamma}_1 Q_1 )^{-1}
\nonumber \\
&&
-
\frac{1}{\lambda^4 z (\log z)^2}( Q_1 {\mit \Gamma}_1 Q_1 )^{-1}
(Q_1+ \lambda^2 Q_1 A_1 Q_1 + \lambda^2  \pi i Q_1 {\mit \Gamma}_1 Q_1 )
( Q_1 {\mit \Gamma}_1 Q_1 )^{-1}
\biggr\|
\nonumber \\
&&
+
\Biggl\|
\sum_{j=2}^{\infty}
(\tilde{E}_{11}(z))^j
\frac{1}{\lambda^2 z \log z}( Q_1 {\mit \Gamma}_1 Q_1 )^{-1}
\Biggr\|
+
O(1)
\label{eqn:1stSmall77d}\\
%&=&
%\biggl\|
%\frac{1}{\lambda^2z^2  (\log z )^2}
%( Q_1 {\mit \Gamma}_1 Q_1 )^{-1}
%\bigl\{ \lambda^2 Q_1 \{ A(z) -z A_1-\log z [{\mit \Gamma}(z) -z {\mit \Gamma}_1 ]
%+ \pi i[ {\mit \Gamma}(z) -z {\mit \Gamma}_1 ]\} Q_1
%\bigr\}
%(Q_1 {\mit \Gamma}_1 Q_1 )^{-1}
%\biggr\|
%\nonumber \\
%O((\log z )^{-1})+
%O( z^{-1}  (\log z )^{-3} )+O(1)
%\label{eqn:1stSmall77e}\\
&=&
O( z^{-1}  (\log z )^{-3} ), %+O(1),
\label{eqn:1stSmall77f}
\end{eqnarray}
where in the second inequality we used Eq. (\ref{eqn:5.C.1.90a}) and that
%The last term in Eq. (\ref{eqn:1stSmall77c}) behaves for small $z$ as
\begin{equation}
\| (Q_{11} -E_{11}^{-1}E_{10} E_{00}^{-1} E_{01})^{-1}E_{11}^{-1} -E_{11}^{-1} \|
%\sum_{j=1}^{\infty}
%\| E_{11}^{-1} E_{10} E_{00}^{-1} E_{01} \|^j
%\| E_{11}^{-1} \|
%=
\leq
\frac{\| E_{11}^{-1} E_{10} E_{00}^{-1} E_{01} \|
\| E_{11}^{-1} \|}{1-\| E_{11}^{-1} E_{10} E_{00}^{-1} E_{01} \| }
=O(1),
\label{eqn:1stSmall78}
\end{equation}
as $z \to 0$.
Substituting Eqs. (\ref{eqn:1stSmall60}) and (\ref{eqn:1stSmall77f})
into Eq. (\ref{eqn:1stSmall50}), we can obtain the estimation (\ref{eqn:1stSmall7}) for $r=0$.

For $r=1$, we can obtain
\begin{eqnarray}
&&
\Biggl\|
\frac{d}{dz}
\biggl[
Q_1 \tilde{R} (z) Q_1
-
\frac{1}{\lambda^2 z \log z}( Q_1 {\mit \Gamma}_1 Q_1 )^{-1}
\nonumber \\
&&
-
\frac{1}{\lambda^4 z (\log z)^2}( Q_1 {\mit \Gamma}_1 Q_1 )^{-1}
(Q_1+ \lambda^2 Q_1 A_1 Q_1 + \lambda^2  \pi i Q_1 {\mit \Gamma}_1 Q_1 )
( Q_1 {\mit \Gamma}_1 Q_1 )^{-1}
\biggr]
\Biggr\|
\label{eqn:1stSmall80a}\\
&\leq&
\Biggl\|
\frac{d}{dz}
\biggl[
E_{11}^{-1}
-
\frac{1}{\lambda^2 z \log z}( Q_1 {\mit \Gamma}_1 Q_1 )^{-1}
\nonumber \\
&&
-
\frac{1}{\lambda^4 z (\log z)^2}( Q_1 {\mit \Gamma}_1 Q_1 )^{-1}
(Q_1+ \lambda^2 Q_1 A_1 Q_1 + \lambda^2  \pi i Q_1 {\mit \Gamma}_1 Q_1 )
( Q_1 {\mit \Gamma}_1 Q_1 )^{-1}
\biggr]
\Biggr\|
\nonumber \\
&&
+\biggl\| \frac{d}{dz}
\bigl\{ (Q_{11} -E_{11}^{-1}E_{10} E_{00}^{-1} E_{01})^{-1}E_{11}^{-1}
-E_{11}^{-1}
\bigr\}
\biggr\|
\label{eqn:1stSmall80b}\\
%&\leq&
%\biggl\|
%E_{11}^{-1}
%\left\{ \frac{d}{dz}
%Q_1 \left\{
%z +\lambda^2 [ A(z) -\log z {\mit \Gamma}(z) + \pi i {\mit \Gamma}(z) ]
%\right\} Q_1 \right\}
%E_{11}^{-1}
%\nonumber \\
%&&
%-
%\frac{d}{dz}
%\frac{1}{\lambda^2 z \log z}( Q_1 {\mit \Gamma}_1 Q_1 )^{-1}
%-
%\frac{d}{dz}
%\frac{1}{\lambda^4 z (\log z)^2}( Q_1 {\mit \Gamma}_1 Q_1 )^{-1}
%(Q_1+ \lambda^2 Q_1 A_1 Q_1 + \lambda^2  \pi i Q_1 {\mit \Gamma}_1 Q_1 )
%( Q_1 {\mit \Gamma}_1 Q_1 )^{-1}
%\biggr\|
%\nonumber \\
%&&
%+\biggl\|
%\bigl\{ (Q_{11} -E_{11}^{-1}E_{10} E_{00}^{-1} E_{01})^{-1} -Q_{11} \bigr\}
%\frac{dE_{11}^{-1}}{dz}
%\biggr\|
%+\biggl\| E_{11}^{-1}
%\frac{d}{dz}
%(Q_{11} -E_{11}^{-1}E_{10} E_{00}^{-1} E_{01})^{-1}
%\biggr\|
%\label{eqn:1stSmall80c}\\
&\leq&
\biggl\|
E_{11}^{-1}
\left\{ \frac{d}{dz}
Q_1 \left\{
z +\lambda^2 [ A(z) + \pi i {\mit \Gamma}(z) ]
\right\} Q_1 \right\}
E_{11}^{-1}
\nonumber \\
&&
-
\frac{1}{\lambda^4 z^2 (\log z)^2}( Q_1 {\mit \Gamma}_1 Q_1 )^{-1}
(Q_1+ \lambda^2 Q_1 A_1 Q_1 + \lambda^2  \pi i Q_1 {\mit \Gamma}_1 Q_1 )
( Q_1 {\mit \Gamma}_1 Q_1 )^{-1}
\biggr\|
\nonumber \\
&&
+
\biggl\|
-E_{11}^{-1}
\biggl\{
\frac{d}{dz}
\lambda^2 (\log z) Q_1 {\mit \Gamma}(z) Q_1
\biggr\}
E_{11}^{-1}
-\frac{d}{dz}
\frac{1}{\lambda^2 z \log z}( Q_1 {\mit \Gamma}_1 Q_1 )^{-1}
\nonumber \\
&&
-
z
\biggl\{
\frac{d}{dz}
\frac{1}{\lambda^4 z^2 (\log z)^2}
\biggr\}( Q_1 {\mit \Gamma}_1 Q_1 )^{-1}
(Q_1+ \lambda^2 Q_1 A_1 Q_1 + \lambda^2  \pi i Q_1 {\mit \Gamma}_1 Q_1 )
( Q_1 {\mit \Gamma}_1 Q_1 )^{-1}
\biggr\|
\nonumber \\
&&
+\biggl\|
\bigl\{ (Q_{11} -E_{11}^{-1}E_{10} E_{00}^{-1} E_{01})^{-1} -Q_{11} \bigr\}
\frac{dE_{11}^{-1}}{dz}
\biggr\|
\nonumber \\
&&
+
\biggl\|
\left\{
\frac{d}{dz}
(Q_{11} -E_{11}^{-1}E_{10} E_{00}^{-1} E_{01})^{-1}
\right\}
E_{11}^{-1}
\biggr\|
\label{eqn:1stSmall80d}\\
&=&O(z^{-2}(\log z)^{-3}),
\label{eqn:1stSmall80e}
\end{eqnarray}
where we used the expression for $dE_{11}^{-1}/dz$
in Eq. (\ref{eqn:1stSmall73a}).
%instead of Eq. (\ref{eqn:5.C.1.90a}).
Actually, the first term in Eq. (\ref{eqn:1stSmall80d}) is estimated as
\begin{eqnarray}
&&\hspace*{-13mm}
\biggl\|
E_{11}^{-1}
\left\{ \frac{d}{dz}
Q_1 \left\{
z +\lambda^2 [ A(z) + \pi i {\mit \Gamma}(z) ]
\right\} Q_1 \right\}
E_{11}^{-1}
\nonumber \\
&&\hspace*{-13mm}
-
\frac{1}{\lambda^4 z^2 (\log z)^2}( Q_1 {\mit \Gamma}_1 Q_1 )^{-1}
(Q_1+ \lambda^2 Q_1 A_1 Q_1 + \lambda^2  \pi i Q_1 {\mit \Gamma}_1 Q_1 )
( Q_1 {\mit \Gamma}_1 Q_1 )^{-1}
\biggr\|
\nonumber \\
&&\hspace*{-13mm}
\leq
\biggl\|
E_{11}^{-1}-\frac{1}{\lambda^2 z \log z}( Q_1 {\mit \Gamma}_1 Q_1 )^{-1}
\biggr\|
\biggl\|
\frac{d}{dz}
Q_1 \left\{
z +\lambda^2 [ A(z) + \pi i {\mit \Gamma}(z) ]
\right\} Q_1
\biggr\|
\| E_{11}^{-1} \|
\nonumber \\
&&\hspace*{-13mm}
~~~+
\frac{\| ( Q_1 {\mit \Gamma}_1 Q_1 )^{-1} \| }{\lambda^2 |z \log z|}
\biggl\|
\frac{d}{dz}
Q_1 \left\{
z +\lambda^2 [ A(z) + \pi i {\mit \Gamma}(z) ]
\right\} Q_1
\biggr\|
\biggl\|
E_{11}^{-1}-\frac{1}{\lambda^2 z \log z}( Q_1 {\mit \Gamma}_1 Q_1 )^{-1}
\biggr\|
\nonumber \\
&&\hspace*{-13mm}
~~~+
\frac{\|( Q_1 {\mit \Gamma}_1 Q_1 )^{-1} \|^2}{\lambda^4 z^2 (\log z)^2 }
\biggl\|
\frac{d}{dz}
Q_1 \left\{
z +\lambda^2 [ A(z) + \pi i {\mit \Gamma}(z) ]
\right\} Q_1
-
Q_1 (1+ \lambda^2 A_1 + \lambda^2  \pi i {\mit \Gamma}_1 )Q_1
\biggr\|
\label{eqn:1stSmall90a}\\
%&=&
%O(z^{-2}(\log z)^{-3})+O(z^{-2}(\log z)^{-3})+O(z^{-1}(\log z)^{-2})
%\nonumber \\
&&\hspace*{-13mm}
=
O(z^{-2}(\log z)^{-3}),
\label{eqn:1stSmall90b}
\end{eqnarray}
as $z \to 0$.
On the other hand, the second term in Eq. (\ref{eqn:1stSmall80d}) is
slightly complicated to evaluate:
\begin{eqnarray}
&&
\biggl\|
-E_{11}^{-1}
\biggl\{
\frac{d}{dz}
\lambda^2 (\log z) Q_1 {\mit \Gamma}(z) Q_1
\biggr\}
E_{11}^{-1}
-\frac{d}{dz}
\frac{1}{\lambda^2 z \log z}( Q_1 {\mit \Gamma}_1 Q_1 )^{-1}
\nonumber \\
&&
-
z \biggl\{
\frac{d}{dz}
\frac{1}{\lambda^4 z^2 (\log z)^2}
\biggr\}( Q_1 {\mit \Gamma}_1 Q_1 )^{-1}
(Q_1+ \lambda^2 Q_1 A_1 Q_1 + \lambda^2  \pi i Q_1 {\mit \Gamma}_1 Q_1 )
( Q_1 {\mit \Gamma}_1 Q_1 )^{-1}
\biggr\|
%\nonumber \\
%&&
\label{eqn:1stSmall95a}\\
&\leq&
\biggl\|
-\frac{1}{\lambda^2 z \log z}( Q_1 {\mit \Gamma}_1 Q_1 )^{-1}
\biggl\{
\frac{d}{dz}
\lambda^2 (\log z) Q_1 {\mit \Gamma}(z) Q_1
\biggr\}
\frac{1}{\lambda^2 z \log z}( Q_1 {\mit \Gamma}_1 Q_1 )^{-1}
%\biggr\|
\nonumber \\
&&
-\frac{d}{dz}
\frac{1}{\lambda^2 z \log z}( Q_1 {\mit \Gamma}_1 Q_1 )^{-1}
\nonumber \\
&&
-\frac{1}{\lambda^2 z \log z}( Q_1 {\mit \Gamma}_1 Q_1 )^{-1}
\biggl\{
\frac{d}{dz}
\lambda^2 (\log z) Q_1 {\mit \Gamma}(z) Q_1
\biggr\}
\tilde{E}_{11}\frac{1}{\lambda^2 z \log z}( Q_1 {\mit \Gamma}_1 Q_1 )^{-1}
\nonumber \\
&&
-\tilde{E}_{11}\frac{1}{\lambda^2 z \log z}( Q_1 {\mit \Gamma}_1 Q_1 )^{-1}
\biggl\{
\frac{d}{dz}
\lambda^2 (\log z) Q_1 {\mit \Gamma}(z) Q_1
\biggr\}
\frac{1}{\lambda^2 z \log z}( Q_1 {\mit \Gamma}_1 Q_1 )^{-1}
\nonumber \\
&&
+
z
\frac{2}{(\lambda^2 z \log z )^3}
\biggl(
\frac{d}{dz}
\lambda^2 z \log z
\biggr)
( Q_1 {\mit \Gamma}_1 Q_1 )^{-1}
(Q_1+ \lambda^2 Q_1 A_1 Q_1 + \lambda^2  \pi i Q_1 {\mit \Gamma}_1 Q_1 )
( Q_1 {\mit \Gamma}_1 Q_1 )^{-1}
\biggr\|
\nonumber \\
&&
+
\biggl\|
\frac{1}{\lambda^2 z \log z}( Q_1 {\mit \Gamma}_1 Q_1 )^{-1}
\biggl\{
\frac{d}{dz}
\lambda^2 (\log z) Q_1 {\mit \Gamma}(z) Q_1
\biggr\}
\sum_{j=2}^{\infty}\tilde{E}_{11}^j \frac{1}{\lambda^2 z \log z}( Q_1 {\mit \Gamma}_1 Q_1 )^{-1}
\biggr\|
\nonumber \\
&&
+
\biggl\|
\tilde{E}_{11} \frac{1}{\lambda^2 z \log z}( Q_1 {\mit \Gamma}_1 Q_1 )^{-1}
\biggl\{
\frac{d}{dz}
\lambda^2 (\log z) Q_1 {\mit \Gamma}(z) Q_1
\biggr\}
\sum_{j=1}^{\infty}\tilde{E}_{11}^j \frac{1}{\lambda^2 z \log z}( Q_1 {\mit \Gamma}_1 Q_1 )^{-1}
\biggr\|
\nonumber \\
&&
+
\biggl\|
\sum_{j=2}^{\infty}\tilde{E}_{11}^j \frac{1}{\lambda^2 z \log z}( Q_1 {\mit \Gamma}_1 Q_1 )^{-1}
\biggl\{
\frac{d}{dz}
\lambda^2 (\log z) Q_1 {\mit \Gamma}(z) Q_1
\biggr\}
E_{11}^{-1}
\biggr\|
\label{eqn:1stSmall95b}\\
%\end{eqnarray}
%\begin{eqnarray}
&\leq&
\biggl\|
-\frac{1}{\lambda^2 z \log z}( Q_1 {\mit \Gamma}_1 Q_1 )^{-1}
\biggl\{
\frac{d}{dz}
\lambda^2 \log z Q_1 {\mit \Gamma}(z) Q_1
\biggr\}
\frac{1}{\lambda^2 z \log z}( Q_1 {\mit \Gamma}_1 Q_1 )^{-1}
\nonumber \\
&&
-\frac{d}{dz}
\frac{1}{\lambda^2 z \log z}( Q_1 {\mit \Gamma}_1 Q_1 )^{-1}
\biggr\|
\nonumber \\
&&
+
\biggl\|
-\frac{1}{\lambda^2 z \log z}( Q_1 {\mit \Gamma}_1 Q_1 )^{-1}
\biggl\{
\frac{d}{dz}
\lambda^2 (\log z) Q_1 {\mit \Gamma}(z) Q_1
\biggr\}
\tilde{E}_{11}
\frac{1}{\lambda^2 z \log z}( Q_1 {\mit \Gamma}_1 Q_1 )^{-1}
\nonumber \\
&&
+
z \frac{1}{(\lambda^2 z \log z )^3}
\biggl(
\frac{d}{dz}
\lambda^2 z \log z
\biggr)
( Q_1 {\mit \Gamma}_1 Q_1 )^{-1}
z (Q_1+ \lambda^2 Q_1 A_1 Q_1 + \lambda^2  \pi i Q_1 {\mit \Gamma}_1 Q_1 )
( Q_1 {\mit \Gamma}_1 Q_1 )^{-1}
\biggr\|
\nonumber \\
&&
+
\biggl\|
-\tilde{E}_{11}
\frac{1}{\lambda^2 z \log z}( Q_1 {\mit \Gamma}_1 Q_1 )^{-1}
\biggl\{
\frac{d}{dz}
\lambda^2 (\log z) Q_1 {\mit \Gamma}(z) Q_1
\biggr\}
\frac{1}{\lambda^2 z \log z}( Q_1 {\mit \Gamma}_1 Q_1 )^{-1}
\nonumber \\
&&
+
z
\frac{1}{(\lambda^2 z \log z )^3}
\biggl(
\frac{d}{dz}
\lambda^2 z \log z
\biggr) ( Q_1 {\mit \Gamma}_1 Q_1 )^{-1}
(Q_1+ \lambda^2 Q_1 A_1 Q_1 + \lambda^2  \pi i Q_1 {\mit \Gamma}_1 Q_1 )
( Q_1 {\mit \Gamma}_1 Q_1 )^{-1}
\biggr\|
\nonumber \\
&&
+O(z^{-2}(\log z)^{-3} )
\label{eqn:1stSmall95c} \\
&=&O(z^{-2}(\log z)^{-3} ).
\label{eqn:1stSmall95d}
\end{eqnarray}
In fact, the first term in Eq. (\ref{eqn:1stSmall95c}) reads
\begin{eqnarray}
&&
\biggl\|
-\frac{1}{\lambda^2 z \log z}( Q_1 {\mit \Gamma}_1 Q_1 )^{-1}
\biggl\{
\frac{d}{dz}
\lambda^2 \log z Q_1 {\mit \Gamma}(z) Q_1
\biggr\}
\frac{1}{\lambda^2 z \log z}( Q_1 {\mit \Gamma}_1 Q_1 )^{-1}
\nonumber\\
&&
-\frac{d}{dz}
\frac{1}{\lambda^2 z \log z}( Q_1 {\mit \Gamma}_1 Q_1 )^{-1}
\biggr\|
\nonumber\\
&\leq&
\biggl\|
-(Q_1 {\mit \Gamma}_1 Q_1 )^{-1}
\biggl\{
\frac{d}{dz}
\lambda^2 (\log z) Q_1 {\mit \Gamma}(z) Q_1
\biggr\}
+\frac{d}{dz}
\lambda^2 z (\log z) Q_1
\biggr\|
\frac{\| ( Q_1 {\mit \Gamma}_1 Q_1 )^{-1} \|}{(\lambda^2z  \log z)^{2}}
\label{eqn:1stSmall96b}\\
&\leq&
\lambda^2
\left\{
\biggl\|
-(Q_1 {\mit \Gamma}_1 Q_1 )^{-1}
Q_1 \frac{{\mit \Gamma}(z)}{z} Q_1
+
Q_1
\biggr\|
+
\biggl\|
-(Q_1 {\mit \Gamma}_1 Q_1 )^{-1}
Q_1 \frac{d{\mit \Gamma}(z)}{dz} Q_1
+
Q_1
\biggr\|
(\log z)
\right\}
\nonumber \\
&&
\times
\| ( Q_1 {\mit \Gamma}_1 Q_1 )^{-1} \|
\frac{1}{\lambda^2 z \log z}
\label{eqn:1stSmall96c}\\
&=&
%O(z^{-1}(\log z)^{-2} ) +O((z \log z)^{-1} )=
O((z \log z)^{-1}).
\label{eqn:1stSmall96d}
\end{eqnarray}
The second term in Eq. (\ref{eqn:1stSmall95c}) also reads
\begin{eqnarray}
&&
\biggl\|
-\frac{1}{\lambda^2 z \log z}( Q_1 {\mit \Gamma}_1 Q_1 )^{-1}
\biggl\{
\frac{d}{dz}
\lambda^2 (\log z) Q_1 {\mit \Gamma}(z) Q_1
\biggr\}
\tilde{E}_{11}
\frac{1}{\lambda^2 z \log z}( Q_1 {\mit \Gamma}_1 Q_1 )^{-1}
\nonumber \\
&&
+
z \frac{1}{(\lambda^2 z \log z )^3}
\biggl(
\frac{d}{dz}
\lambda^2 z \log z
\biggr)
( Q_1 {\mit \Gamma}_1 Q_1 )^{-1}
(Q_1+ \lambda^2 Q_1 A_1 Q_1 + \lambda^2  \pi i Q_1 {\mit \Gamma}_1 Q_1 )
( Q_1 {\mit \Gamma}_1 Q_1 )^{-1}
\biggr\|
\nonumber\\
&\leq&
\|(Q_1 {\mit \Gamma}_1 Q_1 )^{-1}\|
\biggl\|
-
\biggl\{
\frac{d}{dz}
\lambda^2 (\log z) Q_1 {\mit \Gamma}(z) Q_1
\biggr\}
\nonumber \\
&&
\times
(Q_1 {\mit \Gamma}_1 Q_1 )^{-1}
\bigl\{ z Q_1
+\lambda^2 Q_1 \{ A(z) -\log z [{\mit \Gamma}(z) -z {\mit \Gamma}_1 ]
+ \pi i {\mit \Gamma}(z) \} Q_1
\bigr\}
\nonumber \\
&&
+
z \biggl(
\frac{d}{dz}
\lambda^2 z \log z
\biggr)
(Q_1+ \lambda^2 Q_1 A_1 Q_1 + \lambda^2  \pi i Q_1 {\mit \Gamma}_1 Q_1 )
\biggr\|
\frac{\|( Q_1 {\mit \Gamma}_1 Q_1 )^{-1} \| }{(\lambda^2 z \log z )^{3}}
\label{eqn:1stSmall97b}\\
&\leq&
\|(Q_1 {\mit \Gamma}_1 Q_1 )^{-1}\|
\biggl\|
\biggl\{
\frac{d}{dz}
\lambda^2 (\log z) Q_1 {\mit \Gamma}(z) Q_1
\biggr\}
(Q_1 {\mit \Gamma}_1 Q_1 )^{-1}
\biggr\|
\nonumber \\
&&
\times
\bigl\|
-
\bigl\{ z Q_1
+\lambda^2 Q_1 \{ A(z) -\log z [{\mit \Gamma}(z) -z {\mit \Gamma}_1 ]
+ \pi i {\mit \Gamma}(z) \} Q_1
\bigr\}
\nonumber \\
&&
+
z (Q_1+ \lambda^2 Q_1 A_1 Q_1 + \lambda^2  \pi i Q_1 {\mit \Gamma}_1 Q_1 )
\bigr\|
\frac{\|( Q_1 {\mit \Gamma}_1 Q_1 )^{-1} \|}{ (\lambda^2 z \log z )^{3}}
\nonumber \\
&&
+
\|(Q_1 {\mit \Gamma}_1 Q_1 )^{-1}\|
\biggl\|
-
\biggl\{
\frac{d}{dz}
\lambda^2 (\log z) Q_1 {\mit \Gamma}(z) Q_1
\biggr\}
(Q_1 {\mit \Gamma}_1 Q_1 )^{-1}
+
\biggl(
\frac{d}{dz}
\lambda^2 z \log z
\biggr)
Q_1
\biggr\|
\nonumber \\
&&
\times
\bigl\|
z (Q_1+ \lambda^2 Q_1 A_1 Q_1 + \lambda^2  \pi i Q_1 {\mit \Gamma}_1 Q_1 )
\bigr\|
\frac{\|( Q_1 {\mit \Gamma}_1 Q_1 )^{-1} \|}{ (\lambda^2 z \log z )^{3}}
\label{eqn:1stSmall97c}\\
&=&
%O(z^{-3+2}(\log z)^{-3+1+1})+O(z^{-3+1+1}(\log z)^{-3+1})
%=
%O(z^{-1}(\log z)^{-1})+O(z^{-1}(\log z)^{-2})
%=
O((z \log z)^{-1}),
\label{eqn:1stSmall97d}
\end{eqnarray}
as $z \to 0$.
The third term gives the same contribution to the order as the second one does.
Furthermore, the last term in Eq. (\ref{eqn:1stSmall95c}) comes from the estimations of
the second, third, and last terms in Eq. (\ref{eqn:1stSmall95b}),
where each contributes the same order as $O(z^{-2}(\log z)^{-3} )$.
Therefore, Eq. (\ref{eqn:1stSmall95d}) is proved.

On the other hand, the third term in Eq. (\ref{eqn:1stSmall80d}) reads
\begin{equation}
\biggl\|
\bigl\{ (Q_{11} -E_{11}^{-1}E_{10} E_{00}^{-1} E_{01})^{-1} -Q_{11} \bigr\}
\frac{dE_{11}^{-1}}{dz}
\biggr\|
%\nonumber \\
%&\leq&
\leq
\frac
{
\bigl\|
E_{11}^{-1}E_{10} E_{00}^{-1} E_{01}
\bigr\|}
{
1-\bigl\|
E_{11}^{-1}E_{10} E_{00}^{-1} E_{01}
\bigr\|
}
\biggl\|
\frac{dE_{11}^{-1}}{dz}
\biggr\|
=O(z^{-1}),
\label{eqn:1stSmall100b}
\end{equation}
as $z \to 0$,
where Eqs. (\ref{eqn:5.C.1.90b}), (\ref{eqn:5.C.1.100}), (\ref{eqn:5.C.1.130}),
and (\ref{eqn:1stSmall73c}) are used.
In the same way, the last term in Eq. (\ref{eqn:1stSmall80d}) reads
\begin{eqnarray}
&&
\hspace*{-5mm}
\biggl\| E_{11}^{-1}
\frac{d}{dz}
(Q_{11} -E_{11}^{-1}E_{10} E_{00}^{-1} E_{01})^{-1}
\biggr\|
\nonumber\\
&\leq&
\| E_{11}^{-1} \|
\biggl\|
(Q_{11} -E_{11}^{-1}E_{10} E_{00}^{-1} E_{01})^{-1}
\biggr\|^2
\biggl[
\biggl\|
\frac{dE_{11}^{-1}}{dz}
E_{10} E_{00}^{-1} E_{01}
\biggr\|
\nonumber \\
&&
+\biggl\|
E_{11}^{-1}\frac{dE_{10} }{dz}
E_{00}^{-1} E_{01}
\biggr\|
+\biggl\|
E_{11}^{-1}E_{10} E_{00}^{-1}\frac{d E_{01}}{dz}
\biggr\|
+\biggl\|
E_{11}^{-1}E_{10} \frac{d E_{00}^{-1}}{dz} E_{01}
\biggr\|
\biggr]
\label{eqn:1stSmall110a}\\
&=&
O(z^{-1} \log z ),
%[O((\log z)^{1})+O((\log z)^{1})+O((\log z)^{1})+O(z (\log z)^{2})]
\label{eqn:1stSmall110b}
\end{eqnarray}
as $z \to 0$.
By substituting Eqs. (\ref{eqn:1stSmall90b}), (\ref{eqn:1stSmall95d}), (\ref{eqn:1stSmall100b}),
and (\ref{eqn:1stSmall110b}) into Eq. (\ref{eqn:1stSmall80d}),
one finally obtains Eq. (\ref{eqn:1stSmall80e}).
We can now show Eq. (\ref{eqn:1stSmall7}) for $r=1$ by setting Eqs. (\ref{eqn:1stSmall70})
and (\ref{eqn:1stSmall80e}) into Eq. (\ref{eqn:1stSmall50}).
\qed

If we start with expression (\ref{eqn:5.C.1.125b}),
we obtain the following lemma instead of Lemma \ref{lm:1stremainder+}.

\begin{lm} \label{lm:1stremainder-}
: Assume that $0$ is an exceptional point of the first kind for $H$.
Then the $0$-th and the first derivative of $\tilde{R} (z)$
are approximated by those of a finite series
\begin{eqnarray}
&&
\frac{1}{\lambda^2 z (\log z -2\pi i)}( Q_1 {\mit \Gamma}_1 Q_1 )^{-1}
\nonumber\\
&&
+
\frac{1}{\lambda^4 z (\log z -2\pi i)^2}( Q_1 {\mit \Gamma}_1 Q_1 )^{-1}
(Q_1+ \lambda^2 Q_1 A_1 Q_1 - \lambda^2  \pi i Q_1 {\mit \Gamma}_1 Q_1 )
( Q_1 {\mit \Gamma}_1 Q_1 )^{-1} ,
\label{eqn:1stSmall5-}
\end{eqnarray}
that is, it is shown that
\begin{eqnarray}
&&\hspace{-10mm}
\biggl\|
\frac{d^r}{dz^r}
\biggl[
\tilde{R}(z)
-
\frac{1}{\lambda^2 z (\log z -2\pi i)}( Q_1 {\mit \Gamma}_1 Q_1 )^{-1}
\nonumber \\
&&
-
\frac{1}{\lambda^4 z (\log z -2\pi i)^2}( Q_1 {\mit \Gamma}_1 Q_1 )^{-1}
(Q_1+ \lambda^2 Q_1 A_1 Q_1 - \lambda^2  \pi i Q_1 {\mit \Gamma}_1 Q_1 )
( Q_1 {\mit \Gamma}_1 Q_1 )^{-1}
\biggr]
\biggr\|
\nonumber \\
&=&
%\left\{
%\begin{array}{cc}
O(z^{-1} (\log z -2\pi i)^{-3}) ~~\mbox{for}~~r=0,~~~\mbox{or}~~~
%& (r=0) \\
O(z^{-2} (\log z -2\pi i)^{-3}) ~~\mbox{for}~~r=1,
%& (r=1)
%\end{array}
%\right. ,
\label{eqn:1stSmall7-}
\end{eqnarray}
as $z \to 0$.
\end{lm}

%%%%%%%%%%%%%%%%%%%%%%%%%%%%%%%%%%%%%%%%%%%%%%%%%%%%%%%%%%%%%%%%%%%%%%%%%%%%%%%%

\section{The reduced time evolution operator}
\label{sec:6}

In this section, we show that the reduced time evolution operator
%defined below
is expressed by the Fourier transform of the imaginary part of the reduced resolvent
both in the regular case and the exceptional case of the first kind.
We here define the reduced time evolution operator by the $N\times N$ matrix $\tilde{U}(t)$
of the components
$\tilde{U}_{mn}(t):=\langle m |Pe^{-itH}P| n \rangle$,
where $P=E((0, \infty ))$ and $\{ E(B) | B \in \mathbb{B} \}$ is
the spectral measure of $H$, which is a family of the projection operator.
$\mathbb{B}$ is the Borel field of $\mathbb{R}$.

\begin{lm}
\label{lm:reduced time evolution operator }
:
We assume that Eq. (\ref{eqn:.110}) holds so that there is no positive eigenvalue.
Then, for the system with the rational form-factor (\ref{eqn:formfactor1}),
it holds that
\begin{equation}
\tilde{U}(t)
=
\frac{1}{\pi}
\int_{(0, \infty )}
e^{-it\omega} {\rm Im} \tilde{R}^+ (\omega) d\omega
=
\lim_{r \to +0}
\frac{1}{\pi}
\int_r^\infty
e^{-it\omega} {\rm Im} \tilde{R}^+ (\omega) d\omega ,
\label{eqn:6.120}
\end{equation}
both in the regular case and the exceptional case of the first kind,
where
\begin{eqnarray}
{\rm Im} \tilde{R}^+ (\omega)
:=
\frac{1}{2 i} [\tilde{R}^+(\omega) -\tilde{R}^-(\omega)] ,
\label{eqn:6.65}
\end{eqnarray}
which is sometimes called the spectral density.
\end{lm}

{\sl Proof} :
Let us remember that the matrix $\tilde{U}(t)$ is expressed by the spectral measure as
\begin{equation}
\tilde{U}_{mn}(t)
=
\int_{(0, \infty )}
e^{-it\lambda }
d \langle m |E(\lambda )| n \rangle
=
\int_{(0, \infty )}
e^{-it\lambda }
d \tilde{E}_{mn}(\lambda ) ,
\label{eqn:6.110}
\end{equation}
where $\tilde{E}(B)$ is the matrix of the components $\langle m |E(B)|n \rangle $.
Therefore, what we first should do is to clarify the relation between
${\rm Im} \tilde{R}^+ (\omega)$ and $\tilde{E}(\lambda )$.
Resorting to Stone's formula between $E(B)$ and $R(z)$,
we clearly see
\begin{equation}
\frac{1}{2}
[\tilde{E}([a,b])+\tilde{E}((a,b))]
=\lim_{\epsilon \to +0}
\frac{1}{2\pi i} \int_a^b [\tilde{R}(\omega +i\epsilon)
-\tilde{R}(\omega -i\epsilon)] d\omega,
\label{eqn:6.30}
\end{equation}
for $a, b \in \mathbb{R}$ with $a<b$.
Under the assumption (\ref{eqn:.110}),
Lebesgue's dominated convergence theorem and the proof of Lemma \ref{lm:100} tell us that
the exchange between the limit and the integration in Eq. (\ref{eqn:6.30}) is allowed
for $[a,b]\subset (0,\infty)$.
%to obtain
%\begin{equation}
%\frac{1}{2}
%[\tilde{E}([a,b])+\tilde{E}((a,b))]
%=
%\frac{1}{2\pi i} \int_a^b [\tilde{R}^+ (\omega)
%-\tilde{R}^- (\omega)] d\omega ,
%\label{eqn:6.40}
%\end{equation}
%for all $a, b$ with $b>a>0$.
If $[a,b] \subset (-\infty, 0)\backslash \sigma(H)$,
then $\tilde{R}^{\pm} (\omega) =\tilde{R} (\omega) $, and thus
$\tilde{E}([a,b])=\tilde{E}((a,b))=0$.
In addition, by the continuity of $\tilde{R}^{\pm}(\omega)$, Eq. (\ref{eqn:6.30}) tells us
that $\tilde{E}(\{ a\} )=0$ for all $a>0$, which leads to
\begin{equation}
\tilde{E}((a,b))
=\tilde{E}([a,b])=
\frac{1}{\pi} \int_a^b {\rm Im} \tilde{R}^+ (\omega) d\omega ,
\label{eqn:6.60}
\end{equation}
for all $a, b$ with $b>a>0$.

Let us now consider the regular case
and in particular the validity of the expression (\ref{eqn:6.60})
for the interval including the origin.
In this case, 
$\tilde{R}^\pm (0):=\lim_{\omega \to +0}\tilde{R}^\pm (\omega)$ exists to be finite.
Furthermore, $\lim_{\omega \to \infty}\tilde{R}^\pm (\omega)=0$
from Lemma \ref{lm:Large-omega}.
Thus, $\tilde{R}^\pm (\omega)$ is uniformly continuous on $(0, \infty)$.
Therefore, we can take the limit of Eq. (\ref{eqn:6.60}) as $a\to +0$ to obtain 
$\lim_{a\to +0}E([a,b])=E((0,b])$.
We next see that all components of
${\rm Im} \tilde{R}^+ (\omega)$
are integrable, i.e., belong to $L^1 ((0, \infty ))$.
Suppose that $| \psi \rangle \in \mathbb{C}^N$, then
$\langle \psi | \tilde{E}((0, \lambda )) | \psi \rangle$
is positive and a monotonically increasing function of $\lambda$,
and it is also differentiable in this case. Thus Eq. (\ref{eqn:6.60}) tells us that
$\bra{\psi} {\rm Im} \tilde{R}^+ (\omega) \ket{\psi} \geq 0$. In addition,
\begin{equation}
\| \psi \|^2
\geq \lim_{\lambda \to \infty }
\langle \psi | \tilde{E}((0, \lambda )) | \psi \rangle
= \lim_{\lambda \to \infty }
\frac{1}{\pi} \int_0^\lambda
\bra{\psi} {\rm Im} \tilde{R}^+ (\omega) \ket{\psi} d\omega
=
\frac{1}{\pi} \int_0^\infty
\bra{\psi} {\rm Im} \tilde{R}^+ (\omega) \ket{\psi} d\omega .
\label{eqn:6.90}
\end{equation}
Hence, from the monotonic convergence theorem,
we see that $\bra{\psi} {\rm Im} \tilde{R}^+ (\omega) \ket{\psi} \in L^1 ((0, \infty ))$.
From this fact and the use of the polarization identity,
we can prove that all components of the matrix
${\rm Im} \tilde{R}^+ (\omega)$ are integrable.
Thus, extending the rhs of Eq. (\ref{eqn:6.60}) to arbitrary
$B \in \{ B \in \mathbb{B}| B \subset (0,\infty) \}$,
$\int_B {\rm Im} \tilde{R}^+ (\omega) d\omega$ defines a measure.
We can now see from Eq. (\ref{eqn:6.60}) and from E. Hopf's extension theorem that
\begin{equation}
\tilde{E}(B)
=
\frac{1}{\pi} \int_B
{\rm Im} \tilde{R}^+ (\omega) d\omega
\label{eqn:6.100}
\end{equation}
holds for all $B \in \{ B \in \mathbb{B}| B \subset (0,\infty) \}$.
%As is usual, this is understood in a symbolical sense.
%
%
Note that this expression means that
%the subspace $\mathbb{C}^N \oplus \{ 0\}$ of ${\cal H}$
%is absolutely continuous one, i.e.,
the restriction of $\tilde{E}_{mn}(B) $ to
$\{ B \in \mathbb{B}| B \subset (0,\infty) \}$
is absolutely continuous.
Therefore, rewriting of $\tilde{U}(t)$ in Eq. (\ref{eqn:6.110}) into (\ref{eqn:6.120})
is straightforward.

In the exceptional case of the first kind,
from the assumption (\ref{eqn:.110}), $\tilde{R}^\pm (\omega)$ is continuous on $(0, \infty)$,
while %$\tilde{R}^\pm (0)$ diverges like
$\tilde{R}^\pm (\omega)=O((\omega \log \omega)^{-1})$ as $\omega \to +0$,
so that it is not integrable around $0$.
See, Eq. (\ref{eqn:5.C.1.120d}).
However, ${\rm Im} \tilde{R}^+ (\omega)$
is of the order $O(\omega^{-1} (\log \omega)^{-2})$ from Lemmas \ref{lm:1stremainder+} and
\ref{lm:1stremainder-},
and thus it is integrable around $0$.
Hence, Eq. (\ref{eqn:6.100}) holds again, and
Eq. (\ref{eqn:6.120}) is valid for this exceptional case.
\qed

We remark that in the case of no negative eigenvalues (point spectrum) of $H$, 
\cite{Miyamoto(2005)}
%which yield dressed eigenstates,
the restriction of $\tilde{U}(t)$
to the continuous energy spectrum is removed because in such a case $P=I$ the identity.
Furthermore, the connection between $\tilde{U}(t)$ and the observables is easily found, e.g.,
$|\bra{\psi}\tilde{U}(t)\ket{\psi}|^2/\| P\ket{\psi} \|^4$
for $| \psi \rangle \in \mathbb{C}^N$ (or $\mathbb{C}^N \oplus \{ 0\}$)
is the survival probability of $P\ket{\psi}$
which is the probability of finding the system in the state $P\ket{\psi}$
at the later time $t$,
where $P\ket{\psi}$ is just the decaying component of the initial state $\ket{\psi}$.

%%%%%%%%%%%%%%%%%%%%%%%%%%%%%%%%%%%%%%%%%%%%%%%%%%%%%%%%%%%%%%%%%%%%%%%%%%%%%%%%

\section{The asymptotic expansion of the reduced time evolution operator}
\label{sec:7}

We can finally show the asymptotic formula for $\tilde{U}(t)$ at long times
for the rational form factors satisfying our assumptions.
In the following, we assume that Eq. (\ref{eqn:.110}) holds, i.e.,
there is no positive eigenvalue.
However, this is not explicitly mentioned in the statements of the theorems.
Let us first consider the regular case. For this purpose,
according to Lemma \ref{lm:rffimremainder},
we introduce the remainder $F(\omega)$ in the following way:
\begin{equation}
\frac{1}{\pi}{\rm Im} \tilde{R}^+ (\omega)
=
\lambda^2
(K(0))^{-1}
\omega^{n_b} ({\mit \Gamma}_{n_b} +\omega {\mit \Gamma}_{n_b +1} + \omega^2 {\mit \Gamma}_{n_b +2} )
(K(0))^{-1} +F(\omega),
\label{eqn:rffimbothLong110}
\end{equation}
for $\omega >0$.

\begin{thm}%[Jensen and Kato, Lemma 10.1]
\label{thm:rffimLong}
: Assume that $0$ is a regular point for $H$.
For a system with the rational form factor (\ref{eqn:formfactor1}) characterized by
the positive integers $n_a$ and $n_b$ that satisfy
that $n_b \geq 2$ and $n_a =1$,
%that $2\leq n_b \leq n_a$ or that $n_a < n_b $,
the reduced time evolution operator $\tilde{U}(t)$ behaves asymptotically as
\begin{equation}
\tilde{U}(t)
=
\lambda^2
\frac{\Gamma(1+n_b)}{(it)^{n_b+1}}
(K(0))^{-1}
{\mit \Gamma}_{n_b} (K(0))^{-1}
+O(t^{-n_b-2}) ,
\label{eqn:rffimbothLong50}
\end{equation}
as $t\to \infty$.
When $n_b=1$ and $n_a \geq 1$, the error term is replaced by
$O(t^{-3}\log t)$.
\end{thm}

{\sl Proof} :
%In order to prove this theorem,
We first summarize the several properties of ${\rm Im} \tilde{R}^+ (\omega) $.
By Lemma \ref{lm:rffimremainder}, we see that the remainder $F(\omega)$ in
Eq. (\ref{eqn:rffimbothLong110}) is arbitrary-times differentiable.
Particularly it holds that
$\lim_{\omega \to 0} d^r F(\omega) / d \omega^r = 0$
for all $r \leq n_b$, and
\begin{eqnarray}
\left\| \frac{d^{n_b+1} F(\omega)}{d\omega^{n_b+1}} \right\|
&&=
%
%\left\{
%\begin{array}{cc}
%O(\omega^{2}) & (n_b \geq 4)\\
%O(\omega^{n_b-1}\log \omega) & (n_b \leq 3)
%\end{array}
%\right.
%
O(\log \omega), ~
\mbox{or} ~
O(1)
%
%\left\{
%\begin{array}{cc}
%O(\omega^{2}) & (n_a \geq 3)\\
%O(\omega^{n_a-1}) & (n_a \leq 2)
%\end{array}
%\right.
%
,
\label{eqn:rffimbothLong120}
\end{eqnarray}
for $n_b=1$ and $n_a \geq 1$, or $n_b \geq 2$ and $n_a =1$, respectively, and
%for $n_b \leq n_a$ or $n_a < n_b $, respectively, and
\begin{equation}
\left\| \frac{d^{n_b+2} F(\omega)}{d\omega^{n_b+2}} \right\|
=
O(\omega^{-1}),
\mbox{ or }
\left\{
\begin{array}{cc}
O(1) & (n_b \geq 3)\\
O(\log \omega) & (n_b = 2)
\end{array}
\right. ,
\label{eqn:rffimbothLong130}
\end{equation}
for $n_b=1$ and $n_a \geq 1$, or $n_b \geq 2$ and $n_a =1$, respectively,
%for $n_b \leq n_a$ or $n_a < n_b $, respectively,
as $\omega \to 0$.
On the other hand, we see from Lemma \ref{lm:Large-omega} that
$ (d / d \omega )^r {\rm Im} \tilde{R}^+ (\omega) = O(\omega^{-r-1})$
as $\omega \rightarrow \infty $.
In particular, if $m \geq 1$,
$(d /d \omega )^m {\rm Im} \tilde{R}^+ (\omega)$ is integrable
on $[\delta , \infty )$ for an arbitrary $\delta >0$.

Let us now split the integral in Eq. (\ref{eqn:6.120})
into two parts by writing
\begin{equation}
{\rm Im} \tilde{R}^+ (\omega ) = \phi (\omega ) {\rm Im} \tilde{R}^+ (\omega )
+ (1- \phi (\omega ) ) {\rm Im} \tilde{R}^+ (\omega ) ,
\label{eqn:rffimbothLong135}
\end{equation}
where $\phi \in C_0 ^{\infty} ([0,\infty ) )$ and satisfies
$\phi (\omega ) =1$ in a neighborhood of $\omega = 0$.
Such a function is realized by $f (\omega ) = 1- \int_0 ^\omega
g(x) d x$, where $g(x) = h(x)/\int_{\bf R} h(x) d x$ and
%is the normalized one of
$h(x) = \exp(-1/[a^2-(x-d)^2 ])$ ~($|x-d|<a$) or $0$ ($|x-d| \geq a$)
with $d>a>0$.

From Lemma 10.1 in Ref. \makebox(0,1){\Large \cite{Jensen(1979)}}\hspace{2mm}
and the above-mentioned discussion, we see that
$(1- \phi (\omega ) ) {\rm Im} \tilde{R}^+ (\omega ) $ has a contribution of
$O(t^{-m})$ to $\tilde{U}(t)$ for an arbitrary $m\geq 1$,
i.e., this term decays faster than any negative power of $t$.

On the other hand, the contribution of
$\phi (\omega ) {\rm Im} \tilde{R}^+ (\omega ) $ to $\tilde{U}(t)$
gives %from (\ref{eqn:4.10})
the main part of the asymptotic expansion.
Then, the coefficient of ${\mit \Gamma}_{n_b}$, ${\mit \Gamma}_{n_b +1}$, and ${\mit \Gamma}_{n_b +2}$
is given by the form \cite{Copson}
\begin{eqnarray}
%\fl
\int_0 ^{\infty }
\phi (\omega ) \omega^q e^{-i t \omega } ~ d \omega
%& = &
%i^j \frac{d^j}{d t^j } \int_0 ^{\infty }
%\phi (\omega ) \omega^{-1/2} e^{-i t \omega } ~ d \omega \\
& = &
\sum_{k=0}^{N-1}
\frac{1}{(it)^{k+1}}
\left. \frac{d^k \omega^q \phi(\omega)}{d\omega^k} \right|_{\omega=0}
+R_N (t)
%\nonumber \\
%& = &
=
\frac{\Gamma(1+q)}{(it)^{1+q}}
+R_N (t) ,
\label{eqn:rffimbothLong140}
\end{eqnarray}
for all $N\geq 1+q$, where $q$ takes the value $n_b$, $n_b+1$, or $n_b+2$.
We here used that
\begin{equation}
\frac{d^k \omega^q \phi(\omega)}{d\omega^k}\biggr|_{\omega=0}
=
\sum_{j=0}^{\min \{ k,q \}} {{k}\choose{j}}
\frac{d^j \omega^q }{d\omega^j }\biggr|_{\omega=0}
\frac{d^{k-j} \phi(\omega)}{d\omega^{k-j}}\biggr|_{\omega=0}
%=q! \delta_{kq}
=\Gamma(1+q) \delta_{kq},
\label{eqn:rffimbothLong145}
\end{equation}
where $\Gamma(1+n)=\int_0^\infty x^n e^{-x} dx$ is the gamma function.
In addition, the remainder $R_N (t)$ is bounded above by
\begin{equation}
|R_N (t)|
\leq
\frac{1}{t^N} \left| \int_{0}^{\infty}
\frac{d^N \omega^q \phi(\omega)}{d\omega^N} e^{-i\omega t} d\omega \right|
=
o(t^{-N}).
\label{eqn:rffimbothLong160}
\end{equation}
Note that
since all
derivatives of $\phi(\omega)$ vanish in the neighborhood of $\omega =0$,
%and $\phi \in C_0 ^{\infty} ([0,\infty ) )$,
Eq. (\ref{eqn:rffimbothLong140}) is valid for all $N\geq q+1$ and thus
$R_N (t)$ decays faster than any %inverse
negative power of $t$.
Furthermore, we understand, by applying Eq. (\ref{eqn:A.50a}) in
Lemma\ \ref{lm:Jensen and Kato, Lemma 10.2} directly to $\phi (\omega ) F(\omega) $
with the discussion in the first part of this section, that
%if $s, s^{\prime} > 3n+3/2 $,
the contribution of
the Fourier transform of the remainder $\phi (\omega ) F(\omega) $
to $\tilde{U}(t)$ is
\begin{equation}
%
%\left\{
%\begin{array}{cc}
%O(t^{-n_b-2}) & (n_b \geq 2) \\
%O(t^{-3}\log t) & (n_b =1)
%\end{array}
%\right.
%
O(t^{-3}\log t)~~
\mbox{or} ~~
O(t^{-n_b-2}) ,
\label{eqn:rffimbothLong180}
\end{equation}
for $n_b=1$ and $n_a \geq 1$, or $n_b \geq 2$ and $n_a =1$, respectively, as $\omega \to 0$,
where we used the formula of the indefinite integral that
$\int (\log \omega)^2 d\omega =\omega [(\log \omega)^2 -2\log \omega +2]$.
Summarizing the above-noted results, we finally obtain that
%for $2p\neq 1$
%$s, s^{\prime } > \max \{ 3n+3/2, 5/2\}$,
\begin{eqnarray}
&&\hspace{-10mm}
%\fl
\left\| \tilde{U}(t) -
\lambda^2
(K(0))^{-1}
\left[
\frac{\Gamma(1+n_b)}{(it)^{n_b+1}} {\mit \Gamma}_{n_b} + \frac{\Gamma(2+n_b)}{(it)^{n_b+2}} {\mit \Gamma}_{n_b +1} +\frac{\Gamma(3+n_b)}{(it)^{n_b+3}} {\mit \Gamma}_{n_b +2}
\right]
(K(0))^{-1}
\right\|
\nonumber  \\
%\fl
&&\hspace{-10mm}\leq
\left\|
\frac{1}{\pi}\int_0^{\infty}
(1-\phi(\omega)){\rm Im} \tilde{R}^+ (\omega ) e^{-i t \omega } d \omega
\right\|
+
\left\|
\frac{1}{\pi}\int_0^{\infty}
\phi(\omega) F(\omega) e^{-i t \omega } d\omega
\right\| +O(t^{-N})
\nonumber  \\
&&\hspace{-10mm}=
%
%\left\{
%\begin{array}{cc}
%O(t^{-n_b-2}) & (n_b \geq 2) \\
%O(t^{-3}\log t) & (n_b =1)
%\end{array}
%\right.
%
O(t^{-3}\log t)~~
\mbox{or} ~~
O(t^{-n_b-2}) ,
\label{eqn:rffimbothLong150}
\end{eqnarray}
for $n_b=1$ and $n_a \geq 1$, or $n_b \geq 2$ and $n_a =1$, respectively,
%for $n_b \leq n_a$ or $n_a < n_b $, respectively,
as $t\to \infty$, where $O(t^{-N})$ is due to the contribution from $R_N (t)$.
%For $2p=1$, $O(t^{-1-2p})$ is changed to $O(t^{-2}\log t)$.
This is just the asymptotic expansion of $\tilde{U}(t)$
in the statement. %Eq. (\ref{eqn:Long50}).
\qed

It is worth noting that if we resort to Lemma \ref{lm:rffremainder},
instead of Eqs. (\ref{eqn:rffimbothLong120}) and (\ref{eqn:rffimbothLong130}),
we have
\begin{eqnarray}
\hspace*{-5mm}
\frac{d^r F(\omega)}{d\omega^r}
&&=
%
%\left\{
%\begin{array}{cc}
%O(\omega^{2}) & (n_b \geq 4)\\
%O(\omega^{n_b-1}\log \omega) & (n_b \leq 3)
%\end{array}
%\right.
%
O(\omega^{2-r} (\log \omega)^{1+\theta (2-r)}), ~
\mbox{or} ~
\left\{
\begin{array}{cc}
O(\omega^{[2-r]^+}) & (0 \leq r \leq n_b)\\
O(\omega^{n_b +1-r} (\log \omega)^{\theta (n_b +1 -r)}) & (r \geq n_b +1)
\end{array}
\right.
,
\label{eqn:rffimbothLong170}
\end{eqnarray}
for $n_b=1$ and $n_a \geq 1$, or $n_b \geq 2$ and $n_a =1$, respectively.
However, in the latter case, we see that the Fourier transform of $\phi(\omega)F(\omega)$
gives the contribution of the order $O(t^{-n_b -1})$,
which is just the same order as that coming from the dominant one.
Hence, we can only obtain an useless estimation.

We next show the asymptotic formula for $\tilde{U}(t)$ at long times 
for a system with an exceptional point of the first kind. To this end,
we write ${\rm Im} \tilde{R}^+(\omega)$ with the remainder $F(\omega)$
again as follows:
\begin{equation}
\frac{1}{\pi}{\rm Im} \tilde{R}^+ (\omega)
=
\frac{1}{\lambda^2 \omega (\log \omega)^2} (Q_1 {\mit \Gamma}_1 Q_1)^{-1} +F(\omega) .
%(\lambda^2 \omega (\log \omega)^2)^{-1} (Q_1 {\mit \Gamma}_1 Q_1)^{-1} +F(\omega) ,
\label{eqn:rffimLong1st110}
\end{equation}

\begin{thm}%[Jensen and Kato, Lemma 10.1]
\label{thm:rffLong1st}
: Assume that $0$ is an exceptional point of the first kind for $H$,
which necessarily imposes that $n_b =1$.
Then, the reduced time evolution operator $\tilde{U}(t)$
for the rational form factor (\ref{eqn:formfactor1}) behaves asymptotically as
\begin{equation}
\tilde{U}(t)
=
\frac{1}{\lambda^2 \log t}
(Q_1 {\mit \Gamma}_1 Q_1)^{-1} +O((\log t)^{-2}),
\label{eqn:rffimLong1st50}
\end{equation}
as $t \to \infty$.
%, for all $n_b \geq 2$.
%For $n_b=1$ the error term is replaced by $O(t^{-3}(\log t)^2)$.
\end{thm}

{\sl Proof} :
%In order to prove this theorem,
Let us first look over the some properties of ${\rm Im} \tilde{R}^+ (\omega) $ again.
By Lemmas \ref{lm:1stremainder+} and \ref{lm:1stremainder-},
we see that the remainder $F(\omega)$ in
Eq. (\ref{eqn:rffimLong1st110}) is arbitrary-times differentiable,
satisfies that $F(\omega) = O(\omega^{-1} (\log \omega)^{-3})$,
and $d F(\omega) / d \omega = O(\omega^{-2} (\log \omega)^{-3}) $,
as $\omega \to +0$.
On the other hand, we see from Lemma \ref{lm:Large-omega} that
$ (d / d \omega )^r {\rm Im} \tilde{R}^+ (\omega) = O(\omega^{-r-1})$
as $\omega \rightarrow \infty $.
In particular, if $m \geq 1$,
$(d /d \omega )^m {\rm Im} \tilde{R}^+ (\omega)$ is integrable
on $[\delta , \infty )$ for an arbitrary $\delta >0$.

We now split the integral in Eq. (\ref{eqn:6.120})
into two parts as in Eq. (\ref{eqn:rffimbothLong135}) again using
the $C_0 ^{\infty}$-function $\phi(\omega )$.
From Lemma 10.1 in Ref. \makebox(0,1){\Large \cite{Jensen(1979)}}\hspace{2mm}
and the discussion mentioned above,
$(1- \phi (\omega ) ) {\rm Im} \tilde{R}^+ (\omega ) $ has a contribution of
$O(t^{-m})$ to $\tilde{U}(t)$ for an arbitrary $m\geq 1$.
On the other hand, the contribution of
$\phi (\omega ) {\rm Im} \tilde{R}^+ (\omega ) $ to $\tilde{U}(t)$
gives %from (\ref{eqn:4.10})
the dominant part of the asymptotic expansion.
Then, the dominant time dependence of the asymptote of $\tilde{U}(t)$ follows from
Lemma \ref{lm:asymptote of inverse-lagarithmic-fourier-integral}, that is,
\begin{eqnarray}
%\fl
\int_0 ^{\infty }
\phi (\omega ) (\omega (\log \omega)^2)^{-1} e^{-i t \omega } ~ d \omega
=
(\log t)^{-1} +O((\log t)^{-2} ).
\label{eqn:rffimLong1st120}
\end{eqnarray}
Furthermore,
%we understand, from Lemma\ \ref{lm:Jensen and Kato, Lemma 10.2}
%and the discussion in the first of this section, that
the contribution of the Fourier transform of the remainder $\phi (\omega ) F(\omega) $
to $\tilde{U}(t)$ can be estimated
by the similar manner to Lemma\ \ref{lm:asymptote of inverse-lagarithmic-fourier-integral},
rather than Lemma\ \ref{lm:Jensen and Kato, Lemma 10.2}.
By setting $\sigma (\omega)=F(\omega) e^{-it\omega}$
instead in the proof of Lemma \ref{lm:asymptote of inverse-lagarithmic-fourier-integral},
we can apply it to this case,
and we have
\begin{equation}
\int_0^\infty F(\omega) \phi(\omega) e^{-it\omega} d\omega
=
-\lim_{\omega \to +0}(\hat{I} \sigma)(\omega) %\phi(\omega) |_{\omega=0}
+(-1)^N
\int_0^\infty (\hat{I}^N \sigma)(\omega) \frac{d^N \phi(\omega)}{d\omega^N} d\omega .
\label{eqn:rffimLong1st155}
\end{equation}
Then, corresponding to Eq. (\ref{eqn:B.530}), we have
\begin{equation}
\|(\hat{I}\sigma)(\omega)\|
=
\biggl\|
i e^{-it\omega} \int_0^\infty F(\omega-i\eta) e^{-t\eta} d\eta
\biggr\|
+
E(t)
\leq
C
\int_0^\infty \bigr| (\omega-i\eta)^{-1} [\log (\omega-i\eta)]^{-3} \bigr| e^{-t\eta} d\eta
+
E(t),
\label{eqn:rffimLong1st160}
\end{equation}
with an appropriate constant $C$.
Note that in this procedure, $\tilde{R}^+ (\omega)$ is analytically continued to
the lower plane of the second Riemann sheet, while $\tilde{R}^- (\omega)$ still remains
in the lower plane of the first Riemann sheet.
Then, both are ensured to contribute the remainders
of the same order to $F(\omega-i\eta)$ in the above integral
from Lemmas \ref{lm:1stremainder+} and \ref{lm:1stremainder-}.
The remainder term $E(t)$ in Eq. (\ref{eqn:rffimLong1st160}) 
that gives the order of $O(e^{-\gamma t})$ for some $\gamma >0$ 
is responsible for the possible poles
of $\tilde{R}^+ (\omega)$ continued to the second Riemann sheet,
the number of which are guaranteed to be finite from
the analytic Fredholm theorem \cite{the analytic Fredholm theorem}
and Lemma \ref{lm:Large-omega} for
the continued  $\tilde{R}^+ (\omega)$.
Thus, it follows from Lemma \ref{lm:asymptote of inverse-lagarithmic-fourier-integral}
that $\lim_{\omega \to +0} \|(\hat{I}\sigma)(\omega)\| =O((\log t)^{-2})$.
By the same argument as in Lemma \ref{lm:asymptote of inverse-lagarithmic-fourier-integral},
one also sees that the remainder term in Eq. (\ref{eqn:rffimLong1st155})
is of the order of $O(t^{-N+1})$.
Summarizing these arguments, we can finish the proof of the theorem.
\qed

%%%%%%%%%%%%%%%%%%%%%%%%%%%%%%%%%%%%%%%%%%%%%%%%%%%%%%%%%%%%%%%%%%%%%

\section{Concluding remarks}
\label{sec:8}

We have rigorously derived the asymptotic formula of the reduced time evolution
operator for the $N$-level Friedrichs model in the context of the zero energy resonance 
\cite{Jensen(1979)}
%in a mathematically rigorous way,
both for the regular case and the exceptional case of the first kind.
Then, in the latter case, the logarithmically slow decay proportional to $(\log t)^{-1}$
has been found,
and the expansion coefficient has been explicitly presented
by the projection operator associated with the zero energy eigenstates
of the total Hamiltonian, which is an extended state not belonging to the Hilbert space.
We note that the decay involving the logarithmic function 
expressed by $t^{-j}(\log t)^{k}$ ($j=1, 2, \ldots$ and $k=0, \pm 1, \ldots$)
can occur in the short range potential systems in the even dimensional space. \cite{Murata(1982)}
%though they are not regarded as unstable systems.
It should be noted that a realization of the exceptional cases require 
the parameters, e.g., the coupling constant $\lambda$, 
to take such special values 
that the matrix $K(0)$ in Eq. (\ref{eqn:.336}) has a zero eigenvalue. 
In addition, some of the form factors $v_n(\omega)$ have to behave as $|v_n(\omega)|^2 \sim c_n \omega$ 
around $\omega=0$. 
In other words, if all of them behave as 
$|v_n(\omega)|^2 \sim c_n \omega^{q_n}$ with $q_n \geq 2$, 
the exceptional case of the first kind never occurs 
though that of the second kind could happen. 
%See also Eqs. (\ref{eqn:.340}) and (\ref{eqn:.350}). 
These circumstances explain how the exceptional cases are surely exceptional. 
The presented results also enable us to calculate the asymptotic formula for the survival
probability of an arbitrary initial state $\ket{\psi}$ localized over the $N$ discrete levels.
If we choose the special initial state to satisfy
${\mit \Gamma}_{n_b} (K(0))^{-1}\ket{\psi}=0$ in Eq. (\ref{eqn:rffimbothLong50})
or $(Q_1 {\mit \Gamma}_1 Q_1)^{-1}\ket{\psi}=0$ in Eq. (\ref{eqn:rffimLong1st50}),
our estimations are useless and other decay laws could appear. \cite{Miyamoto(2004)}
The long time behavior of the reduced time evolution operator
for the exceptional case of the second and the third kind are not examined.
As is expected, in the former case, the non decaying component
associated with the localized zero energy eigenstate will appear
due to the divergent behavior of $Q_2\tilde{R}(z)Q_2=O(z^{-1})$ in Eq. (\ref{eqn:5.C.2.120d}).
%On the other hand, in order to extract the long time behavior of the decaying component,
%we should estimate not only $Q_2 \tilde{R}(z)Q_2$ but also the other terms
%$Q_k \tilde{R}(z)Q_l$ for $k, l=0, 2$ in detail.
The latter case can occur in the $N$-level cases of the model only for $N\geq 3$,
which yields a more complicated situation.
In the whole of the paper, we assumed that there is no bound eigenstate
with a positive eigenenergy. This situation is actually realized in the weak coupling cases.
\cite{Davies(1974),Miyamoto(2005)} However,
its compatibility with the existence of the extended zero energy eigenstate
is still not clear in the multilevel cases (except the single level case).
The emergence of the logarithmic decay $(\log t)^{-1}$ is just due to
the logarithmic energy dependence of the self energy $S(\omega)$
and it comes from the assumption (\ref{eqn:formfactor1}) 
where $|v_n(\omega)|^2 \sim c_n \omega^{q_n}$ with a positive integer $q_n$ is required. 
%about the zero energy behavior of the form factors.
%in Eq. (\ref{eqn:formfactor1}).
Therefore, if we choose another type of form factor,
it is not necessary for such a slow decay to occur even in the exceptional case.
\cite{Kofman(1994),Lewenstein(2000),Nakazato(2003)}
However, we stress that our assumption is often satisfied by
actual systems. \cite{Facchi(1998),Antoniou(2001)}
The experimental realization of the exceptional case
%either for the power decay or the logarithmic decay
requires
the setup of parameters like $\omega_1 \simeq \lambda^2 \Lambda $, 
where $\Lambda$ is a typical cutoff constant. 
This seems in a strong coupling region to be naturally satisfied
\cite{Jittoh(2005),Garcia-Calderon(2001)}, 
and hence it could be suggested to invoke the artificial quantum structures for a realization.

%%%%%%%%%%%%%%%%%%%%%%%%%%%%%%%%%%%%%%%%%%%%%%%%%%%%%%%%%%%%%%%%%%%%

%\begin{acknowledgments}
\section*{Acknowledgments}

The author would like to thank Professor I.\ Ohba
and Professor H.\ Nakazato for useful comments.
He also would like to express his gratitude to the organizers
of the International Workshop TQMFA2005, Palermo, Italy,
November 11-13, 2005,
{\sl New Trends in Quantum Mechanics: Fundamental Aspects and Applications}.
Discussions during the YITP workshop YITP-W-05-21 on ``Fundamental Problems and Applications of
Quantum Field Theory'' were useful in completing this work.
This research is partly supported by a Grant for The 21st Century COE Program
%(Physics of Self-Organization Systems)
at Waseda University from the Ministry of Education, Culture,
Sports, Science and Technology, Japan.

%\end{acknowledgments}

%%%%%%%%%%%%%%%%%%%%%%%%%%%%%%%%%%%%%%%%%%%%%%%%%%%%%%%%%%%%%%%%%%%%

\appendix

\section{Characteristics of self energy for the rational form factor}
%\section*{Appendix}
%\section{}

\begin{lm}%[Jensen and Kato, Lemma 10.1]
\label{lm:formfactor1}
Suppose that $\eta (\omega)$ is a rational function, i.e.,
it is expressed by $\eta (\omega) =\pi (\omega) / \rho (\omega)$,
where $\pi (\omega)$ and $\rho (\omega)$ are the polynomials
of the degree $m$ and $n$, respectively. Furthermore, we assume that
$n\geq m+1$ and $\rho (z)$ has no zeros in $[0, \infty )$. Then
\begin{equation}
\int_{0}^{\infty}
\frac{\eta (\omega)}{\omega -\zeta }
d\omega
=
\frac{P_{n-1}(\zeta)-\pi (\zeta) \log (-\zeta) }{\rho(\zeta)},
\label{eqn:B.30}
\end{equation}
for all $\zeta \in \mathbb{C}\backslash
([0, \infty ) \cup \{ a_k \}_{k=1}^{N})$,
where $a_k$ is a pole of $m_k$-th order of $\eta(z)$,
$N$ is the number of such poles, and
$P_n (\zeta)$ is a polynomial of $\zeta$ of the degree not greater than $n$ .
For $\zeta=|\zeta|e^{i\theta}$ with $0 \leq \theta \leq 2 \pi$,
we define $-\zeta =|\zeta|e^{i\phi }$ with $-\pi \leq \phi \leq  \pi$.

\end{lm}

{\sl Proof} :
From the fundamental theorems for the complex functions, it holds that
\begin{equation}
\int_{0}^{\infty}
\frac{\eta (\omega)}{\omega -\zeta }
d\omega
=
-\sum_{k=1}^{N+1} {\rm Res }
\left(
\frac{\eta (z)}{z -\zeta } \log (-z), z=a_k
\right) ,
\label{eqn:B.40}
\end{equation}
where $a_{N+1}=\zeta$.
Then the residue at $z=a_k$ for $k\leq N$ is deduced to explicitly
\begin{eqnarray}
&&\hspace*{-5mm}
%{\rm Res }
%\left(
%\frac{\eta (z)}{z -\zeta } \log (-z), z=a_k
%\right)
%\nonumber \\
%=
%\frac{1}{(m_k -1)!}
%\frac{d^{m_k -1}}{dz^{m_k -1}}
%\left[
%(z-a_k)^{m_k} \frac{\eta (z)}{z -\zeta } \log (-z)
%\right]_{z=a_k}
%\nonumber \\
%&=&
\frac{1}{(m_k -1)!}
\sum_{j=0}^{m_k -1}
{m_k -1 \choose j}
\left[
\left(
\frac{d^{m_k -1-j}}{dz^{m_k -1-j}}
(z-a_k)^{m_k} \frac{\eta (z)}{z -\zeta }
\right)
\left(
\frac{d^j}{dz^j}
\log (-z)
\right)
\right]_{z=a_k}
\nonumber \\
&=&
-\sum_{j=1}^{m_k -1}
\frac{P_{m_k -1-j}(\zeta)}{(a_k -\zeta )^{m_k-j}}
-
\frac{P_{m_k -1}(\zeta)}{(a_k -\zeta )^{m_k}}
\log (-a_k)  .
\label{eqn:B.50}
\end{eqnarray}
For $z=\zeta$, which is a simple pole, the residue becomes
\begin{equation}
{\rm Res }
\left(
\frac{\eta (z)}{z -\zeta } \log (-z), z=\zeta
\right)
=
\eta (\zeta) \log (-\zeta) .
\label{eqn:B.60}
\end{equation}
Therefore, by setting Eqs. (\ref{eqn:B.50}) and (\ref{eqn:B.60}) into Eq. (\ref{eqn:B.40}), 
one obtains Eq. (\ref{eqn:B.30}), and the proof is completed.
\qed

\begin{lm}\label{lm:1st_order}
Suppose that the function $\eta (\omega)$ belonging to $L^1 ([0, \infty ))$ is of the form
\begin{equation}
\eta (\omega ) := \omega^p r(\omega ),
\label{eqn:B.130}
\end{equation}
where $p>1$ and $r(\omega)$ is a $C^1$-function defined in $[0,\infty)$.
Then it holds that both $\eta (\omega)/\omega$ and
$\eta (\omega)/\omega^2  \in L^1 ([0, \infty ))$, and
\begin{equation}
\lim_{E \to +0} \frac{1}{E}
\left[
P\int_{0}^{\infty} \frac{\eta (\omega)}{\omega -E} d\omega
-
\int_{0}^{\infty} \frac{\eta (\omega)}{\omega} d\omega
\right]
=
\int_{0}^{\infty} \frac{\eta (\omega)}{\omega^2} d\omega .
\label{eqn:B.140}
\end{equation}
\end{lm}

{\sl Proof} : From the proof of Proposition 3.2.2 in
Ref. \makebox(0,1){\Large \cite{Exner(1985)}}\hspace{3mm} ,
the principal value of the integral on the rhs is written by the absolutely
integrable function as follows
\begin{equation}
P\int_{0}^{\infty} \frac{\eta (\omega)}{\omega -E} d\omega
=\int_{0}^{\infty} \frac{\eta (\omega)-\eta (E)\varphi_{\delta}(\omega-E)}{\omega -E} d\omega ,
\label{eqn:B.150}
\end{equation}
for all $E>0$, where $\varphi_{\delta}(\omega)$ is a $C_0^{\infty}$-function
with support $[-\delta , \delta ]$ ($0< \delta < E$), even with respect to the origin, and such that $\varphi_{\delta}(0)=1$.
In the following, we choose such that
$\varphi_{\delta}(\omega)=\exp[1-1/(1-(\omega/\delta)^2)]$
for $\omega \in (-\delta , \delta )$ or $0$ otherwise, and $\delta=E/2$.
Then, the proof of Eq. (\ref{eqn:B.140}) is equivalent to that of
%On the other hand,
%since from the assumption (\ref{eqn:B.130}) $\eta (\omega) /\omega$ is absolutely integrable,
%the first equality in Eq. (\ref{eqn:B.140}) is obvious.
%Therefore, it is sufficient to show that
\begin{equation}
\lim_{E \to +0}
\int_{0}^{\infty}
\left\{
\frac{1}{E}
\left[
\frac{\eta (\omega)}{\omega}
-\frac{\eta (\omega)-\eta (E)\varphi_{\delta}(\omega-E)}{\omega -E}
\right]
+
\frac{\eta (\omega)}{\omega^2}
\right\}
d\omega =0,
\label{eqn:B.160}
\end{equation}
which will be shown in the following.
We note that the above-mentioned integrand %denoted by $h(\omega, E)$
can be rewritten as
\begin{eqnarray}
%&&%\hspace*{-5mm}
%\frac{1}{E}
%\left[
%\frac{\eta (\omega)}{\omega}
%-\frac{\eta (\omega)-\eta (E)\varphi_{\delta}(\omega-E)}{\omega -E}
%\right]
%+
%\frac{\eta (\omega)}{\omega^2}
%\nonumber\\
&&
%=
-E\frac{\eta (\omega)}{\omega^2 (\omega-E)}
+\frac{\eta (E)\varphi_{\delta}(\omega-E)}{E(\omega -E)}
\label{eqn:B.170} \\
&&
=
\frac{\eta (E)\varphi_{\delta}(\omega-E)}{E\omega}
-\frac{E \eta (\omega)-\omega \eta (E)\varphi_{\delta}(\omega-E)}{\omega^2 (\omega -E)}.
\label{eqn:B.180}
\end{eqnarray}
We also put $I_1 =(0, E/2]$, $I_2=(E/2, 3E/2)$, and $I_3=[3E/2, \infty)$.

Let us first consider the case that $\omega \in I_1 \cup I_3$.
Then, since $\varphi_{\delta} (\omega-E)=0$, we can use Eq. (\ref{eqn:B.170}) to
estimate the integrand in Eq. (\ref{eqn:B.160}), which reads
\begin{equation}
\left|
E\frac{\eta (\omega)}{\omega^2 (\omega-E)}
\right|
\leq
2 \left|
\frac{\eta (\omega)}{\omega^2}
\right| ,
\label{eqn:B.190}
\end{equation}
where the rhs is absolutely integrable and independent of $E$.
Furthermore, it follows that
$\lim_{E \to +0} E \chi_{I_1 \cup I_3} (\omega ) \eta (\omega)/[\omega(\omega-E)] =0$ for every $\omega \in (0, \infty )$, where $ \chi_{I_1 \cup I_3} (\omega ) = 1$ ($\omega \in I_1 \cup I_3$) or
$0$ ($\omega \in I_2$), being the characteristic function.
Thus, by the dominated convergence theorem, we can see that
\begin{equation}
\lim_{E \to +0}
\int_{I_1 \cup I_3}
E\frac{\eta (\omega)}{\omega^2 (\omega-E)}
d\omega
=
\lim_{E \to +0}
\int_{0}^{\infty}
E \chi_{I_1 \cup I_3} (\omega ) \frac{\eta (\omega)}{\omega^2 (\omega-E)}
d\omega =0.
\label{eqn:B.200}
\end{equation}

Next, for $\omega \in I_2$, we can use Eq. (\ref{eqn:B.180}).
The integration of the first term of Eq. (\ref{eqn:B.180}) is estimated  by
\begin{equation}
\left|
\int_{I_2}
\frac{\eta (E)\varphi_{\delta}(\omega-E)}{E\omega}
d\omega
\right|
\leq
\frac{\eta (E)}{E^2/2}
\int_{I_2}
\varphi_{\delta}(\omega-E)
d\omega
=
\frac{\eta (E)}{E}
\int_{-1}^{1}
\varphi_{1}(x)
dx \to 0, %\mbox{ as } E \to +0.
\label{eqn:B.210}
\end{equation}
as $E \to +0$, because $\eta (\omega)=O(\omega^p )$ where $p>1$.
The second term of Eq. (\ref{eqn:B.180}) is also estimated with the decomposition
\begin{equation}
|E\eta(\omega)-\omega \eta(E)\varphi_{\delta}(\omega-E)|
\leq
|(E-\omega)\eta(\omega)|+|\omega (\eta(\omega) -\eta(E))|
+|\omega \eta(E)||1-\varphi_{\delta}(\omega-E)| .
\label{eqn:B.220}
\end{equation}
%where $C(E)=\sup_{E/2 \leq \omega \leq 3E/2} |d\eta(\omega)/d\omega|$.
The integral corresponding the first term on the rhs of the above is evaluated as
\begin{equation}
\int_{I_2}
\frac{|(E-\omega )\eta (\omega)|}{\omega^2 |\omega -E|} d\omega \to 0,
%\mbox{ as } E \to +0,
\label{eqn:B.235b}
\end{equation}
as $E \to +0$, because of the fact $\eta (\omega)/\omega^2 \in L^1 ([0, \infty))$.
The integral corresponding the second term is also evaluated as
\begin{equation}
%&&\hspace*{-10mm}
\int_{I_2} \frac{\omega|\eta (\omega)-\eta (E)|}{\omega^2 |\omega -E|} d\omega
\leq
(\ln 3) \sup_{\omega \in I_2} \left| \eta^{\prime}(\omega) \right|
%\label{eqn:B.230a} \\
%\nonumber\\
%&&\hspace*{-10mm}
%\leq
%(\ln 3)
%\left[
%p E^{p-1}
%\max\{ ({\textstyle \frac{1}{2}})^{p-1}, ({\textstyle \frac{3}{2}})^{p-1}\}
%\sup_{\omega \in I_1 \cup I_2} \left| r(\omega) \right|
%+
%\left(\frac{3E}{2}\right)^{p}
%\sup_{\omega \in I_1 \cup I_2} \left| r^{\prime}(\omega) \right|
%\right]
%\label{eqn:a110} \\
%&\rightarrow&0 ~~~(E \to +0),
\to 0,
%\nonumber \\
%&&
\label{eqn:B.230b}
\end{equation}
as $E \to +0$, because of the assumption on $\eta (\omega)$, 
where %the prime on $\eta^{\prime} (\omega)$
$\eta^{\prime} (\omega)$ is the derivative of $\eta (\omega)$.
The integral corresponding the last term on the rhs 
of Eq. (\ref{eqn:B.220}) is also estimated as
\begin{equation}
\int_{I_2}
\frac{\omega|\eta (E)||1-\varphi_{\delta}(\omega-E)|}{\omega^2 |\omega -E|} d\omega
\leq
%(\ln 3) |\eta (E)| \sup_{\omega \in I_2} \left| \varphi_{\delta}^{\prime} (\omega-E) \right|
%=
(\ln 3) \frac{|\eta (E)|}{\delta} \sup_{|x| \leq 1} \left| \varphi_{1}^{\prime}(x) \right|
\to
0,
\label{eqn:a140}
\end{equation}
as $E \to +0$. Thus, we see from Eqs. (\ref{eqn:B.235b}), (\ref{eqn:B.230b}), and (\ref{eqn:a140}),
\begin{equation}
\lim_{E \to +0}
\int_{I_2}
\frac{E \eta (\omega)-\omega \eta (E)\varphi_{\delta}(\omega-E)}{\omega^2 (\omega -E)}
d\omega=0.
\label{eqn:B.240b}
\end{equation}
Equations (\ref{eqn:B.200}), (\ref{eqn:B.210}), and (\ref{eqn:B.240b})
mean the completion of the proof of Eq. (\ref{eqn:B.160}). \qed

\section{Asymptotic expansion of the Fourier integrals}

We have to estimate the integrals of the form
$%\begin{equation}
U(t)
=
\int_0^{\infty}
e^{-it\lambda}
F(\lambda)d\lambda
%\label{eqn:A.10}
$ %\end{equation}
in which $F(\lambda)=0$ identically either for small $\lambda >0$ or for large
$\lambda$ where $F$ is supposed to take values
in an arbitrary Banach space ${\bf B}$.
The following lemma is essentially the same as Lemma 10.2 in
Ref. \makebox(0,1){\Large \cite{Jensen(1979)}}\hspace{2mm}.

\begin{lm}%[Jensen and Kato, Lemma 10.2]
\label{lm:Jensen and Kato, Lemma 10.2}
: Suppose that $F(\lambda)=0$ for $\lambda > a >0$,
($F \in C^{k+1} (\delta, \infty; {\bf B})$),
$F^{(k+1)} \in L^1 (\delta, \infty; {\bf B})$ for any $\delta>0$
and for an integer $k \geq 0$, and that
%$F^{(k+1)} (\lambda)=O(\lambda^{\theta-2})$ as $\lambda \to +0$
%for some $\theta \in (0,1)$. Assume further that
$F^{(j)} (0)=0$ for $j\leq k-1$. Then
\begin{equation}
\| U(t) \|
\leq
\frac{1}{t^k}
\left(
\int_{0}^{2\pi/t} \| F^{(k)}(\lambda) \| d\lambda
+
\frac{\pi}{2t}\int_{\pi/t}^{a}
\sup_{\mu \in [\lambda, \lambda+\pi/t]} \| F^{(k+1)} (\mu) \|d\lambda
\right),
\label{eqn:A.50a}
\end{equation}
for all $t>\pi/a$. Here $F^{(k)} (\lambda)$ denotes the $k$-th derivative of
$F(\lambda)$ and so forth.
\end{lm}

{\sl Proof} :
By extending $F$ by $F(\lambda)=0$ to
$\lambda<0$, we obtain a function $F$ on $(-\infty,\infty)$ with
$F^{(k)} \in L^1 (-\infty,\infty;{\bf B})$.
Then we have that
\begin{eqnarray}
\hspace*{-8mm}
\int_{-\infty}^{\infty}
\| F^{(k)}(\lambda+h) -F^{(k)} (\lambda) \| d\lambda
&=&
\left(
\int_{-\infty}^{h}
+\int_{h}^{a}
\right)
\| F^{(k)}(\lambda+h) -F^{(k)} (\lambda) \| d\lambda
\label{eqn:A100c} \\
&\leq&
2\int_{0}^{2h} \| F^{(k)}(\lambda) \| d\lambda
+\int_{h}^{a} d\lambda \int_{\lambda}^{\lambda+h} \| F^{(k+1)} (\mu) \|d\mu
\label{eqn:A100} \\
&\leq&
2\int_{0}^{2h} \| F^{(k)}(\lambda) \| d\lambda
+h\int_{h}^{a} \sup_{\mu \in [\lambda, \lambda+h]} \| F^{(k+1)} (\mu) \|d\lambda .
\label{eqn:A100b}
\end{eqnarray}
By noting that $e^{-it\lambda}=-e^{-it(\lambda -\pi/t)}$,
one sees that
the lhs of Eq. (\ref{eqn:A100c}) is just an upper bound of
the Fourier transform of $2F^{(k)}$.
Remember that the Fourier transform of $F^{(k)}$ is equal to $(it)^k U(t)$
under the assumption of the lemma, i.e., $F^{(j)} (0)=0$ for $j\leq k-1$,
then the desired result follows.
\qed

\begin{lm}
\label{lm:asymptote of inverse-lagarithmic-fourier-integral}
Suppose that $\phi \in C_0 ^{\infty} ([0,\infty ) )$ and satisfies
$\phi (\omega ) =1$ in a neighborhood of $\omega = 0$. 
%and ${\rm supp}\phi \subset [0, \omega_0]$ for $\omega_0 <1/(e^q \sqrt{2})$.
It then holds that for any positive integer $q \geq 2$ and $N \geq 1$, 
\begin{eqnarray}
\int_0^\infty \omega^{-1} (\log \omega)^{-q} \phi(\omega) e^{-it\omega} d\omega
&=&
\frac{(-1)^{q}}{q-1}
\sum_{j=0}^{N-1}
{q+j-2 \choose j}
(\log t)^{1-q-j}
\left( \frac{d}{d\nu} -\frac{\pi}{2}i \right)^j \Gamma(\nu)
|_{\nu=1}
\nonumber \\
&&
+O((\log t)^{1-q-N})
\label{eqn:B.500a}%\\
%&=&
%\frac{(-1)^{q}}{q-1} (\log t)^{1-q} +O((\log t)^{-q} ),
%\label{eqn:B.500b}
\end{eqnarray}
as $t \to \infty$. %for an arbitrary integer $N\geq 1$,
%where $\Gamma^{(k)}(0)=\lim_{\nu \to +0}d^k\Gamma(\nu)/d \nu^k$.
\end{lm}

{\sl Proof} :  We first put $\sigma (\omega)=\omega^{-1} (\log \omega)^{-q} e^{-it\omega}$ and
introduce the indefinite integral operator $\hat{I}$ as \cite{Copson}
\begin{equation}
(\hat{I}\sigma)(\omega)=\int_c^\omega s^{-1} (\log s)^{-q} e^{-its} ds,
\label{eqn:B.410a}
\end{equation}
where $c$ is an arbitrary complex number. We can also recursively show
\begin{equation}
(\hat{I}^k \sigma)(\omega)
=\frac{1}{(k-1)!}\int_c^\omega (\omega-s)^{k-1} s^{-1} (\log s)^{-q} e^{-its} ds.
\label{eqn:B.410b}
\end{equation}
Then, repeating the partial integration, we obtain
\begin{equation}
\int_0^\infty \omega^{-1} (\log \omega)^{-q} \phi(\omega) e^{-it\omega} d\omega
%&=&
%\sum_{k=0}^{N-1}
%\left[ (-1)^k (\hat{I}^{k+1} \sigma)(\omega)\frac{d^k \phi(\omega)}{d\omega^k} \right]_0^\infty
%+(-1)^N
%\int_0^\infty (\hat{I}^N \sigma)(\omega) \frac{d^N \phi(\omega)}{d\omega^N} d\omega
%\label{eqn:B.420a}\\
%&=&
=
-(\hat{I} \sigma)(\omega)\phi(\omega) |_{\omega=0}
+R_N (t),
\label{eqn:B.420b}
\end{equation}
for all $N\geq 1$ with
\begin{equation}
R_N (t)=(-1)^N
\int_0^\infty (\hat{I}^N \sigma)(\omega) d^N \phi(\omega)/d\omega^N d\omega,
\label{eqn:B.425}
\end{equation}
where we used the fact that
$d^k \phi(\omega)/d\omega^k =\delta_{k0}$ at $\omega=0$ for any $k \geq 0$, and
$d^k \phi(\omega)/d\omega^k =0$ at $\omega=\infty$ for any $k \geq 0$.
We now choose $c=\omega-i\infty$ and change the variable as $s:=\omega-i\eta$, which leads to
\begin{equation}
(\hat{I}\sigma)(\omega)
=i e^{-it\omega} \int_0^\infty (\omega-i\eta)^{-1} [\log (\omega-i\eta)]^{-q} e^{-t\eta} d\eta.
\label{eqn:B.530}
\end{equation}
Then, we can use the dominated convergence theorem to obtain
\begin{equation}
\lim_{\omega \to +0}
(\hat{I}\sigma)(\omega)
=i \int_0^\infty (-i\eta)^{-1} [\log (-i\eta)]^{-q} e^{-t\eta} d\eta.
\label{eqn:B.535}
\end{equation}
Here we used the fact that there is a positive number $\omega_0 <1/(e^{q}\sqrt{2})$ such that 
\begin{eqnarray}
|(\omega-i\eta)^{-1} [\log (\omega-i\eta)]^{-q} e^{-t\eta} |
&\leq&
\left\{
\begin{array}{cc}
\eta^{-1} |\log \eta |^{-q} e^{-t\eta} &
(0<\eta <\omega_0)\\
C e^{-t\eta} &(\omega_0 \leq \eta )
\end{array}
\right.
\nonumber\\
&\leq&
\bigl[\chi_{\eta\leq \omega_0}(\eta) \eta^{-1} |\log \eta |^{-q} +C\bigr]e^{-t\eta} ,
\label{eqn:B.537b}
\end{eqnarray}
for all $0<\omega \leq\omega_0$ and all $0<\eta<\infty$,
where $\chi_{\eta\leq \omega_0}(\eta)=1$ for $\eta\leq \omega_0$ or $0$ otherwise,
and $C$ is an appropriate constant.
The existence of such a $C$ is ensured by the fact that
$\log (\omega-i\eta)$ has no zeros in the rectangular region
$\{ \omega-i\eta | 0<\omega\leq\omega_0, \omega_0\leq \eta<\infty \}$,
and its modulus diverges as $\eta \to \infty$.
The function on the rhs of Eq. (\ref{eqn:B.537b}) is integrable,
so that the use of the dominated convergence theorem is valid.
%We also see that $\lim_{t \to \infty}\lim_{\omega \to +0}(\hat{I}\sigma)(\omega)=0$,
%because $|(-i\eta)^{-1} [\log (-i\eta)]^q e^{-t\eta}|
%\leq [\chi_{\eta\leq 1}(\eta)\eta^{-1} [\log \eta]^q
%+\chi_{\eta> 1}(\eta)(\pi/2)^q ]e^{-t_0\eta}$
%for all $\eta>0$ and $t>t_0>0$, where the function on the rhs is integrable.
To evaluate the asymptotic behavior of Eq. (\ref{eqn:B.535}),
putting $\epsilon =(\log t)^{-1}$ and
$t\eta=\xi $, one obtains
\begin{eqnarray}
&&\hspace*{-10mm}
\lim_{\omega \to +0}
(\hat{I}\sigma)(\omega)
=
i (-\epsilon)^{q}
\int_0^\infty \frac{-i\epsilon^{-1}}{1-q}
\left(\frac{d}{d\xi}[1-\epsilon \log (-i\xi)]^{1-q} \right)e^{-\xi} d\xi
\label{eqn:B.540a}\\
&&\hspace*{-10mm}
=
- \frac{(-\epsilon)^{q-1}}{1-q}
\int_0^\infty 
[1-\epsilon \log (-i\xi)]^{1-q} e^{-\xi} d\xi
\label{eqn:B.540b}\\
%&&\hspace*{-10mm}
%=
%- \frac{(-\epsilon)^{q-1}}{1-q}
%\int_0^\infty
%\biggl[
%1+
%\sum_{j=1}^{N-1}
%\frac{(q-1)q\cdots (q-1+j-1)}{j!}\epsilon^j
%[\log (-i\xi)]^j
%\nonumber\\
%&&\hspace*{-5mm}
%+
%\epsilon^N
%\frac{(q-1)q\cdots (q-1+N-1)}{(N-1)!}
%\int_0^1
%\frac{(1-u)^{N-1}}{[1-\epsilon u \log (-i\xi)]^{q+N}}
%[\log (-i\xi)]^N
%du
%\biggr]
%e^{-\xi} d\xi
%\label{eqn:B.540c}\\
&&\hspace*{-10mm}
=
- \frac{(-\epsilon)^{q-1}}{1-q}
\biggl\{
\sum_{j=0}^{N-1}
{q+j-2 \choose j}
\epsilon^j
\left( \frac{d}{d\nu} -\frac{\pi}{2}i \right)^j \Gamma(\nu)
%\sum_{k=0}^{j}{j \choose k}\left( \frac{-\pi i}{2} \right)^{j-k}\Gamma^{(k)}(\nu)
|_{\nu=1}
\nonumber\\
&&\hspace*{-5mm}
+
\epsilon^N
N{q+N-2 \choose N}
\int_0^\infty 
\left[
\int_0^1 \frac{(1-u)^{N-1}}{[1-\epsilon u \log (-i\xi)]^{q+N}} du
\right]
[\log (-i\xi)]^N
e^{-\xi} d\xi
\biggr\}
%\nonumber\\
%&&
\label{eqn:B.540e}
\end{eqnarray}
where we used the formulas that 
$f(\epsilon)=\sum_{j=0}^{N-1}\epsilon^j f^{(j)}(0)/j!
+
[\epsilon^N/(N-1)!]\int_0^1 (1-u)^{N-1} f^{(N)}(\epsilon u)du$ 
with $f(\epsilon)=[1-\epsilon \log (-i\xi)]^{1-q}$, 
and 
$\int_{0}^{\infty} x^{\nu-1} e^{-\mu x} (\log x)^j dx
=\partial^j[\mu^{-\nu}\Gamma(\nu)]/\partial \nu^j$. 
The last integral %in the remainder term 
in Eq. (\ref{eqn:B.540e}) turns out to be finite because we have%it is estimated as
\begin{equation}
\left|
\int_0^\infty
\left[
\int_0^1 \frac{(1-u)^{N-1}}{[1-\epsilon u \log (-i\xi)]^{q+N}} du
\right]
[\log (-i\xi)]^N
e^{-\xi} d\xi
\right|
\leq
\int_0^\infty
\frac{[(\pi/2)^2+(\log \xi)^2]^{q/2+N}}{(\pi/2)^{q+N}} e^{-\xi} d\xi,
\label{eqn:B.545c}
\end{equation}
where we used that
\begin{equation}
|1-\epsilon u \log (-i\xi)|^2
=
(1-\epsilon u \log \xi)^2 +(\epsilon u \pi/2)^2
%\label{eqn:B.547a}\\
%&=&
%[(\log \xi)^2+(\pi/2)^2 ]\left[ \epsilon u -\frac{\log \xi}{(\log \xi)^2+(\pi/2)^2 } \right]^2
%-\frac{(\log \xi)^2}{(\log \xi)^2+(\pi/2)^2 }+1
%\label{eqn:B.547b}\\
%&\geq&
\geq
\frac{(\pi/2)^2}{(\pi/2)^2+(\log \xi)^2} .
\label{eqn:B.547c}
\end{equation}
Let us now evaluate the upperbound of $|(\hat{I}^N \sigma)(\omega)|$.
%in Eq. (\ref{eqn:B.450a}) for $p=-1$ and $q\leq -2$:
Using the estimation (\ref{eqn:B.537b}), we have
\begin{eqnarray}
|(\hat{I}^N \sigma)(\omega)|
&=&
\frac{1}{(N-1)!}
\left|
\int_0^\infty (i\eta)^{N-1}(\omega-i\eta)^{-1} [\log (\omega-i\eta)]^{-q} e^{-t\eta} d\eta
\right|
\label{eqn:B.550a}\\
%&\leq&
%\frac{1}{(N-1)!}
%\int_0^\infty \eta^{N-1}
%[\chi_{\eta\leq \omega_0}(\eta) \eta^{-1} |\log \eta |^{-q} +C]e^{-t\eta}d\eta
%\label{eqn:B.550b}\\
&\leq&
\frac{1}{(N-1)!}
\int_0^\infty
[\eta^{N-2} |\log \omega_0|^{-q} +C\eta^{N-1} ]e^{-t\eta}d\eta
=O(t^{-N+1}).
\label{eqn:B.550bb}
%&=&
%\frac{1}{(N-1)!}
%\left[\frac{(N-2)!|\log \omega_0|^{-|q|}}{t^{N-1}}+\frac{(N-1)!C}{t^{N}} \right].
%\label{eqn:B.550bbb}
\end{eqnarray}
Substituting this result into the error term $R_N (t)$ in Eq. (\ref{eqn:B.425}), we see that
$R_N (t)=O(t^{-N+1})$ for any integer $N\geq2$.
This proves the statement of the lemma.
\qed

%%%%%%%%%%%%%%%%%%%%%%%%%%%%%%%%%%%%%%%%%%%%%%%%%%%%%%%%%%%%%%%%%%%%


\begin{thebibliography}{99}
\bibitem{innerproduct}
As is usual, the inner product between $\ket{c}$ and $\ket{d} \in \mathbb{C}^N$
is defined by $\braket{d}{c}:=\sum_{n=1}^{N} d_n^* c_n$,
and then the norm of $\ket{c}$ is defined by $\| c \|=\|\ket{c}\|=\sqrt{\braket{c}{c}}$.
Furthermore,
the inner product between $\ket{f}$ and
$\ket{g} \in L^2 ((0,\infty ))$ is also given by
$\braket{g}{f}:=\int_0^{\infty} g^* (\omega ) f(\omega) d\omega $ and
the norm of $\ket{f}$ is $\| f \|=\|\ket{f}\|=\sqrt{\braket{f}{f}}$.
\bibitem{threshold}
For a clarity of the discussion, 
we here choose the threshold of the energy continuum to be zero; 
however an extension to general cases could be easily done.
\end{thebibliography}
\end{document}